\documentclass[a4paper,11pt]{article}
\pdfoutput=1 
\usepackage{jheppub}
\usepackage{amsmath}
\usepackage{amsfonts}
\usepackage{amssymb}
\usepackage{slashed}
\usepackage{bm}
\usepackage{graphicx}%
\usepackage[T1]{fontenc} 
\usepackage{mathrsfs}
\usepackage{yfonts}

\DeclareMathAlphabet{\mathpzc}{OT1}{pzc}{m}{it}

\newcommand{\ie}{i\epsilon}

\newcommand{\dhd}{{\textstyle d}
\lower.03ex\hbox{\kern-0.38em$^{\scriptstyle-}$}\kern-0.05em{}}
\newcommand{\dbar}{{\textstyle \delta}
\lower.03ex\hbox{\kern-0.38em$^{\scriptstyle-}$}\kern-0.05em{}}
\newcommand{\half}{{1\over 2}}
\newcommand{\bu}{{\bullet}}

\newcommand{\bare}{{\bar e}}
\newcommand{\barf}{{\bar f}}
\newcommand{\barg}{{\bar g}}
\newcommand{\barh}{{\bar h}}
\newcommand{\barj}{{\bar j}}

\newcommand{\bsi}{{\bar \psi}}

\newcommand{\bhi}{{\bar \chi}}

\newcommand{\barB}{{\bar B}}

\newcommand{\Bsi}{{\bar \Psi}}
\newcommand{\Bxi}{{\bar \Xi}}

\newcommand{\cald}{{\cal D}}

\newcommand{\calh}{{\cal H}}

\newcommand{\calo}{{\cal O}}

\newcommand{\calW}{{\cal W}}

\newcommand{\hatp}{{\hat p}}

\newcommand{\hatA}{{\hat A}}
\newcommand{\hatB}{{\hat B}}

\newcommand{\hatF}{{\hat F}}

\newcommand{\hpsi}{{\hat \psi}} 
\newcommand{\hsi}{{\hat \psi}}
\newcommand{\hbsi}{\hat {\bar\psi}}

\newcommand{\tilj}{{\tilde j}} 

\newcommand{\tilq}{{\tilde q}}

\newcommand{\tilA}{{\tilde A}}
\newcommand{\tilB}{{\tilde B}}
\newcommand{\tilC}{{\tilde C}}

\newcommand{\tilF}{{\tilde F}}

\newcommand{\tilJ}{{\tilde J}}

\newcommand{\tilS}{{\tilde S}}

\newcommand{\tsi}{\tilde {\psi}} 
 
\newcommand{\tipsi}{\tilde {\psi}} 
\newcommand{\tigma}{\tilde {\sigma}}

\newcommand{\tartial}{\tilde {\partial}}

\newcommand{\brA}{\breve {A}} 
\newcommand{\breB}{\breve {B}}

\newcommand{\cheV}{\check {V}} 
\newcommand{\cheW}{\check {W}} 
\newcommand{\chvigma}{\check\varsigma}

\newcommand{\vigma}{\varsigma}

\newcommand{\notk}{{\not\! k}}
\newcommand{\notp}{{\not\! p}}

\newcommand{\notA}{{\not\! \!A}}
\newcommand{\notB}{{\not\! \!B}}
\newcommand{\notD}{{\not \!\!D}}

\newcommand{\dd}{{\Delta\Delta}}
\pdfoutput=1
\abstract{
The  Drell-Yan hadronic tensor for electromagnetic (EM) current  is calculated in the Sudakov region $s\gg Q^2\gg q_\perp^2$ 
with ${1\over Q^2}$ accuracy, first at the tree level and then with the double-log  accuracy.
It is demonstrated that in the leading order in $N_c$ the higher-twist quark-quark-gluon TMDs reduce to 
leading-twist TMDs due to QCD equation of motion. The resulting tensor for unpolarized hadrons is 
 EM gauge-invariant and depends on two leading-twist TMDs: $f_1$ responsible for total DY cross section,
 and Boer-Mulders function $h_1^\perp$.  The order-of-magnitude estimates of angular distributions for 
 DY process seem to agree with LHC results at corresponding kinematics.}
\arxivnumber{}
\affiliation{ Physics Department, Old Dominion University, Norfolk, VA 23529, USA and Thomas Jefferson National Accelerator Facility, Newport News, VA 23606, USA}

\emailAdd{balitsky@jlab.org}
\begin{document}

\title{\boldmath Gauge-invariant TMD factorization for Drell-Yan hadronic tensor at small $x$}
\author{I. Balitsky }
\preprint{JLAB-THY-20-3294}
\maketitle

 8
\flushbottom

\section{Introduction\label{aba:sec1}}

\bigskip

The Drell-Yan (DY) process of production of lepton pairs with large invariant
mass in hadronic collisions \cite{Drell:1970wh} is one of the most important tools to study  QCD. From 
experimental viewpoint,  it is 
 a unique source of information
about partonic structure of hadrons \cite{Peng:2014hta}. On the theoretical side, it serves as  a testing ground for
 factorization approaches in various kinematics regions, like the classical collinear factorization \cite{Radyushkin:1977tq, Politzer:1977fi, Kajantie:1977fq, Altarelli:1977kt,   Ellis:1978ty, Efremov:1978xm},  TMD factorization
 \cite{Collins:2011zzd, Collins:1981uw, Collins:1984kg, Ji:2004wu, GarciaEchevarria:2011rb}, and
SCET \cite{Bauer:2001yt,Bauer:2002nz,Becher:2007ty, Becher:2010tm}. 

The differential cross section of DY process is determined by the product of leptonic tensor and hadronic tensor. 
The hadronic tensor $W_{\mu\nu}$  is defined as 
\begin{eqnarray}
\hspace{-1mm}
W_{\mu\nu}(p_A,p_B,q)~&\stackrel{\rm def}{=}&~{1\over (2\pi)^4}\sum_X\!\int\! d^4x~e^{-iqx}
\langle p_A,p_B|J_\mu(x)|X\rangle\langle X|J_\nu(0) |p_A,p_B\rangle
\nonumber\\
~&=&~{1\over (2\pi)^4}\!\int\! d^4x~e^{-iqx}
\langle p_A,p_B|J_\mu(x)J_\nu(0) |p_A,p_B\rangle  
\label{W}.
\end{eqnarray}
where $p_A,p_B$ are hadron momenta, $q$ is the momentum of DY pair, $\sum_X$ denotes the sum over full set of ``out''  states and $J_\mu$ is either electromagnetic  or $Z$-boson current. In this paper I consider only the case of electromagnetic current, the Z-boson case
will be studied in a separate publication.
For unpolarized hadrons, the hadronic tensor $W_{\mu\nu}$  is parametrized by 4 functions, for example in Collins-Soper frame \cite{Collins:1977iv}
\begin{eqnarray}
&&\hspace{-1mm}
W_{\mu\nu}~=~-\big(g_{\mu\nu}-{q_\mu q_\nu\over q^2}\big)(W_T+W_{\dd})-2X_\mu X_\nu W_\dd
\nonumber\\
&&\hspace{33mm}
+~Z_\mu Z_\nu(W_L-W_T-W_\dd)-(X_\mu Z_\nu+X_\nu Z_\mu)W_\Delta
\label{Ws}
\end{eqnarray}
where  $X$, $Z$ are unit vectors orthogonal to $q$ and to each other (their explicit form is presented in Sect. \ref{sec:4DYs}). 

Conventionally, the analysis of hadronic tensor (\ref{W}) in the Sudakov region $q^2\equiv Q^2\gg q_\perp^2$ is performed by 
using TMD factorization. For example,  functions $W_T$ and $W_\dd$ can be represented in a standard
 TMD-factorized way \cite{Collins:2011zzd, Collins:2014jpa} 
\begin{eqnarray}
&&\hspace{-2mm}
W_i~=~\sum_{\rm flavors}e_f^2\!\int\! d^2k_\perp
\cald_{f/A}^{(i)}(x_A,k_\perp)\cald_{f/B}^{(i)}(x_B,q_\perp-k_\perp)C_i(q,k_\perp)
\nonumber\\
&&\hspace{-2mm}
+~{\rm power ~corrections}~+~{\rm Y-terms}
\label{TMDf}
\end{eqnarray}
where $\cald_{f/A}(x_A,k_\perp)$ is the TMD density of  a parton $f$  in hadron $A$
with fraction of momentum $x_A$ and transverse momentum $k_\perp$, $\cald_{f/B}(x_B,q_\perp-k_\perp)$ is a similar quantity for hadron $B$, and  
 coefficient functions $C_i(q,k)$ are determined by the cross section $\sigma(ff\rightarrow \mu^+\mu^-)$  of production of DY pair of invariant mass $q^2$ in the scattering of two partons. 
 
 There is, however,  a problem with Eq. (\ref{TMDf}) for the functions $W_L$ and $W_\Delta$.  
 The reason is that while $W_T$ and $W_\dd$ are determined by 
leading-twist quark TMDs, $W_L$ and $W_\Delta$ start from terms ${q_\perp\over Q}$ and $\sim{q_\perp^2\over Q^2}$determined
by quark-quark-gluon TMDs. The power corrections 
$\sim{q_\perp\over Q}$ were found in Ref. \cite{Mulders:1995dh} more than two decades ago but there was no calculation of power corrections  
$\sim{q_\perp^2\over Q^2}$ until recently. Also, the leading-twist contribution is not gauge invariant.
\footnote{Hereafter gauge invariance of hadronic tensor means electromagnetic (EM) gauge invariance, namely 
that $q^\mu W_{\mu\nu}~=~0$.} 
It is well known from DVCS studies  that check of EM gauge invariance sometimes involves
cancellation of contributions of different twists (see e.g. \cite{Ji:1998pc,Guichon:1998xv, Anikin:2000em,Penttinen:2000dg,Belitsky:2000vx,Radyushkin:2000jy,Radyushkin:2000ap}) so the fact that we need power corrections to check $q^\mu W_{\mu\nu}~=~0$
should not come as a surprise. Still, the absence of gauge invariance may cause discomfort in practical applications of TMD factorization. 

In a recent paper \cite{Balitsky:2017gis} A. Tarasov  and the author calculated power corrections $\sim {q_\perp^2\over Q^2}$ to total DY cross section production which are determined by quark-quark-gluon operators. In this paper I present the result of calculation of symmetric part of $W_{\mu\nu}(q)$ for unpolarized hadrons at large $s\gg Q^2\gg q_\perp^2$ relevant for DY experiments at LHC. The method of calculation is based on the rapidity factorization approach  developed in
  Refs. \cite{Balitsky:2017flc,Balitsky:2017gis}.
The calculations will be performed  in the leading order in perturbation theory, first at the tree level and then in the double-logarithmic
approximation for coefficient functions $C_i(q,k)$. In this paper I consider 
only the production of leptons by virtual photon and  leave the case of Z-boson production  for future publication.

To find all functions in Eq. (\ref{Ws}) we need to have gauge-invariant expression for $W_{\mu\nu}$ in terms of TMDs.
As noted above,  only  $W_T$ and $W_\dd$ come
from leading-twist quark-antiquark TMD while two other structures come from higher-twist quark-antiquark-gluon TMDs.
Fortunately, in the leading
order in $N_c$ the latter are related to the former by QCD equations of motion (\cite{Balitsky:2017gis}, see also Ref. \cite{Mulders:1995dh}). 
Moreover, in the small-$x$ region 
$x_A,x_B\ll 1$ all structures can be expressed by just two leading-twist TMDs - $f_1(x,k_\perp)$ (responsible for the total cross section) and $h_1^\perp(x,k_\perp)$ (the Boer-Mulders function \cite {Boer:1997nt}). The results for four functions in Eq. (\ref{Ws}), presented in next Section, are of the type of Eq. (\ref{TMDf}) with TMDs $f_1(x,k_\perp)$ and/or $h_1^\perp(x,k_\perp)$ and  tree-level coefficient functions constructed of $q$ and $k_\perp$.

The paper is organized as follows.  
In section \ref{sec:Wmunu} I present the resulting gauge-invariant expression for $W_{\mu\nu}$ 
up to ${1\over Q^2}$ terms which is calculated in the rest of the paper.  
In section  \ref{sec:funt}   the TMD factorization is derived
from the rapidity factorization  of  the double functional integral for a cross section of particle production. 
In section \ref{sect:power}  I explain the method of
calculation of power corrections based on approximate solution of classical Yang-Mills equations. 
Using this method, DY hadronic tensor for small $x$  is calculated in Sections  \ref{sec:lhtc}, \ref{sec:tw3first}, and \ref{2ndtype}.
Section \ref{sec:results}  contains results of calculations and  order-of-magnitude estimate of angular coefficients of DY cross section.
The matching of obtained TMDs and coefficient functions $C_i$ in the double-log approximation is discussed in Sect. \ref{sec:match} 
and the last section \ref{sec:coutlook} is devoted to conclusions and outlook.
The necessary technical details can be found in appendices.

\section{Gauge-invariant hadronic tensor \label{sec:Wmunu}}
To set up the stage,   in this Section I present the final result for tree-level DY hadronic tensor. It is determined by two leading-twist TMDs: 
the function $f_1^f(x,k_\perp)$ responsible for the total DY cross section and Boer-Mulders time-odd function $h^\perp_{1}(x,k_\perp)$ 
(the explicit definition of these functions is presented in the Appendix \ref{sec:paramlt}).
The result reads
\begin{equation}
\hspace{-1mm}
W_{\mu\nu}(q)~=~{1\over N_c}\sum_f e_f^2\!\int\!d^2k_\perp
\Big[ F^f(q,k_\perp)W_{\mu\nu}^F(q,k_\perp)
+H^f(q,k_\perp)W_{\mu\nu}^H(q,k_\perp)\Big]
\label{wmunu}
\end{equation}
where $e_f$ are electric charges of quarks, $q=x_A p_A+x_B p_B+q_\perp$ and 
\begin{eqnarray}
&&\hspace{-11mm}
F^f(q,k_\perp)~=~f_1^f\big(x_A,k_\perp\big)\barf_1^f\big(x_B,(q-k)_\perp\big)~+~f_1^f\leftrightarrow\barf_1^f
\nonumber\\
&&\hspace{-11mm}
H^f(q,k_\perp)~=~h^\perp_{1}\big(x_A,k_\perp\big)\barh^\perp_{1}\big(x_B,(q-k)_\perp\big) ~+~h_{1f}^\perp\leftrightarrow\barh_{1f}^\perp
\label{FH}
\end{eqnarray}
The gauge-invariant structures
$W^F_{\mu\nu}$ and $W^H_{\mu\nu}$ are given by
\begin{eqnarray}
&&\hspace{-11mm}W^F_{\mu\nu}(q,k_\perp)~=~-g_{\mu\nu}^\perp+{2(k,q-k)_\perp\over Q^2}g_{\mu\nu}^\parallel
+{2\over Q^2}\big[x_Ap_{A\mu}k^\perp_\nu+x_Bp_{B\mu}(q-k)^\perp_\nu+\mu\leftrightarrow\nu\big]
\nonumber\\
&&\hspace{-11mm}
+{4x_A^2p_{A\mu}p_{A\nu}\over Q^4}k_\perp^2+{4x_B^2p_{B\mu}p_{B\nu}\over Q^4}(q-k)_\perp^2,
\nonumber\\
&&\hspace{-11mm}
m^2W^H_{\mu\nu}(q,k_\perp)~=~-\big[k^\perp_\mu(q-k)^\perp_\nu+k^\perp_\nu(q-k)^\perp_\mu+g_{\mu\nu}^\perp(k,q-k)_\perp\big]-2{g^\parallel_{\mu\nu}\over Q^2}k_\perp^2(q-k)_\perp^2
\nonumber\\
&&\hspace{-11mm}
-~2x_A\big[p_{A\mu}(q-k)^\perp_\nu+\mu\leftrightarrow \nu\big]{k_\perp^2 \over Q^2}
-2x_B\big[p_{B\mu}k^\perp_\nu+\mu\leftrightarrow \nu\big]{(q-k)_\perp^2 \over Q^2}
\nonumber\\
&&\hspace{-11mm}
-~{4x_A^2p_{A\mu}p_{A\nu}\over Q^4}k_\perp^2(k,q-k)_\perp-{4x_B^2p_{B\mu}p_{B\nu}\over Q^4}(q-k)_\perp^2(k,q-k)_\perp\Big\}
\label{wfwh}
\end{eqnarray}
where $g_{\mu\nu}^\perp$ and $g_{\mu\nu}^\parallel$ are transverse and longitudinal parts of metric tensor (the explicit form of our notations is specified in the next Section, see the paragraph including Eq. (\ref{delta})).
It is easy to check that $q^\mu W^F_{\mu\nu}~=~0$ and  $q^\mu W^H_{\mu\nu}~=~0$. As we will see below,  in some of the structures the corrections to Eq. (\ref{wmunu}) are of order $O(x_A)$ and $O(x_B)$ while in others on the top of that
there are corrections $\sim O\big({1\over N_c}\big)$ times some other higher-twist TMDs discussed in Ref. \cite{Balitsky:2017gis}.
It should be also noted that $W^F$ part coincides with the result obtained in Refs. \cite{Nefedov:2018vyt,Nefedov:2020ugj} using 
parton Reggeization approach to DY process \cite{Nefedov:2012cq}.

In the rest of the paper I will  derive the above equations and discuss  their accuracy. Let me mention upfront that since the approximations made in Eq. (\ref{wmunu}) are $x_A,x_B\ll 1$ 
and $q_\perp^2\ll Q^2\simeq x_A x_Bs$,  I 
hope that the results of this paper can be used for studies of DY process at LHC with $Q^2\sim$ 100GeV or less. 
\footnote{The reader should not be confused by using small-$x$ approximation at LHC with $Q\sim100$GeV. One should distinguish between small-$x$ approximation 
and small-$x$ resummation.
In  the kinematics discussed in this paper $x_A\sim x_B\sim 0.1$ so the small-$x$ resummation of 
$\alpha_s\ln x_{A(B)}$ is unnecessary, or better to say, should be done on the par with Sudakov resummation of $\alpha_s\ln{Q^2/q_\perp^2}$, see e.g. Ref. \cite{Marzani:2015oyb}.   On the other hand, if one has an expression like  $f_1(x_A,k_\perp)+x_A f^\perp(x_A,k_\perp)$, one can safely neglect the second term, see the discussion in Appendix \ref{sec:qqgparam}.}
Last but not least, the derivation of
 the above equations is lengthly so the readers interested in final formulas for structures $W_i$ and the discussion of approximations can go directly to Sect. \ref{sec:results}.

\section{TMD factorization from rapidity factorization \label{sec:funt}}

As was mentioned in the Introduction,  to find the TMD formulas of Eq. (\ref{TMDf}) type
I use the  rapidity factorization approach to developed in Refs. \cite{Balitsky:2017flc,Balitsky:2017gis}.
Let me quickly summarize basic ideas of this approach. 
The sum over full set of ``out''  states in Eq. (\ref{W})  can be represented by a double functional integral
\begin{eqnarray}
&&\hspace{-2mm}
(2\pi)^4W_{\mu\nu}(p_A,p_B,q)~=~\sum_X\!\int\! d^4x~e^{-iqx}
\langle p_A,p_B|J_\mu(x)|X\rangle\langle X|J_\nu(0)|p_A,p_B\rangle
\label{dablfun}\\
&&\hspace{-2mm}
=~\lim_{t_i\rightarrow -\infty}^{t_f\rightarrow\infty}\!\!\int \! d^4x~ e^{-iqx}
\!\int^{\tilA(t_f)=A(t_f)}\!\!  D\tilA_\mu DA_\mu \!\int^{\tsi(t_f)=\psi(t_f)}\! D\tilde{\bar\psi}D\tilde{\psi} D\bsi D\psi 
~\Psi^\ast_{p_A}(\vec{\tilA}(t_i),\tipsi(t_i))
\nonumber\\
&&\hspace{-2mm}
\times~\Psi^\ast_{p_B}(\vec{\tilA}(t_i),\tipsi(t_i))e^{-iS_{\rm QCD}(\tilA,\tipsi)}e^{iS_{\rm QCD}(A,\psi)}
\tilde{J}_\mu(x)J_\nu(0)\Psi_{p_A}(\vec{A}(t_i),\psi(t_i))\Psi_{p_B}(\vec{A}(t_i),\psi(t_i)).
\nonumber
\end{eqnarray}
where $J_\mu=\sum_{\rm flavors} e_f\bsi_f\gamma_\mu\psi_f$ is the electromagnetic current.
In this double functional integral the amplitude $\langle X|J_\mu(0)|p_A,p_B\rangle$ is given by the integral over $\psi,A$ fields 
whereas the complex conjugate amplitude $\langle p_A,p_B|J^\mu(x)|X\rangle$ is  represented by the integral over $\tsi,\tilA$ fields. 
Also, $\Psi_p(\vec{A}(t_i),\psi(t_i))$ denotes the proton wave function at the initial time $t_i$ and the boundary conditions
$\tilA(t_f)=A(t_f)$ and $\tsi(t_f)=\psi(t_f)$ reflect the sum over all states $X$, cf. Refs. \cite{Balitsky:1988fi,Balitsky:1990ck,Balitsky:1991yz}.

We use 
Sudakov variables $p=\alpha p_1+\beta p_2+p_\perp$, where $p_1$ and $p_2$ are light-like vectors close to $p_A$ and $p_B$ so that 
$p_A=p_1+{m^2\over s}p_2$ and $p_A=p_1+{m^2\over s}p_2$ with $m$ being the proton mass.
Also, we use the notations $x_\bu\equiv x_\mu p_1^\mu$ and $x_\ast\equiv x_\mu p_2^\mu$ 
for the dimensionless light-cone coordinates ($x_\ast=\sqrt{s\over 2}x_+$ and $x_\bu=\sqrt{s\over 2}x_-$). Our metric is $g^{\mu\nu}~=~(1,-1,-1,-1)$ 
which we will frequently rewrite as a sum of longitudinal part and transverse part: 
\begin{equation}
g^{\mu\nu}~=~g_\parallel^{\mu\nu}+g^{\mu\nu}_\perp~=~{2\over s}\big(p_1^\mu p_2^\nu+p_2^\mu p_1^\nu)+g_\perp^{\mu\nu}
\label{delta}
\end{equation}
Consequently,  $p\cdot q~=~(\alpha_p\beta_q+\alpha_q\beta_p){s\over 2}-(p,q)_\perp$ where $(p,q)_\perp\equiv -p_iq^i$. 
Throughout the paper, the sum over the Latin indices $i$, $j$, ... runs over two transverse components while the sum over Greek indices $\mu$, $\nu$, ... runs over four components as usual.

\begin{figure}[htb]
\begin{center}

\vspace{-4mm}
\includegraphics[width=155mm]{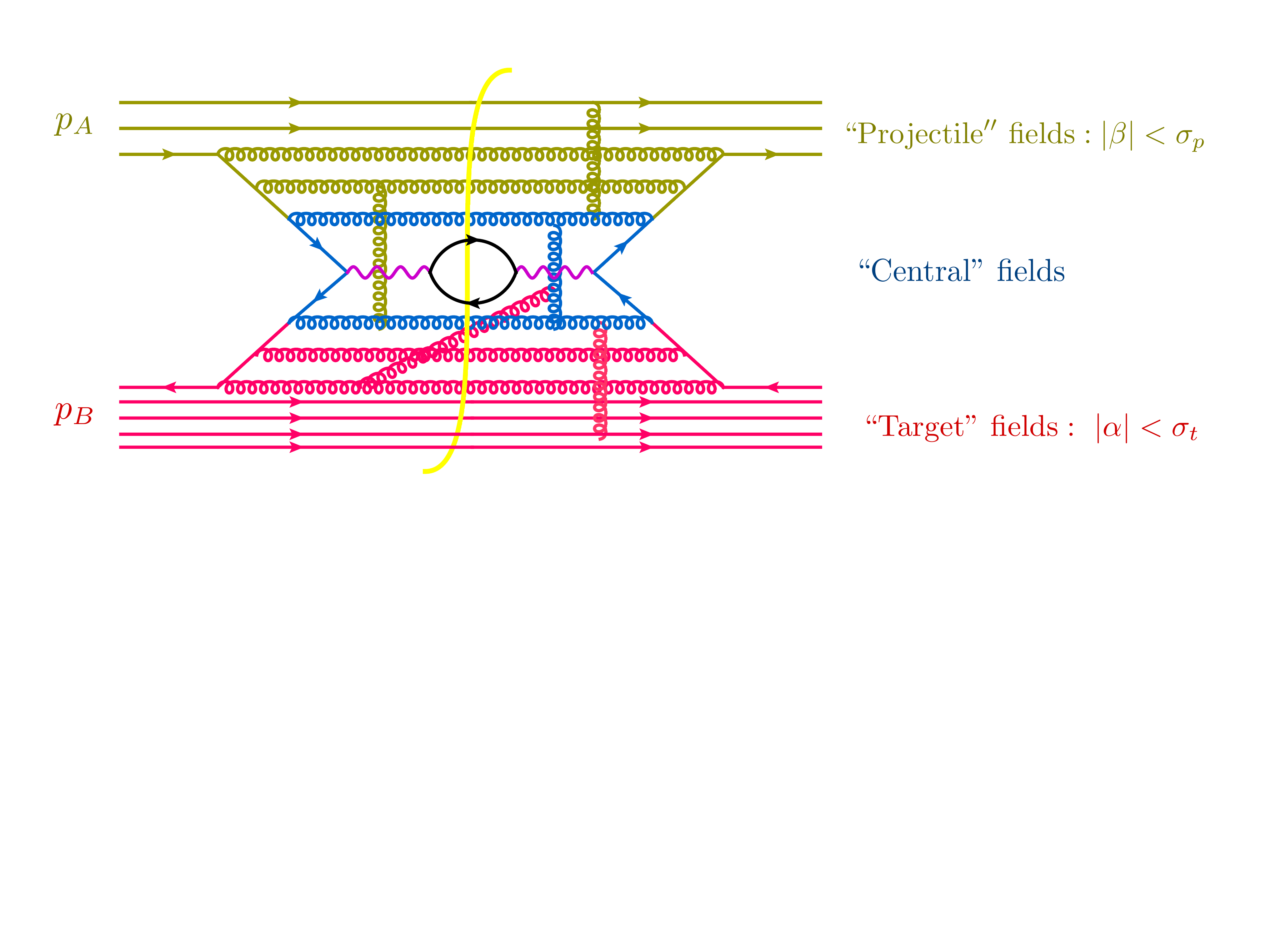}
\end{center}

\vspace{-64mm}
\caption{Rapidity factorization for DY particle production \label{fig:2}}
\end{figure}

Following Ref. \cite{Balitsky:2017flc}  we separate quark and gluon fields in the functional integral (\ref{dablfun}) into three sectors (see figure \ref{fig:2}): 
``projectile'' fields $A_\mu, \psi_A$ 
with $|\beta|<\sigma_p$, 
``target'' fields $B_\mu, \psi_B$ with $|\alpha|<\sigma_t$ and ``central rapidity'' fields $C_\mu,\psi_C$ with $|\alpha|>\sigma_t$ and $|\beta|>\sigma_p$, 
see Fig. \ref{fig:2}. 
\footnote{Although the kinematics is best suited for LHC collider, I call $A$ hadron  ``projectile''  and $B$  hadron 
``target'' for convenience.}
Our goal is to integrate over central fields and get
the amplitude in the factorized form, i.e. as a product of functional integrals over $A$ fields representing projectile matrix elements (TMDs of the projectile) 
and functional integrals over $B$ fields representing target matrix elements (TMDs of the target).
In the spirit of background-field method, we ``freeze'' projectile and target fields and get a sum of diagrams in these external fields. 
Since  $|\beta|<\sigma_p$ in the projectile fields and $|\alpha|<\sigma_t$  in the target fields, at the  tree level 
one can set with power accuracy $\beta=0$ for the  projectile fields and $\alpha=0$ for the target fields - the corrections will
be $O\big({m^2\over\sigma_p s}\big)$ and  $O\big({m^2\over\sigma_t s}\big)$.
Beyond the tree level, the integration over $C$ fields produces
 logarithms of the cutoffs $\sigma_p$ and $\sigma_t$ which match  the corresponding
logs in TMDs of the projectile and the target, see the discussion in Sect. \ref{sec:match}

From integrals over projectile and target fields in the above equation we see that 
the functional integral over $C$ fields  should
be done in the background of $A$ and $B$ fields satisfying 
\begin{equation}
\tilA(t_f)~=~A(t_f),~~~\tipsi_A(t_f)~=~\psi_A(t_f)~~{\rm and} ~~~\tilB(t_f)~=~B(t_f),~~~\tipsi_B(t_f)~=~\psi_B(t_f).
\label{baukon}
\end{equation}
Combining this with our approximation
that at the tree level $\beta=0$ for $A$, $\tilA$ fields and $\alpha=0$ for $B$, $\tilB$ fields, 
which corresponds to  $A=A(x_\bu,x_\perp),~\tilA=\tilA(x_\bu,x_\perp)$ and $B=B(x_\ast,x_\perp),~\tilB=\tilB(x_\ast,x_\perp)$, 
we see that for the purpose of calculation of the functional integral over central fields we can set 
\begin{eqnarray}
&&
A(x_\bu,x_\perp)=\tilA(x_\bu,x_\perp),~~~~\psi_A(x_\bu,x_\perp)=\tipsi_A(x_\bu,x_\perp)
\nonumber\\
&&
{\rm and}
\nonumber\\
&&
B(x_\ast,x_\perp)=\tilB(x_\ast,x_\perp),~~~~\psi_B(x_\ast,x_\perp)=\tipsi_B(x_\ast,x_\perp).
\label{baukond}
\end{eqnarray}
In other words, since $A$, $\psi$ and $\tilA$, $\tipsi$ do not depend on $x_\ast$, if they coincide at $x_\ast=\infty$ they coincide everywhere.
Similarly, since $B$, $\psi_B$ and $\tilB$, $\tipsi_B$ do not depend on $x_\bu$, if they coincide at $x_\bu=\infty$ they should be equal.

Summarizing, we see that at the tree level in our approximation
\begin{eqnarray}
&&\hspace{-1mm}
\int\!DC_\mu\! \int^{\tilC(t_f)=C(t_f)} D\tilC_\mu \!\int\!D\bsi_C D\psi_C \int^{\tsi_C(t_f)=\psi_C(t_f)}~D\tilde{\bar\psi}_C D\tsi_C ~ \tilJ_\mu (x)J_\nu(0)~e^{-i\tilS_C
+iS_C}
\nonumber\\
&&\hspace{-1mm}
=~
\calo(q,x;A,\psi_A;B,\psi_B),
\label{funtc}
\end{eqnarray}
where now $S_C~=~S_{\rm QCD}(C + A+B, \psi_C + \psi_A + \psi_B)-S_{\rm QCD}(A, \psi_A)-S_{\rm QCD}(B, \psi_B)$ and 
$\tilS_C~=~S_{\rm QCD}(\tilC+A+B, \tilde{\psi}_C + \psi_A + \psi_B)-S_{\rm QCD}(A, \psi_A)-S_{\rm QCD}(B, \psi_B)$. It is well known that 
in the tree approximation the double functional integral (\ref{funtc}) is given by a set of 
retarded Green functions in the background fields \cite{Gelis:2003vh,Gelis:2006yv,Gelis:2007kn} (see also appendix A of ref. \cite{Balitsky:2017flc} for the proof).
Since the double functional integral (\ref{funtc}) is given by a set of retarded Green functions 
in the background fields $A$ and $B$, the calculation of the tree-level contribution to $\bsi\gamma_\mu\psi$  
in the r.h.s. of Eq. (\ref{funtc}) is equivalent to solving YM equation for 
$\psi(x)$ and $A_\mu(x)$ with initial condition that the solution has the same asymptotics at $t\rightarrow -\infty$ 
as the superposition of incoming projectile and  target background fields. 

The hadronic tensor  (\ref{W}) can now be represented as
\begin{eqnarray}
&&\hspace{-1mm}  
W_{\mu\nu}(p_A,p_B,q)~=~{1\over (2\pi)^4}\!\int \! d^4x ~e^{-iqx}
 \langle p_A|\langle p_B| \hat\calo_{\mu\nu}(q,x;\hatA,\hsi_A;\hatB,\hsi_B)|p_A\rangle |p_B\rangle,   
\label{W4}
\end{eqnarray}
where $\hat \calo_{\mu\nu}(q,x;\hatA,\hsi_A;\hatB,\hsi_B)$ should be expanded in a series in $\hatA$, $\hsi_A$, $\hatB$, $\hsi_B$ operators
and evaluated between the corresponding (projectile or target) states: if
\begin{equation}
\hat\calo_{\mu\nu}(q,x;\hatA,\hsi_A;\hatB,\hsi_B)
~=~\sum_{m,n}\! \int\! dz_mdz'_n c_{m,n}(q,x)\hat\Phi_A(z_m)\hat\Phi_B(z'_n)
\end{equation}
where $c_{m,n}$ are coefficients and $\Phi$ can be any of $A_\mu$, $\psi$ or $\bsi$ with appropriate Lorentz indices. We get then
\begin{equation}
\hspace{-1mm}  
W_{\mu\nu}~=~\frac{1}{(2\pi)^4}\!\int \! d^4x  e^{-iqx}
\sum_{m,n}\! \int\! dz_m c_{m,n}(q,x)
\langle p_A|\hat\Phi_A(z_m)|p_A\rangle\!\int\! dz'_n\langle p_B| \hat\Phi_B(z'_n)|p_B\rangle.   
\label{W5}
\end{equation}
As we will demonstrate below, the relevant operators $\hat\Phi_A$ and $\hat\Phi_B$ are  quark and gluon fields with Wilson-line type gauge links 
collinear to either $p_2$ for $A$ fields or  $p_1$ for $B$ fields.

\section{Power corrections and solution of classical YM equations \label{sect:power}}
\subsection{Power counting for background fields}
As we discussed in previous section, to get the hadronic tensor in the form  (\ref{W4}) we need to calculate 
the functional integral (\ref{funtc}) in the background of the fields (\ref{baukond}). Since we integrate over fields 
(\ref{baukond}) afterwards, we may assume that they satisfy Yang-Mills equations
\footnote{As was mentioned above, for the purpose of calculation of integral over $C$ fields the projectile and target fields are
``frozen''.}
\begin{eqnarray}
&&\hspace{-1mm}
i\slashed{D}_A \psi_A~=~0,~~~D_A^\nu A_{\mu\nu}^a~=~g^2\sum_f\bar\psi^f_A\gamma_\mu t^a\psi^f_A,
\nonumber\\
&&\hspace{-1mm}
i\slashed{D}_B \psi_B~=~0,~~~D_B^\nu B_{\mu\nu}^a~=~g^2\sum_f\bar\psi^f_B\gamma_\mu t^a\psi^f_B,  
\label{YMs}
\end{eqnarray}
where $A_{\mu\nu}\equiv\partial_\mu A_\nu-\partial_\nu A_\mu-i[A_\mu, A_\nu]$, 
$D_A^\mu\equiv (\partial^\mu-i[A^\mu,)$ and similarly for $B$ fields. 
\footnote{Since we are dealing with tree approximation and quark equations of motion, it is convenient to include coupling constant $g$
in the definition of gluon fields.}

It is convenient to choose a gauge where $A_\ast=0$ for projectile fields and $B_\bu=0$ for target fields.
(The existence of such gauge was proved in appendix B of Ref. \cite{Balitsky:2017flc} by explicit construction.)
The relative 
strength of Lorentz components of projectile and target fields in this gauge was found in ref. \cite{Balitsky:2017flc}
\begin{eqnarray}
&&\hspace{-1mm}
\slashed{p}_1\psi_A(x_\bu,x_\perp)~\sim~m_\perp^{5/2}, ~~~\gamma_i\psi_A(x_\bu,x_\perp)~\sim~m_\perp^{3/2}, ~~~~~
\slashed{p}_2\psi_A(x_\bu,x_\perp)~\sim~s\sqrt{m_\perp},
\nonumber\\
&&\hspace{-1mm}
\slashed{p}_1\psi_B(x_\ast,x_\perp)~\sim~s\sqrt{m_\perp}, ~~~\gamma_i\psi_B(x_\ast,x_\perp)~\sim~m_\perp^{3/2}, ~~~~~
\slashed{p}_2\psi_B(x_\ast,x_\perp)~\sim~m_\perp^{5/2},
\nonumber\\
&&\hspace{-1mm}
A_\bu(x_\bu,x_\perp)~\sim~B_\ast(x_\ast,x_\perp)~\sim~m_\perp^2,~~~~~~A_i(x_\bu,x_\perp)~\sim~B_i(x_\ast,x_\perp)~\sim m_\perp.
\label{fildz}
\end{eqnarray}
Here $m_\perp$ is a scale of order of $m$ or $q_\perp$.
As discussed in Refs.  \cite{Balitsky:2017flc,Balitsky:2017gis},  our rapidity factorization (\ref{W5})  is applicable 
in the region where $s,Q^2\gg q_\perp^2,m^2$, 
while the relation between $q_\perp^2$ and $m^2$ and between $Q^2$ and $s$ may be arbitrary. 
Correspondingly,  for the purpose of counting of powers of $s$, we do not distinguish between $s$ and $Q^2$ so our
power counting will be correct at any Bjorken $x$. 
The distinction  will come at a later time when we specify to small $x$ and 
disregard ${1\over s}$ in comparison to ${1\over Q^2}$ in final expressions for TMDs and/or coefficient functions.  Similarly, for the 
purpose of power counting we will not distinguish between $m$ and $q_\perp$ so we
introduce $m_\perp$ which may be of order of $m$ or $q_\perp$ depending on matrix element.

Note also that in our gauge
\begin{eqnarray}
&&\hspace{-5mm}
A_i(x_\bu,x_\perp)~=~{2\over s}\!\int_{-\infty}^{x_\bu}\! dx'_\bu ~A_{\ast i}(x'_\bu,x_\perp)
,~~~~
B_i(x_\ast,x_\perp)~=~{2\over s}\!\int_{-\infty}^{x_\ast}\! dx'_\ast ~B_{\bu i}(x'_\ast,x_\perp)
\label{AfromF}
\end{eqnarray}
where $A_{\ast i}\equiv F^{(A)}_{\ast i}$ and $B_{\bu i}\equiv F^{(B)}_{\bu i}$ are field strengths for $A$ and $B$ fields respectively.

Thus, to find TMD factorization formula with power corrections at the tree level we need to calculate the functional integral (\ref{dablfun}) in the background fields of the strength given by eqs. (\ref{fildz}).

\subsection{Approximate solution of classical equations \label{sect:ApprSol} at $q_\perp^2\ll Q^2$.}

\begin{figure}[htb]
\begin{center}
\includegraphics[width=99mm]{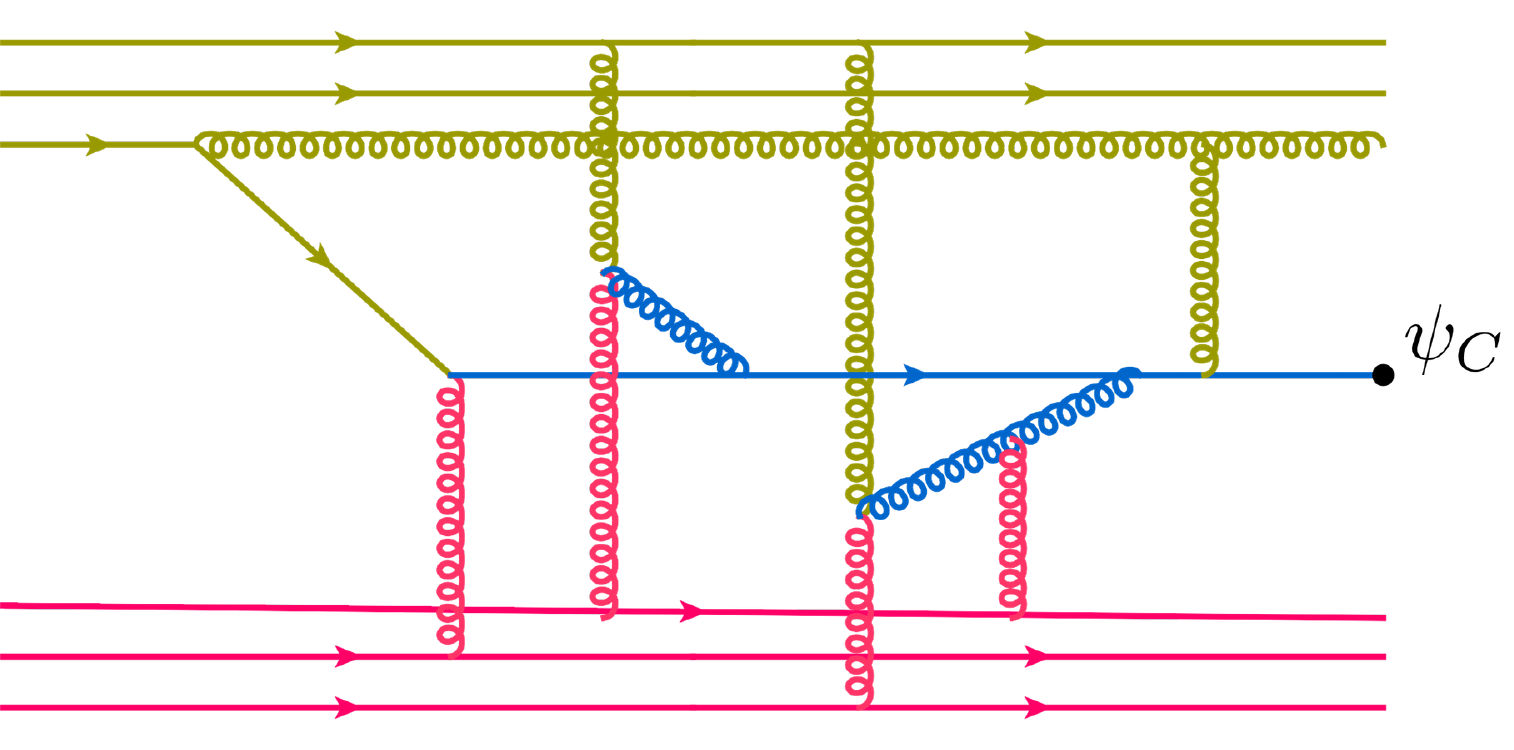}
\end{center}
\caption{Typical diagram for the classical field with projectile/target sources. The Green functions of central fields are given by retarded propagators.   \label{fig:3}}
\end{figure}

As we discussed in section \ref{sec:funt}, the calculation of the functional integral (\ref{funtc}) over $C$-fields 
in the tree approximation reduces to finding fields $C_\mu$ and $\psi_C$ as solutions of Yang-Mills equations for the action
 $S_C~=~S_{\rm QCD}(C+A+B, \psi_C + \psi_A + \psi_B)-S_{\rm QCD}(A, \psi_A)-S_{\rm QCD}(B, \psi_B)$
\begin{eqnarray}
&&\hspace{-1mm}
(i\slashed{\partial}+g\slashed{A}+g\slashed{B}+g\slashed{C})(\psi^f_A+\psi^f_B+\psi^f_C)~=~0,
\label{yd}\\
&&\hspace{-1mm}
D^\nu F_{\mu\nu}^a(A+B+C)~=~g^2\sum_f(\bsi^f_A+\bsi^f_B+\bsi^f_C)\gamma_\mu t^a(\psi^f_A+\psi^f_B+\psi^f_C).
\nonumber
\end{eqnarray}
The solution of eq. (\ref{yd}) which we need corresponds to the sum of set of diagrams
in background field $A + B$ with {\it retarded} Green functions, see figure \ref{fig:3}.
The sum of tree diagrams with retarded Green functions gives fields $C_\mu$ and $\psi_C$ that vanish at $t\rightarrow -\infty$. Thus, we are solving the usual classical YM equations
\footnote{We take into account only $u,d,s,c$ quarks and consider them massless. In principle, one can include ``massless'' $b$-quark for $q_\perp^2\gg m_b^2$.}
\begin{equation}
\mathbb{D}^\nu \mathbb{F}^a_{\mu\nu}~=~g^2\sum_f\bar{\Psi}^f t^a\gamma_\mu \Psi^f,~~~~\slashed{\mathbb{P}}\Psi^f~=~0,
\label{kleqs}
\end{equation}
where
\begin{eqnarray}
&&\mathbb{A}_\mu = C_\mu + A_\mu + B_\mu,\ \ \ \Psi^f = \psi^f_C + \psi^f_A + \psi^f_B,
\nonumber\\
&&\mathbb{P}_\mu~\equiv~i\partial_\mu+C_\mu+A_\mu+B_\mu, ~~~~
\mathbb{F}_{\mu\nu}~=~\partial_\mu \mathbb{A}_\nu-\mu\leftrightarrow\nu-i[\mathbb{A}_\mu, \mathbb{A}_\nu],
\end{eqnarray}
with boundary conditions
\begin{eqnarray}
&&\hspace{-11mm}
\mathbb{A}_\mu(x)\stackrel{x_\ast\rightarrow -\infty}{=}A_\mu(x_\bu,x_\perp),~~~~
\Psi(x)\stackrel{x_\ast\rightarrow -\infty}{=}\psi_A(x_\bu,x_\perp),
\nonumber\\
&&\hspace{-11mm}
\mathbb{A}_\mu(x)\stackrel{x_\bu\rightarrow -\infty}{=}B_\mu(x_\ast,x_\perp),~~~~
\Psi(x)\stackrel{x_\bu\rightarrow -\infty}{=}\psi_B(x_\ast,x_\perp)
\label{inicondi}
\end{eqnarray}
following from $C_\mu,\psi_C\stackrel{t\rightarrow -\infty}{\rightarrow} 0$.
These boundary conditions reflect the fact that at $t\rightarrow -\infty$ we have only incoming hadrons with $A$ and $B$ fields.

As discussed in Ref. \cite{Balitsky:2017flc}, for our case of particle production with ${q_\perp\over Q}\ll 1$ it is possible to find 
the approximate solution of  (\ref{kleqs}) as a series in this small parameter. 
One solves Eqs. (\ref{kleqs}) iteratively,  order by order in perturbation theory,  starting from the 
zero-order approximation in the form of the sum of projectile and target fields
\begin{eqnarray}
&&\hspace{-1mm}
\mathbb{A}_\mu^{[0]}(x)~=~A_\mu(x_\bu,x_\perp)+B_\mu(x_\ast,x_\perp),
\nonumber\\
&&\hspace{-1mm}
\Psi^{[0]}(x)~=~\psi_A(x_\bu,x_\perp)+\psi_B(x_\ast,x_\perp)
\label{trials}
\end{eqnarray}
and improving it by calculation of Feynman diagrams with retarded propagators in the background fields (\ref{trials}).

Let me now explain how the parameter ${m_\perp^2/s}$ comes up in the rapidity-factorization approach
(for details, see Ref. \cite{Balitsky:2017flc}).
When we expand quark and gluon propagators in powers of background fields, 
we get a set of diagrams shown in figure \ref{fig:3}.
 The typical bare gluon propagator in figure \ref{fig:3} is
\begin{equation}
{1\over p^2+i\epsilon p_0}~=~{1\over\alpha\beta s-p_\perp^2+i\epsilon(\alpha+\beta)}.
\label{gluonpropagator}
\end{equation}
In the tree approximation, the  transverse momenta 
in tree diagrams  are determined by further integration over projectile (``A'') and target (``B'') fields 
in eq. (\ref{dablfun}) which converge on  either $q_\perp$ or $m_N$. On the other hand, the integrals over 
 $\alpha$ converge on either $\alpha_q$ or $\alpha\sim 1$ and similarly the characteristic $\beta$'s
 are either $\beta_q$ or $\beta \sim 1$.
Since $\alpha_q\beta_qs=Q_\parallel^2\gg q_\perp^2$, one can expand gluon and quark propagators 
in powers of ${p_\perp^2\over \alpha\beta s}$ 
\begin{eqnarray}
&&
{1\over p^2+i\epsilon p_0}~=~{1\over s(\alpha+i\epsilon)(\beta+i\epsilon)}
\Big(1
+{p_\perp^2/s\over(\alpha+i\epsilon)(\beta+i\epsilon)}+...\Big),
\label{propexpan}\\
&&
{\slashed{p}\over p^2+i\epsilon p_0}~=~{1\over s}\Big({\slashed{p}_1\over \beta+i\epsilon}
+{\slashed{p}_2\over \alpha+i\epsilon}+{\slashed{p}_\perp\over (\alpha+i\epsilon)(\beta+i\epsilon)}\Big)
\Big(1
+{p_\perp^2/s\over(\alpha+i\epsilon)(\beta+i\epsilon)}+...\Big).
\nonumber
\end{eqnarray}
After the expansion (\ref{propexpan}), the dynamics in the transverse space effectively becomes trivial: 
all background fields stand either at $x$ or at $0$.  Note that in this statement is solely a consequence of 
$Q^2\gg q_\perp^2$  and  does not rely on small-$x$ approximation.

\subsection{Power expansion of classical quark fields}

Now we expand the classical  quark fields  in powers of  ${p_\perp^2\over p_\parallel^2}\sim{m_\perp^2\over s}$ 
(the corresponding expansion of classical gluon fields is presented in  Ref. \cite{Balitsky:2017flc}, but we do not need it here). 
As demonstrated in Ref. \cite{Balitsky:2017gis},
expanding it in powers of $p^2_\perp/p_\parallel^2$ we obtain
\begin{eqnarray}
&&\hspace{-1mm}
\Psi(x)~=~\Psi_1(x)+\Psi_2(x)
+\dots,
\label{klfildz}
\end{eqnarray}
where
\begin{eqnarray}
&&\hspace{-5mm}
\Psi_1~=~\psi_A+\Xi_{1},~~~~
\Xi_{1}~=~-{\slashed{p}_2\over s}\gamma^iB_i{1\over \alpha+i\epsilon}\psi_A
~=~{i\over s}\sigma_{\ast i}B^i{1\over \alpha+i\epsilon}\psi_A,
\nonumber\\
&&\hspace{-5mm}
\Bsi_1~=~\bar\psi_A+\Bxi_{1},~~~~
\Bxi_{1}~=~-\big(\bar\psi_A{1\over\alpha-i\epsilon}\big)\gamma^iB_i{\slashed{p}_2\over s}
~=~-{i\over s}\big(\bar\psi_A{1\over\alpha-i\epsilon}\big)B^i\sigma_{\bu i}
\nonumber\\
&&\hspace{-5mm}
\Psi_2~=~\psi_B+\Xi_{2},~~~~
\Xi_{2}~=~-{\slashed{p}_1\over s}\gamma^iA_i{1\over\beta+i\epsilon}\psi_B
~=~{i\over s}\sigma_{\bu i}A^i{1\over\beta+i\epsilon}\psi_B,
\nonumber\\
&&\hspace{-5mm}
\Bsi_2~=~\bar\psi_B+\Bxi_{2},~~~~
\Bxi_{2}~=~-\big(\bar\psi_B{1\over \beta-i\epsilon}\big)\gamma^iA_i{\slashed{p}_1\over s}
~=~-{i\over s}\big(\bar\psi_B{1\over \beta-i\epsilon}\big)A_i\sigma_{\bu i}
\label{fildz0}
\end{eqnarray}
and dots stand for terms subleading in ${q_\perp^2\over Q^2}$ and/or $\alpha_q,\beta_q$ parameters (hereafter we assume the small-$x$ approximation
$\alpha_q,\beta_q\ll 1$ in all calculations).
In this formula
\begin{eqnarray}
&&\hspace{-1mm}
{1\over \alpha+i\epsilon}\psi_A(x_\bu,x_\perp)~\equiv~-i\!\int_{-\infty}^{x_\bu}\! dx'_\bu~\psi_A(x'_\bu,x_\perp),
\nonumber\\
&&\hspace{-1mm}
\Big(\bsi_A{1\over \alpha-i\epsilon}\Big)(x_\bu,x_\perp)~\equiv~i\!\int_{-\infty}^{x_\bu}\! dx'_\bu~\bsi_A(x'_\bu,x_\perp)
\label{3.25}
\end{eqnarray}
and similarly for ${1\over\beta\pm i\epsilon}$. 
For brevity,  in what follows we denote $\big(\bar\psi_A{1\over\alpha}\big)(x)\equiv \big(\bsi_A{1\over \alpha-i\epsilon}\big)(x)$ and
$\big(\bar\psi_B{1\over\beta}\big)(x)\equiv \big(\bsi_B{1\over \beta-i\epsilon}\big)(x)$.
Let us estimate the relative size of corrections $\Xi$ in Eq. (\ref{fildz0}) at small $x$. As we will see, ${1\over\alpha}$ and ${1\over\beta}$ transform
to ${1\over\alpha_q}$ and ${1\over\beta_q}$ in our TMDs so
\begin{eqnarray}
&&\hspace{-5mm}
\Xi_{1}~\sim~\psi_A{m_\perp\over \alpha_q\sqrt{s}}~\sim~\psi_A{q_\perp\over Q},~~~~\Xi_{2}~\sim~\psi_B{m_\perp\over \beta_q\sqrt{s}}~\sim~\psi_B{q_\perp\over Q}
\label{psi0}
\end{eqnarray}
if $\alpha_q\sim\beta_q\sim {Q\over\sqrt{s}}$ (recall that we assume that the DY pair is emitted in the central region of rapidity).
For example, the correction $\sim [\bsi_A\gamma_\mu\Xi_{2}][\bsi_B\gamma_\nu\Xi_{1}]$ will be of order of ${q_\perp^2\over Q^2}$ in comparison to leading-twist contribution $[\bsi_A\gamma_\mu\psi_B][\bsi_B\gamma_\nu\psi_A]$.
\footnote{The reader may wonder why there are no corrections $\sim {q_\perp^2\over Q^2}$ coming from next terms in the expansion (\ref{klfildz})
like $[\bsi_A(x)\gamma_\mu\psi_B(x)][\bsi_B(0)\gamma_\nu{\gamma^i\over s}\slashed{p}_2{1\over \beta}{1\over \alpha}\gamma^j\partial_iB_j\Psi_A(0)]$. The reason is that 
${1\over \beta}$ between $\bsi_B(0)$ and $B_j(0)$  does \underline{not} transform to ${1\over\beta_q}$ and remains $\sim O(1)$, see the discussion 
in the Appendix 8.3.4 of Ref. \cite{Balitsky:2017gis}.
}

\section{Hadronic tensor at  $s\gg Q^2\gg q_\perp^2$\label{sec:lhtc}}

In general, our method is applicable for calculation of power corrections
at any $s,Q^2\gg q_\perp^2,m^2_N$. However, the expressions are greatly simplified 
in the physically interesting case $s\gg Q^2\gg q_\perp^2$ which is considered in this paper.

As we noted above, we take into account only hadronic tensor due to electromagnetic currents of $u,d,s,c$ quarks and consider these quarks to be massless. 
It is convenient to define coordinate-space hadronic tensor multiplied by $N_c{2\over s}$ (and denoted by extra ``check'' mark) as follows
\begin{eqnarray}
&&\hspace{-1mm}
\cheW_{\mu\nu}(p_A,p_B, x)~\equiv~
N_c{2\over s}\langle p_A,p_B|J_\mu(x)J_\nu(0)|p_A,p_B\rangle
\label{defW}\\
&&\hspace{-1mm}
W_{\mu\nu}(p_A, p_B, q)~=~{s/2\over(2\pi)^4N_c}\int\!d^4x ~e^{-iqx} \cheW_{\mu\nu}(p_A,p_B, x).
\nonumber
\end{eqnarray}
For future use, let us also define the hadronic tensor in mixed representation: in momentum longitudinal 
space but in transverse coordinate space
\begin{eqnarray}
&&\hspace{-1mm}
W_{\mu\nu}(p_A, p_B, q)~=~\int\!d^2x_\perp ~e^{i(q,x)_\perp} W_{\mu\nu}(\alpha_q,\beta_q, x_\perp),
\label{defWcoord}\\
&&\hspace{-1mm}
W_{\mu\nu}(\alpha_q,\beta_q,x_\perp)~\equiv~
{1\over(2\pi)^4}{2\over s}\!\int\! dx_\bu dx_\ast ~e^{-i\alpha_qx_\bu-i\beta_q x_\ast}
\langle p_A,p_B|J_\mu(x_\bu,x_\ast,x_\perp)J_\nu(0)|p_A,p_B\rangle.
\nonumber
\end{eqnarray}

After integration over central fields in the tree approximation we obtain
\begin{equation}
\hspace{-1mm}
\cheW_{\mu\nu}(p_A,p_B, x)~\equiv~
N_c{2\over s}\langle A,B|J_\mu(x_\bu,x_\ast,x_\perp)J_\nu(0)|A,B\rangle
\label{4.3}
\end{equation}
where
\begin{eqnarray}
&&\hspace{-1mm}
J^\mu~=~J^\mu_A+J^\mu_B+J^\mu_{AB}+J^\mu_{BA},
\nonumber\\
&&\hspace{-1mm}
J^\mu_A~=~\sum_f e_f\bar\Psi_1^f\gamma^\mu\Psi_1^f,~~~
J^\mu_{AB}~=~\sum_f e_f\bar\Psi_1^f\gamma^\mu\Psi_2^f
\label{jeiz}
\end{eqnarray}
and similarly for $J^\mu_B$ and $J^\mu_{BA}$.  Here $\langle A,B|\calo(\psi_A,A_\mu,\psi_B,B_\mu)|A,B\rangle$ denotes 
double functional integral over $A$ and $B$ fields  which gives matrix elements between projectile and target states of Eq. (\ref{W5}) type.

The leading-twist contribution to $W_{\mu\nu}(q)$  comes only from product $J_{AB}^\mu(x)J_{BA}^\nu(0)$ (or $J_{BA}^\mu(x)J_{AB}^\nu(0)$), 
while power corrections may come also from other terms like $J_A^\mu(x)J_{B}^\nu(0)$. We will consider all terms in turn.

\subsection{Leading-twist contribution and ${1\over Q^2}$ terms from $J_{AB}^\mu(x)J_{BA}^\nu(0)$}
Power expansion of $J_{AB}^\mu(x)J_{BA}^\nu(0)$ reads
\begin{eqnarray}
&&\hspace{-1mm}
\Bsi_1(x)\gamma^\mu\Psi_2(x)\Bsi_2(0)\gamma^\nu\Psi_1(0)
~+~...
\end{eqnarray}
where quark fields are given by Eq. (\ref{fildz0}).
As we mentioned above,  in Ref. \cite{Balitsky:2017gis} it is  demonstrated that terms neglected in the r.h.s.  lead to power corrections $\sim {q_\perp^2\over \alpha_qs}$ or
$\sim {q_\perp^2\over \beta_qs}$ which are much smaller than ${q_\perp^2\over \alpha_q\beta_qs}={q_\perp^2\over Q^2_\|}\sim{q_\perp^2\over Q^2}$ (if DY pair 
is emitted in the central region of rapidity). Note that since we want to calculate the leading power corrections,  we can substitute $Q^2_\parallel$ with $Q^2$.
 In the limit $s\gg Q^2\gg q_\perp^2$ this change of variables can only lead to errors of the order of subleading power terms.
 \footnote{ Except for the leading-twist term where the difference between $Q^2_\parallel$ and $Q^2$ matters.}

As to terms $\sim \Bsi_1(x)\gamma_\mu\Psi_2(x)\Bsi_2(0)\gamma_\nu\Psi_1(0)$, they can be decomposed
using eq.  (\ref{fildz0}) as follows:
\begin{eqnarray}
&&\hspace{-1mm}
\big[\big(\bar\psi_A +\Bxi_{1}\big)(x)\gamma_\mu\big(\psi_B+\Xi_{2}\big)(x)\big]
[\big(\bar\psi_B+\Bxi_{2}\big)(0)\gamma_\nu\big(\psi_A+\Xi_{1}\big)(0)\big]~+~x\leftrightarrow 0
\nonumber\\
&&\hspace{-1mm}
=~[\bar\psi_A(x)\gamma_\mu\psi_B(x)\big]\big[\bar\psi_B(0)\gamma_\nu\psi_A(0)\big]
\label{7lines}\\
&&\hspace{-1mm}
+~[\Bxi_{1}(x)\gamma_\mu\psi_B(x)\big]\big[\bar\psi_B(0)\gamma_\nu\psi_A(0)\big]
+[\bar\psi_A(x)\gamma_\mu\Xi_{2}(x)\big]\big[\bar\psi_B(0)\gamma_\nu\psi_A(0)\big]
\nonumber\\
&&\hspace{-1mm}
+~
[\bar\psi_A(x)\gamma_\mu\psi_B(x)\big]\big[\Bxi_{2}(0)\gamma_\nu\psi_A(0)\big]
+[\bar\psi_A(x)\gamma_\mu\psi_B(x)\big]\big[\bar\psi_B(0)\gamma_\nu\Xi_{1}(0)\big]
\nonumber\\
&&\hspace{-1mm}
+~[\Bxi_{1}(x)\gamma_\mu\psi_B(x)\big]\big[\bar\psi_B(0)\gamma_\nu\Xi_{1}(0)\big]
+[\bar\psi_A(x)\gamma_\mu\Xi_{2}(x)\big]\big[\Bxi_{2}(0)\gamma_\nu\psi_A(0)\big]
\nonumber\\
&&\hspace{-1mm}
+~[\Bxi_{1}(x)\gamma_\mu\psi_B(x)\big]\big[\Bxi_{2}(0)\gamma_\nu\psi_A(0)\big]
+[\bar\psi_A(x)\gamma_\mu\Xi_{2}(x)\big]\big[\bar\psi_B(0)\gamma_\nu\Xi_{1}(0)\big]
\nonumber\\
&&\hspace{-1mm}
+~[\Bxi_{1}(x)\gamma_\mu\Xi_{2}(x)\big]\big[\bar\psi_B(0)\gamma_\nu\psi_A(0)\big]
+[\bar\psi_A(x)\gamma_\mu\psi_{B}(x)\big]\big[\Bxi_{2}(0)\gamma_\nu\Xi_{1}(0)\big]
~+~x\leftrightarrow 0.
\nonumber
\end{eqnarray}
where the square brackets mean trace over Lorentz and color indices. 

First, let us consider the leading-twist term coming from the first term in the r.h.s. of this equation.

\subsection{Leading-twist contribution}
As we mentioned, the leading-twist term comes from $J_{AB}^\mu(x)J_{BA}^\nu(0)$ and \\
$J_{BA}^\mu(x)J_{AB}^\nu(0)$. Using Fierz transformation (\ref{fierz}) one obtains 
\begin{eqnarray}
&&\hspace{-1mm}
{N_c\over s}\big(\big[\bar\psi_A(x_\bu,x_\perp)\gamma_\mu\psi_B(x_\ast,x_\perp)\big]\big[\bar\psi_B(0)\gamma_\nu\psi_A(0)\big]+\mu\leftrightarrow\nu\big)
~+~x\leftrightarrow 0
\nonumber\\
&&\hspace{-1mm}
=~{g_{\mu\nu}\over 2s}\big[-(\bsi_A\psi_A)(\bsi_B\psi_B)+(\bsi_A\gamma_5\psi_A)(\bsi_B\gamma_5\psi_B)
+(\psi_A\gamma_\alpha\psi_A)(\bsi_B\gamma^\alpha\psi_B)
\nonumber\\
&&\hspace{-1mm}
+~(\psi_A\gamma_\alpha\gamma_5\psi_A)(\bsi_B\gamma^\alpha\gamma_5\psi_B)
-\half(\psi_A\sigma^{\alpha\beta}\psi_A)(\bsi_B\sigma_{\alpha\beta}\psi_B)\big]
\nonumber\\
&&\hspace{-1mm}
-~{1\over 2s}[(\psi_A\gamma_\mu\psi_A)(\bsi_B\gamma_\nu\psi_B)+\mu\leftrightarrow\nu]
-{1\over 2s}(\psi_A\gamma_\mu\gamma_5\psi_A)(\bsi_B\gamma_\nu\gamma_5\psi_B)+\mu\leftrightarrow\nu]
\nonumber\\
&&\hspace{-1mm}
+~{1\over 2s}[(\psi_A\sigma_{\nu\alpha}\psi_A)(\bsi_B\sigma_{\mu\alpha}\psi_B)
+(\psi_A\sigma_{\mu\alpha}\psi_A)(\bsi_B\sigma_{\nu\alpha}\psi_B)]~+~x\leftrightarrow 0
\label{ltfierz}
\end{eqnarray}
where all parentheses in the r.h.s. are color singlet.
As usual, after integration over background fields $A$ and $B$ we promote $A$, $\psi_A$ and $B$, $\psi_B$ to operators $\hatA$, $\hat\psi$.
A subtle point is that our operators are not under T-product ordering so one should be careful while changing the order of operators in
 formulas like Fierz transformation. Fortunately, all operators in the r.h.s of Eq. (\ref{ltfierz}) are separated either by space-like intervals or light-like intervals so they commute with each other.

From parametrization of two-quark operators in section \ref{sec:paramlt},
it is clear that the leading-twist contribution to $W_{\mu\nu}(q)$  comes from 
\begin{eqnarray}
&&\hspace{-1mm}
\cheW^{\rm lt}_{\mu\nu}~=~
{1\over 2s}(g_{\mu\nu}g^{\alpha\beta}-\delta_\mu^\alpha\delta_\nu^\beta-\delta_\nu^\alpha\delta_\mu^\beta)
\langle\hbsi(x_\bu,x_\perp)\gamma_\alpha\hsi(0)\rangle_A\langle\hbsi(0)\gamma_\beta\hsi(x_\ast,x_\perp)\rangle_B
\label{cheklt}\\
&&\hspace{-1mm}
+~{1\over 2s}\big(\delta_\mu^\alpha\delta_\nu^\beta+\delta_\nu^\alpha\delta_\mu^\beta-\half g_{\mu\nu}g^{\alpha\beta}\big)
\langle\hbsi(x_\bu,x_\perp)\sigma_{\alpha\xi}\hsi(0)\rangle_A\langle\hbsi(0)\sigma_\beta^{~\xi}\hsi(x_\ast,x_\perp)\rangle_B
~+~x\leftrightarrow 0
\nonumber
\end{eqnarray}
Hereafter we use notations $\langle\calo\rangle_A\equiv\langle p_A|\calo|p_A\rangle$ and 
$\langle\calo\rangle_B\equiv\langle p_B|\calo|p_B\rangle$ for brevity\footnote{
In a general gauge for
projectile and target fields these matrix elements read 
\begin{eqnarray}
&&\hspace{-2mm}
\langle p_A|\hsi_{f}(x)\gamma_\mu \hsi_{f}(0)|p_A\rangle
~=~\langle p_A|\hsi_f(x_\bu,x_\perp)\gamma_\mu[x_\bu,-\infty_\bu]_x[x_\perp,0_\perp]_{-\infty_\bu}[-\infty_\bu,0_\bu]_0\hsi_f(0)|p_A\rangle,
\nonumber\\
&&\hspace{-2mm}
\langle p_B|\hsi_{f}(x)\gamma_\mu \hsi_{f}(0)|p_B\rangle
~=~\langle p_B|\hsi_f(x_\ast,x_\perp)\gamma_\mu[x_\ast,-\infty_\ast]_x[x_\perp,0_\perp]_{-\infty_\ast}[-\infty_\ast,0_\ast]_0\hsi_f(0)|p_B\rangle
\label{gaugelinks}
\end{eqnarray}
and similarly for other operators.
}. The corresponding leading-twist contribution to  to $W_{\mu\nu}(q)$  
has the form \cite{Tangerman:1994eh}
\begin{eqnarray}
&&\hspace{-1mm}W_{\mu\nu}^{\rm lt}(\alpha_q,\beta_q,q_\perp)
~=~{1\over 16\pi^4N_c}\!\int\! dx_\bu dx_\ast d^2x_\perp~e^{-i\alpha_qx_\bu-i\beta_qx_\ast+i(q,x)_\perp}\cheW^{\rm lt}_{\mu\nu}(x)
\nonumber\\
&&\hspace{-1mm}
~=~\sum_f e_f^2{1\over N_c}\!\int\! d^2k_\perp\Big(-g_{\mu\nu}^\perp\big[f_1^f(\alpha_q,k_\perp)\barf_1^f(\beta_q,q_\perp-k_\perp) 
+\barf_1^f(\alpha_q,k_\perp)f_1^f(\beta_q,q_\perp-k_\perp)\big]
\nonumber\\
&&\hspace{17mm}
-~\big[k^\perp_\mu(q-k)^\perp_\nu+k^\perp_\nu(q-k)^\perp_\mu+g_{\mu\nu}^\perp(k,q-k)_\perp\big]
\nonumber\\
&&\hspace{22mm}
\times~\big[h^\perp_{1f}(\alpha_q,k_\perp)\barh^\perp_{1f}(\beta_q,q_\perp-k_\perp)
+\barh^\perp_{1f}(\alpha_q,k_\perp)h^\perp_{1f}(\beta_q,q_\perp-k_\perp)\big]\Big)
\label{WLT}
\end{eqnarray}

Let us discuss other terms proportional to different TMDs in parametrizations in Sect.  \ref{sec:paramlt}. To this end, we write down terms from Eq. (\ref{wfwh}) that we are looking for in Sudakov variables:
\begin{eqnarray}
&&\hspace{-1mm}
g_\perp^{\mu\nu}\Big[1+{q_\perp^2\over \alpha_q\beta_qs}\Big],~~ {q_\perp^\mu q_\perp^\nu\over q_\perp^2}
\Big[1+{q_\perp^2\over \alpha_q\beta_qs}\Big],~~g_\parallel^{\mu\nu}\Big[0+{q_\perp^2\over \alpha_q\beta_qs}\Big],~~
{2\over \alpha_qs}\big(p_2^\mu q_\perp^\nu+\mu\leftrightarrow\nu\big),~~
\nonumber\\
&&\hspace{-1mm}
{2\over \beta_qs}\big(p_1^\mu q_\perp^\nu+\mu\leftrightarrow\nu\big),~~
{p_1^\mu p_1^\nu \over \beta_q^2s^2},~~
{p_2^\mu p_2^\nu\over \alpha_q^2s^2}
\label{pc}
\end{eqnarray}
Here zero in the third term means that the contribution of order one is actually absent. 
As discussed in  Sect.  \ref{sec:paramlt}, all TMDs considered here can have only logarithmic dependence on Bjorken $x$ ($\equiv\alpha_q$ or $\beta_q$)
but \underline{not} the power dependence ${1\over x}$. It is easy to see that other quark-antiquark TMDs give contributions to $W_{\mu}(q)$ which look like terms in Eq. (\ref{pc}) but without extra ${1\over\alpha_q}$ and/or  ${1\over\beta_q}$ so they are power suppressed in low-x regime
$s\gg Q^2$.

Let us also specify the terms which we do not calculate.  Roughly speaking, they correspond to terms in Eq. (\ref{pc}) multiplied by 
${m_\perp^2\over Q^2}$ or by either $\alpha_q$ or $\beta_q$. Our strategy in the next sections is to compare a certain  term in 
$\cheW_{\mu\nu}$ to terms in Eq. (\ref{pc}), and, if it is smaller, neglect, if it is of the same size, calculate. 

\section{Terms coming from $J_{AB}^\mu(x)J_{BA}^\nu(0)$ \label{sec:tw3first}}
We separate terms in Eq. (\ref{7lines}) according to number of gluon fields (contained in  $\Xi$'s ).
\begin{equation}
\cheW_{\mu\nu}~\stackrel{{\rm sym}~\mu,\nu}{=}~\cheW_{\mu\nu}^{\rm lt}+\cheW_{\mu\nu}^{(1)}+\cheW_{\mu\nu}^{(2a)}+\cheW_{\mu\nu}^{(2b)}
+\cheW_{\mu\nu}^{(2c)}
\end{equation}
where leading-twist terms without gluons (quark-antiquark TMDs) were considered in previous Section, and
\begin{eqnarray}
&&\hspace{-1mm}
\cheW_{\mu\nu}^{(1)}(x)~=~
{N_c\over s}\langle A,B|
\big[\bar\psi_A(x)\gamma_\mu\psi_B(x)\big]\big[\bar\psi_B(0)\gamma_\nu\Xi_{1}(0)\big]
\nonumber\\
&&\hspace{-1mm}
+~\big[\Bxi_{1}(x)\gamma_\mu\psi_B(x)\big]\big[\bar\psi_B(0)\gamma_\nu\psi_A(0)\big]
+\big[\bar\psi_A(x)\gamma_\mu\Xi_{2}(x)\big]\big[\bar\psi_B(0)\gamma_\nu\psi_A(0)\big]
\nonumber\\
&&\hspace{-1mm}
+~\big[\bar\psi_A(x)\gamma_\mu\psi_B(x)\big]\big[\Bxi_{2}(0)\gamma_\nu\psi_A(0)\big]
+\mu\leftrightarrow\nu|A,B\rangle~+~x\leftrightarrow 0
\label{kalw1}
\end{eqnarray}
\begin{eqnarray}
&&\hspace{-1mm}
\cheW_{\mu\nu}^{(2a)}(x)~=~
{N_c\over s}\langle A,B|\big[\bar\psi_A(x)\gamma_\mu\Xi_{2}(x)\big]\big[\bar\psi_B(0)\gamma_\nu\Xi_{1}(0)\big]
\nonumber\\
&&\hspace{-1mm}
+~[\Bxi_{1}(x)\gamma_\mu\psi_B(x)\big]\big[\Bxi_{2}(0)\gamma_\nu\psi_A(0)\big]
+\mu\leftrightarrow\nu|A,B\rangle~+~x\leftrightarrow 0
\label{kalw2a}
\end{eqnarray}
\begin{eqnarray}
&&\hspace{-1mm}
\cheW_{\mu\nu}^{(2b)}(x)~=~
{N_c\over s}\langle A,B|
\big[\bar\psi_A(x)\gamma_\mu\Xi_{2}(x)\big]\big[\Bxi_{2}(0)\gamma_\nu\psi_A(0)\big]
\nonumber\\
&&\hspace{-1mm}
+~
\big[\Bxi_{1}(x)\gamma_\mu\psi_B(x)\big]\big[\bar\psi_B(0)\gamma_\nu\Xi_{1}(0)\big]
+\mu\leftrightarrow\nu|A,B\rangle~+~x\leftrightarrow 0
\label{kalw2b}
\end{eqnarray}
and
\begin{eqnarray}
&&\hspace{-1mm}
\cheW_{\mu\nu}^{(2c)}(x)~=~
{N_c\over s}\langle A,B|\big[\Bxi_{1}(x)\gamma_\mu\Xi_{2}(x)\big]\big[\bar\psi_B(0)\gamma_\nu\psi_A(0)\big]
\nonumber\\
&&\hspace{-1mm}
+~\big[\bar\psi_A(x)\gamma_\mu\psi_B(x)\big]\big[\Bxi_{2}(0)\gamma_\nu\Xi_{1}(0)\big]
+\mu\leftrightarrow\nu|A,B\rangle~+~x\leftrightarrow 0
\label{kalw2c}
\end{eqnarray}
The corresponding contributions to $W_{\mu\nu}(q)$ will be denoted $W_{\mu\nu}^{(1)}$,  $W_{\mu\nu}^{(2)a}$,  $W_{\mu\nu}^{(2)b}$, and 
 $W_{\mu\nu}^{(2)c}$, respectively.
We will consider these contributions in turn.

\subsection{Terms with one quark-quark-gluon operator \label{1qqG}}
In this section we consider terms in Eq. (\ref{kalw1}) which will lead to ${2\over \beta_qs}p_1^\mu q_\perp^\nu+\mu\leftrightarrow\nu$ and 
${2\over \alpha_qs}p_2^\mu q_\perp^\nu+\mu\leftrightarrow\nu$ contributions to $W_{\mu\nu}(q)$.

\subsubsection{Term with $\Xi_{1}$ \label{sec:onexifirst}}
Let us start with the last term in Eq. (\ref{kalw1}). The Fierz transformation (\ref{fierz}) yields
\begin{eqnarray}
&&\hspace{-1mm}
\half[\bar\psi_A(x)\gamma_\mu\psi_B(x)\big]\big[\bar\psi_B(0)\gamma_\nu\Xi_{1}(0)\big]~+~\mu\leftrightarrow\nu
\nonumber\\
&&\hspace{-1mm}
=~{g_{\mu\nu}\over 4}\big\{\big[\bar\psi_A^{m}(x){\notp_2\over s}\gamma^i{1\over \alpha}\psi_A^k(0)\big]\big[\bar\psi_B^{n}\barB_i^{nk}(0)\psi_{B}^{m}(x)\big]~-~(\psi_A^k\otimes\psi_B^n\leftrightarrow\gamma_5\psi_A^k\otimes\gamma_5\psi_B^n\big\}
\nonumber\\
&&\hspace{-1mm}
+~{1\over 4}(\delta_\mu^\alpha\delta_\nu^\beta+\delta_\nu^\alpha\delta_\mu^\beta-g_{\mu\nu}g^{\alpha\beta})
\nonumber\\
&&\hspace{-1mm}
\times~\big\{\big[\bar\psi_A^{m}(x)\gamma_\alpha{\notp_2\over s}\gamma^i{1\over \alpha}\psi_A^k(0)\big]
\big[\bar\psi_B^{n}\barB_i^{nk}(0)\gamma_\beta\psi^{m}_{B}(x)\big]
~+~(\gamma_\alpha\otimes\gamma_\beta\leftrightarrow\gamma_\alpha\gamma_5\otimes\gamma_\beta\gamma_5)\big\}
\nonumber\\
&&\hspace{-1mm}
-~{1\over 4}(\delta_\mu^\alpha\delta_\nu^\beta+\delta_\nu^\alpha\delta_\mu^\beta-\half g_{\mu\nu}g^{\alpha\beta})
\big[\bar\psi_A^{m}(x)\sigma_{\alpha\xi}{\notp_2\over s}\gamma^i{1\over \alpha}\psi_A^k(0)\big]
\big[\bar\psi_B^{n}\barB_i^{nk}(0)\sigma_\beta^{~\xi}\psi^{m}_{B}(x)\big]
\label{onexi1}
\end{eqnarray}
where we used Eq. (\ref{fildz0}) $\Xi_{1}(0)=-{\notp_2\over s}\gamma^i\barB_i{1\over \alpha}\psi_A(0)$. Note that all
colors are in the fundamental representation so e.g. $B^{mn}(x)\equiv (t_a)^{mn}B^a(x)$.

Promoting $A$ and $B$ fields to operators and sorting out the color-singlet contributions we get
\footnote{We will keep different notations $A_i$ and  $B_i$ for the projectile and target gluon fields because of the relations
(\ref{abefildz}) and (\ref{abeznaki})
}
\begin{eqnarray}
&&\hspace{-1mm}
\cheW_{1\mu\nu}^{(1)}(x)~=~
{N_c\over s}\langle A,B|[\hbsi_A(x)\gamma_\mu\hsi_B(x)\big]\big[\hbsi_B(0)\gamma_\nu\Xi_{1}(0)\big]~+~\mu\leftrightarrow\nu
|A,B\rangle~+~x\leftrightarrow 0
\nonumber\\
&&\hspace{-1mm}
=~{g_{\mu\nu}\over 2s^2}\big\{\langle\hbsi(x)\notp_2\gamma^i{1\over \alpha}\hsi(0)\rangle_A\langle\hbsi\barB_i(0)\hsi(x)\rangle_B
~-~(\hsi(0)\otimes\hsi(x)\leftrightarrow\gamma_5\hsi(0)\otimes\gamma_5\hsi(x)\big\}
\nonumber\\
&&\hspace{-1mm}
+~{1\over 2s^2}(\delta_\mu^\alpha\delta_\nu^\beta+\delta_\nu^\alpha\delta_\mu^\beta-g_{\mu\nu}g^{\alpha\beta})
\big\{\langle\hbsi(x)\gamma_\alpha\notp_2\gamma_i{1\over \alpha}\hsi(0)\rangle_A
\langle\hbsi B^i(0)\gamma_\beta\hsi(x)\rangle_B
\nonumber\\
&&\hspace{33mm}
~+~(\hsi(0)\otimes\hsi(x)\leftrightarrow\gamma_5\hsi(0)\otimes\gamma_5\hsi(x)\big\}
\label{onexi2}\\
&&\hspace{-1mm}
-~{1\over 2s^2}(\delta_\mu^\alpha\delta_\nu^\beta+\delta_\nu^\alpha\delta_\mu^\beta-\half g_{\mu\nu}g^{\alpha\beta})
\langle\hbsi(x)\sigma_{\alpha\xi}\notp_2\gamma^i{1\over \alpha}\hsi(0)\rangle_A
\langle\hbsi B^i(0)\sigma_\beta^{~\xi}\hsi(x)\rangle_B~+~x\leftrightarrow 0
\nonumber
\end{eqnarray}

It is convenient to treat terms $\sim g_{\mu\nu}$ separately so we define 
$\cheW_{1\mu\nu}^{(1)}(x)~=~\cheW_{\mu\nu}^{(1a)}(x)+\cheW_{1\mu\nu}^{(1b)}(x)$  where 
\begin{eqnarray}
&&\hspace{-1mm}
\cheW_{\mu\nu}^{(1a)}(x)~
\nonumber\\
&&\hspace{-1mm}
=~{g_{\mu\nu}\over 2s^2}\Big(\big\{\langle\bar\psi(x)\notp_2\gamma^i{1\over \alpha}\psi(0)\rangle_A\langle\bar\psi\barB_i(0)\psi(x)]\rangle_B
~-~(\psi(0)\otimes\psi(x)\leftrightarrow\gamma_5\psi(0)\otimes\gamma_5\psi(x)\big\}
\nonumber\\
&&\hspace{15mm}
-~
\big\{\langle\bar\psi(x)\gamma_\alpha\notp_2\gamma_i{1\over \alpha}\psi(0)\rangle_A
\langle\bar\psi B^i(0)\gamma^\alpha\psi(x)\rangle_B
+(\psi(0)\otimes\psi(x)\leftrightarrow\gamma_5\psi(0)\otimes\gamma_5\psi(x)\big\}
\nonumber\\
&&\hspace{33mm}
+~\half 
\langle\bar\psi(x)\sigma_{\alpha\beta}\notp_2\gamma^i{1\over \alpha}\psi(0)\rangle_A
\langle\bar\psi B^i(0)\sigma^{\alpha\beta}\psi(x)\rangle_B~+~x\leftrightarrow 0
\label{onexi3}
\end{eqnarray}
Hereafter we omit ``hat'' notation from from operators: $\langle\calo\rangle_{A,B}\equiv\langle\hat{\calo}\rangle_{A,B}$ for brevity.

Let us now estimate this contribution to $\cheW_{\mu\nu}$. First, recall that $B_i$ is of order of $m_\perp$ (more accurately, it will be $\sim q_i$ after the Fourier transformation, see e.g. Eq. (\ref{11.42}) or Eq. (\ref{paramj})). Next, as demonstrated in Sect. \ref{sec:qqgparam} (see Eqs. (\ref{maelqg1}), (\ref{maelqg2})), 
 ${1\over\alpha}$ in the target matrix element turns to $\pm{1\over\alpha_q}$ after Fourier transformation. Due to this fact we will  replace ${1\over\alpha}$ by ${1\over\alpha_q}$ in our estimates, even in the coordinate space. Similarly, for the estimate of  the target matrix elements we will replaces operator ${1\over\beta}$ by ${1\over\beta_q}$ whenever appropriate.

Now we will demonstrate that three terms in the r.h.s. of Eq. (\ref{onexi3}) are small in comparison to 
terms listed in Eq. (\ref{pc}).  The projectile matrix element in the first term in the r.h.s. of Eq. (\ref{onexi3})  brings factor $s$ (see Eq. (\ref{hmael})) but the target matrix element 
can produce only factor $x_i$ so the first term is $\sim{g_{\mu\nu}m_\perp^2\over\alpha_qs}$ which is smaller than
${g_{\mu\nu}q_\perp^2\over Q^2}$ that we have in Eq. (\ref{pc}) 
(and will calculate in the next Section). As to the second term in  the r.h.s. of Eq. (\ref{onexi2}), 
it can be rewritten as
\begin{eqnarray}
&&\hspace{-1mm}
-~
\big\{\langle\bar\psi(x)\gamma_j\notp_2\gamma_i{1\over \alpha}\psi(0)\rangle_A
\langle\bar\psi B^i(0)\gamma^j\psi(x)\rangle_B
+{2\over s}\langle\bar\psi(x)\notp_1\notp_2\gamma_i{1\over \alpha}\psi(0)\rangle_A
\langle\bar\psi B^i(0)\notp_2\psi(x)\rangle_B
\nonumber\\
&&\hspace{33mm}
+~(\psi(0)\otimes\psi(x)\leftrightarrow\gamma_5\psi(0)\otimes\gamma_5\psi(x)\big\}{g_{\mu\nu}\over 2s^2}
\label{onexi5}
\end{eqnarray}
The projectile matrix element in the first term in  the r.h.s. of this equation brings factor $s$ but, as we discussed above, the target matrix element
cannot produce factor $s$ so this term is again $\sim{g_{\mu\nu}m_\perp^2\over\alpha_qs}\ll {g_{\mu\nu}m_\perp^2\over Q^2}$. 
As to the second term,
converting three $\gamma$-matrices in the projectile matrix element to a combination of $\gamma$'s and $\gamma\gamma_5$'s and looking
at the parametrization of Sect. \ref{sec:paramlt},  we see that ${2\over s}\langle\bar\psi(x)\notp_1\notp_2\gamma_i{1\over \alpha}\psi(0)\rangle_A$ 
is not proportional to $s$. In addition, as discuss in Sect. \ref{sec:paramlt}, the target matrix element 
$\langle\bar\psi B^i(0)\gamma_\mu\psi(x)\rangle_B$ 
knows about $p_1$ only via the direction of Wilson lines so it  can be proportional only to ${p_{1\mu}\over p_1\cdot p_2}$  that does not change at 
rescaling of $p_1$.
Thus, $\langle\bar\psi B^i(0)\notp_2\psi(x)\rangle_B$ is $\sim O(1)$
and therefore the second term in Eq. (\ref{onexi5}) is even smaller than the first one. 
Finally, let us discuss the third term in the r.h.s. of Eq. (\ref{onexi3}). If both $\alpha$ and $\beta$ are transverse
\begin{eqnarray}
&&\hspace{-1mm}
{g_{\mu\nu}\over 4s^2}\langle\bar\psi(x)\sigma_{\alpha\beta}\notp_2\gamma^i{1\over \alpha}\psi(0)\rangle_A
\langle\bar\psi B^i(0)\sigma^{\alpha\beta}\psi(x)\rangle_B \sim{g_{\mu\nu}m_\perp^2\over\alpha_qs}
\label{onexi6}
\end{eqnarray}
similarly to the first term in Eq. (\ref{onexi5}). If both indices are longitudinal, we get
\begin{eqnarray}
&&\hspace{-1mm}
{g_{\mu\nu}\over s^4}\langle\bar\psi(x)\sigma_{\ast\bu}\notp_2\gamma^i{1\over \alpha}\psi(0)\rangle_A
\langle\bar\psi B^i(0)\sigma_{\bu\ast}\psi(x)\rangle_B
\nonumber\\
&&\hspace{33mm}
=~{g_{\mu\nu}\over s^3}\langle\bar\psi(x)\notp_2\gamma^i{1\over \alpha}\psi(0)\rangle_A
\langle\bar\psi B^i(0)\sigma_{\bu\ast}\psi(x)\rangle_B
\label{onexi7}
\end{eqnarray}
The projectile  matrix element brings a factor $s$, but the target one is $\sim O(1)$ due to the reason discussed above, so this
contribution is negligible. Finally, let us consider the case when index $\alpha$ is longitudinal and $\beta$ is transverse
\begin{eqnarray}
&&\hspace{-1mm}
{g_{\mu\nu}\over 2s^3}\langle\bar\psi(x)\sigma_{\bu j}\notp_2\gamma^i{1\over \alpha}\psi(0)\rangle_A
\langle\bar\psi B^i(0)\sigma_\ast^{~j}\psi(x)\rangle_B
\label{onexi8}
\end{eqnarray}
Again, the target matrix element is $\sim O(1)$ while the projectile one 
can bring one factor of $s$ as can be seen from parametrization (\ref{hmael}) by reducing the number of $\gamma$-matrices to two. 
Thus, the contribution (\ref{onexi8}) is negligible and so is the total contribution (\ref{onexi3}).

We get
\begin{eqnarray}
&&\hspace{-1mm}
\cheW_{\mu\nu}^{(1)}(x)~\simeq~\cheW_{\mu\nu}^{(1b)}(x)~=~
\nonumber\\
&&\hspace{-1mm}
=~{1\over 2s^2}
\big\{\langle\bar\psi(x)\gamma_\mu\notp_2\gamma_i{1\over \alpha}\psi(0)\rangle_A
\langle\bar\psi B^i(0)\gamma_\nu\psi(x)\rangle_B
~+~(\psi(0)\otimes\psi(x)\leftrightarrow\gamma_5\psi(0)\otimes\gamma_5\psi(x)\big\}
\nonumber\\
&&\hspace{-1mm}
-~{1\over 2s^2}
\langle\bar\psi(x)\sigma_{\mu\xi}\notp_2\gamma^i{1\over \alpha}\psi(0)\rangle_A
\langle\bar\psi B^i(0)\sigma_\nu^{~\xi}\psi(x)\rangle_B~+~\mu\leftrightarrow\nu~+~x\leftrightarrow 0
\label{onexi9}
\end{eqnarray}
Let us start with  the case when both of the indices $\mu$ and $\nu$ are transverse. It is easy to see that the power counting for the 
first term in the r.h.s. of Eq. (\ref{onexi9}) is the same as for  Eq. (\ref{onexi5}) so it is small. Also, the estimate of the second term
in Eq. (\ref{onexi9}) is similar either to the estimate of Eq. (\ref{onexi6}) or (\ref{onexi8}) so it can be neglected. 

Next, let us consider both $\mu$ and $\nu$ longitudinal.  It is easy to see that multiplication of the r.h.s. of Eq. (\ref{onexi9}) by
$p_2^\mu p_2^\nu$ gives zero so there is no term proportional to $p_1^\mu p_1^\nu$. The term proportional to $p_2^\mu p_2^\nu$
has the form
\begin{eqnarray}
&&\hspace{-1mm}
{4\over s^4}p_{2\mu}p_{2\nu}
\big\{\langle\bar\psi(x)\notp_1\notp_2\gamma_i{1\over \alpha}\psi(0)\rangle_A
\langle\bar\psi B^i(0)\notp_1\psi(x)\rangle_B
~+~(\psi(0)\otimes\psi(x)\leftrightarrow\gamma_5\psi(0)\otimes\gamma_5\psi(x)\big\}
\nonumber\\
&&\hspace{-1mm}
-~{4\over s^4}p_{2\mu}p_{2\nu}
\langle\bar\psi(x)\sigma_{\bu\xi}\notp_2\gamma^i{1\over \alpha}\psi(0)\rangle_A
\langle\bar\psi B^i(0)\sigma_\bu^{~\xi}\psi(x)\rangle_B
\label{onexi10}
\end{eqnarray}
It is easy to see that both projectile and target matrix elements are proportional to the first power of $s$ so the resulting 
estimate is ${p_{2\mu}p_{2\nu}\over \alpha_qs^2}m_\perp^2$ which is $\sim O\big({m_\perp^2\over s}\big)$ in comparison to
the corrseponding term in Eq. (\ref{pc}). If one index is $p_1$ and the other $p_2$ we get
\begin{eqnarray}
&&\hspace{-1mm}
{g^\parallel_{\mu\nu}\over s^3}\Big[\langle\bar\psi(x)\notp_1\notp_2\gamma_i{1\over \alpha}\psi(0)\rangle_A
\langle\bar\psi B^i(0)\notp_2\psi(x)\rangle_B
~+~(\psi(0)\otimes\psi(x)\leftrightarrow\gamma_5\psi(0)\otimes\gamma_5\psi(x)
\label{onexi11}\\
&&\hspace{-1mm}
-~
\langle\bar\psi(x)\sigma_{\bu j}\notp_2\gamma^i{1\over \alpha}\psi(0)\rangle_A
\langle\bar\psi B^i(0)\sigma_\ast^{~j}\psi(x)\rangle_B
-~
{4\over s} \langle\bar\psi(x)\notp_2\gamma^i{1\over \alpha}\psi(0)\rangle_A
\langle\bar\psi B^i(0)\sigma_{\bu\ast}\psi(x)\rangle_B\Big]
\nonumber
\end{eqnarray}
It is easy to see that in all terms  the projectile matrix element is $\sim s$ but the target one is $\sim O(1)$ so 
the corresponding contribution $\sim {g_{\mu\nu}m_\perp^2\over s^2}$ is negligible. 

Finally, let  us consider the case when one of the indices in Eq. (\ref{onexi9}) is longitudinal and one transverse. For example,
let $\mu$ be longitudinal and $\nu$ transverse, the opposite case
will differ by replacement $\mu\leftrightarrow\nu$. 
Using the decomposition of $g^{\mu\nu}$ in longitudinal and transverse part (\ref{delta}) 
we get
\begin{eqnarray}
&&\hspace{-1mm}
\Big({2p_{1\mu} p_2^{\mu'}\over s}+p_1\leftrightarrow p_2\Big)\cheW_{\mu\nu}^{(1b)}(x)~=~
\Big({p_{2\mu} p_1^{\mu'}\over s^3}+p_1\leftrightarrow p_2\Big)
\nonumber\\
&&\hspace{-1mm}
\times~\Big[
\big\{\langle\bar\psi(x)\gamma_{\mu'}\notp_2\gamma_i{1\over \alpha}\psi(0)\rangle_A
\langle\bar\psi B^i(0)\gamma_\nu\psi(x)\rangle_B
~+~(\psi(0)\otimes\psi(x)\leftrightarrow\gamma_5\psi(0)\otimes\gamma_5\psi(x)\big\}
\nonumber\\
&&\hspace{-1mm}
-~
\langle\bar\psi(x)\sigma_{\mu'\xi}\notp_2\gamma^i{1\over \alpha}\psi(0)\rangle_A
\langle\bar\psi B^i(0)\sigma_\nu^{~\xi}\psi(x)\rangle_B~+~\mu'\leftrightarrow\nu\Big]~+~x\leftrightarrow 0
\label{onexi12}
\end{eqnarray}
The term proportional to $p_{2\mu}$ in the r.h.s. can be expressed using Eq. (\ref{gammas1fild}) as follows
\begin{eqnarray}
&&\hspace{-3mm}
{p_{2\mu}\over s^3}
\Big\{\Big[
\langle \bar\psi(x)\gamma_{\nu_\perp}\notp_2\gamma_i{1\over \alpha}\psi(0)\rangle_A
\langle \bar\psi B^i(0)\notp_1\psi(x)\rangle_B
+\langle\bar\psi(x)\notp_1\notp_2\gamma_i{1\over \alpha}\psi(0)\rangle_A
\langle\bar\psi B^i(0)\gamma_{\nu_\perp}\psi(x)\rangle_B
\nonumber\\
&&\hspace{33mm}
~+~(\psi(0)\otimes\psi(x)\leftrightarrow\gamma_5\psi(0)\otimes\gamma_5\psi(x)\Big]
\label{onexi13}\\
&&\hspace{-3mm}
-~\langle \bar\psi(x)\sigma_{\bu\xi}\notp_2\gamma^i{1\over \alpha}\psi(0)\rangle_A
\langle \bar\psi B^i(0)\sigma_{\nu_\perp}^{~\xi}\psi(x)\rangle_B
-\langle \bar\psi(x)\sigma_{\nu_\perp\xi}\notp_2\gamma^i{1\over \alpha}\psi(0) \rangle_A
\langle \bar\psi B^i(0)\sigma_\bu^{~\xi}\psi(x)\rangle_B\Big\}
\nonumber\\
&&\hspace{-3mm}
=~{p_{2\mu}\over s^3}\Big\{\Big[-\langle\bar\psi(x)\notp_2{1\over \alpha}\psi(0) \rangle_A
\langle \bar\psi(0)\notp_1\breB_{\nu_\perp}(0)\psi(x)\rangle_B
\nonumber\\
&&\hspace{-3mm}
+{s\over 2}\langle\bar\psi(x)\gamma^i{1\over \alpha}\psi(0) \rangle_A
\langle \bar\psi(0)\gamma_{\nu_\perp}\breB_i(0)\psi(x)\rangle_B
~+~(\psi(0)\otimes\psi(x)\leftrightarrow\gamma_5\psi(0)\otimes\gamma_5\psi(x))\Big]
\nonumber\\
&&\hspace{-3mm}
+~i\langle\bar\psi(x)\sigma_{\bu j}\sigma_{\ast i}{1\over \alpha}\psi(0) \rangle_A
\langle \bar\psi B^i(0)\sigma_{\nu_\perp}^{~j}\psi(x)\rangle_B
+i\langle\bar\psi(x)\sigma_{\nu_\perp j}\sigma_{\ast i}{1\over \alpha}\psi(0) \rangle_A
\langle \bar\psi B^i(0)\sigma_\bu^{~j}\psi(x)\rangle_B
\nonumber\\
&&\hspace{-3mm}
-~\langle\bar\psi(x)\sigma_{\ast i}{1\over \alpha}\psi(0) \rangle_A
\langle \bar\psi B^i(0)\sigma_{\bu \nu_\perp}\psi(x)\rangle_B
+{2i\over s}\langle\bar\psi(x)\sigma_{\bu \nu_\perp}\sigma_{\ast i}{1\over \alpha}\psi(0) \rangle_A
\langle \bar\psi B^i(0)\sigma_{\ast\bu}\psi(x)\rangle_B\Big\}
\nonumber
\end{eqnarray}
Hereafter we use notation $\breB_i\equiv B_i-i\tilB_i\gamma_5$.

Let us at evaluate two the most important contributions. The first is
\begin{eqnarray}
&&\hspace{-1mm}
-{p_{2\mu}\over s^3}\langle\bar\psi(x)\notp_2{1\over \alpha}\psi(0) \rangle_A
\langle \bar\psi(0)\notp_1\breB_{\nu_\perp}(0)\psi(x)\rangle_B
\nonumber\\
&&\hspace{33mm}
=~{p_{2\mu}\over s^3}\langle\bar\psi(x)\notp_2{1\over \alpha}\psi(0) \rangle_A
\langle \bar\psi(0)\notB(0)\notp_1\gamma_{\nu_\perp}\psi(x)\rangle_B
\label{onexi14}
\end{eqnarray}
As we shall see below, due to QCD equations of motion $\notB$ in the r.h.s. of this equation can be replaced by transverse momentum of the target TMD $k_\perp$. Also, ${1\over \alpha}$ will be replaced by ${1\over\alpha_q}$ so from the parametrizations (\ref{Amael}) and (\ref{barbmael}) we see that
\begin{eqnarray}
&&\hspace{-1mm}
~{p_{2\mu}\over s^3}\langle\bar\psi(x)\notp_2{1\over \alpha}\psi(0) \rangle_A
\langle \bar\psi(0)\notB(0)\notp_1\gamma_{\nu_\perp}\psi(x)\rangle_B~\sim ~{p_{2\mu}\over \alpha_q s}k_\nu f\barf
\label{onexi15}
\end{eqnarray}
which is of order of fourth term in Eq. (\ref{pc}). 
The second relevant term is
\begin{eqnarray}
&&\hspace{-1mm}
{ip_{2\mu}\over s^3}\langle\bar\psi(x)\sigma_{\nu_\perp j}\sigma_{\ast i}{1\over \alpha}\psi(0) \rangle_A\langle \bar\psi B^i(0)\sigma_\bu^{~j}\psi(x)\rangle_B
-~{p_{2\mu}\over s^3}\langle\bar\psi(x)\sigma_{\ast i}{1\over \alpha}\psi(0) \rangle_A\langle \bar\psi B^i(0)\sigma_{\bu \nu_\perp}\psi(x)\rangle_B
\nonumber\\
&&\hspace{5mm}
=~{p_{2\mu}\over s^3}\langle\bar\psi(x)\sigma_\ast^{~ j}{1\over \alpha}\psi(0) \rangle_A
\langle \bar\psi(0)[B_\nu(0)\sigma_{\bu j}-\nu\leftrightarrow j]\psi(x)\rangle_B~-~{p_{2\mu}\over s^3}\langle\bar\psi(x)\sigma_{\ast \nu_\perp}{1\over \alpha}\psi(0) \rangle_A
\nonumber\\
&&\hspace{11mm}
\times~
\langle \bar\psi B^j(0)\sigma_{\bu j}\psi(x)\rangle_B
~=~{ip_{2\mu}\over s^3}\langle\bar\psi(x)\sigma_{\ast \nu_\perp}{1\over \alpha}\psi(0) \rangle_A
\langle \bar\psi(0) \notB(0)\notp_1\psi(x)\rangle_B
\label{onexi16}
\end{eqnarray}
where we used formula (\ref{sigmasigmas}) 
and the fact that for unpolarized protons 
\begin{equation}
\langle p|\bar\psi(0)[A_i(0)\sigma_{\bu j}-i\leftrightarrow j]\psi(x)|p\rangle=0
\label{formula2}
\end{equation}
from parity conservation.
\footnote{A rigorous argument goes like that: the matrix element (\ref{formula2})  can be rewritten as
$\epsilon_{\nu_\perp j}\epsilon_{kl}\langle \bar\psi(0)[A_k(0)\sigma_{\bu l}\psi(x)\rangle
~=~\epsilon_{j\nu_\perp}\langle \bar\psi(0)\notA(0)\notp_1\gamma_5\psi(x)\rangle$.
As demonstrated in Sect. \ref{sec:qqgparam}, $\notA$ in this formula can be replaced by $\notk_\perp$ so the contribution is
proportional to matrix element $k^i\langle \bar\psi(0)i\sigma_{\bu i}\gamma_5\psi(x)\rangle=k^i\epsilon_{ij}
\langle \bar\psi(0)\sigma_{\bu j}\psi(x)\rangle$ which vanishes as seen from the parametrization (\ref{hmael}).
\label{zanulil}
} 
Again, ${1\over\alpha}$ 
will turn to ${1\over\alpha_q}$ and $\notB$ can be replaced by $\notk_\perp$ for the target, so (\ref{onexi7}) is of order of 
\begin{eqnarray}
&&\hspace{-1mm}
{ip_{2\mu}\over s^3}\langle\bar\psi(x)\sigma_{\ast \nu_\perp}{1\over \alpha}\psi(0) \rangle_A
\langle \bar\psi(0) \notB(0)\notp_1\psi(x)\rangle_B~\sim~{p_{2\mu}\over \alpha_q s}k_\nu h\barh
\label{onexi17}
\end{eqnarray}
Let us demonstrate that the remaining terms in the r.h.s. of Eq. (\ref{onexi13}) are negligible.
First, term coming from replacement $\psi(0)\otimes\psi(x)\leftrightarrow\gamma_5\psi(0)\otimes\gamma_5\psi(x)$ 
in Eq. (\ref{onexi14}) vanishes since $\langle\bar\psi(x)\notp_2\gamma_5\psi(0) \rangle_A=0$ for unpolarized hadrons, 
see Eq. (\ref{mael5}). Next, term ${p_{2\mu}\over 2s^2}\langle\bar\psi(x)\gamma^i{1\over \alpha}\psi(0) \rangle_A
\langle \bar\psi(0)\gamma_{\nu_\perp}\breB_i(0)\psi(x)\rangle_B$ is small because neither projectile no target matrix elements can bring factor $s$.
Last, using Eq. (\ref{sigmasigmas}) we get
\begin{eqnarray}
&&\hspace{-1mm}
{ip_{2\mu}\over s^3}\big[\langle\bar\psi(x)\sigma_{\bu j}\sigma_{\ast i}{1\over \alpha}\psi(0) \rangle_A
\langle \bar\psi B^i(0)\sigma_{\nu_\perp}^{~j}\psi(x)\rangle_B
\nonumber\\
&&\hspace{22mm} 
+~{2\over s}\langle\bar\psi(x)\sigma_{\bu \nu_\perp}\sigma_{\ast i}{1\over \alpha}\psi(0) \rangle_A
\langle \bar\psi B^i(0)\sigma_{\ast\bu}\psi(x)\rangle_B\big]
\nonumber\\
&&\hspace{-1mm} 
=~{ip_{2\mu}\over 2s^2}\langle\bar\psi(x)\big[g_{ij}+i\epsilon_{ij}\gamma_5 
+i(\sigma_{ij}-{2\over s}g_{ij}\sigma_{\bu\ast})\big]\psi(0) \rangle_A
\langle \bar\psi B^i(0)\sigma_{\nu_\perp}^{~j}\psi(x)\rangle_B
\nonumber\\
&&\hspace{22mm} 
+~{i\over s^3}\langle\bar\psi(x)\big[g_{i\nu_\perp}+i\epsilon_{i\nu_\perp}\gamma_5 
+i(\sigma_{i\nu_\perp}-{2\over s}g_{i\nu_\perp}\sigma_{\bu\ast})\big]\psi(0) \rangle_A
\langle \bar\psi B^i(0)\sigma_{\ast\bu}\psi(x)\rangle_B
\nonumber\\
&&\hspace{-1mm} 
=~{ip_{2\mu}\over 2s^2}\langle\bar\psi(x)\big[1-{2\over s}\sigma_{\bu\ast}\big]\psi(0) \rangle_A
\langle \bar\psi(0)\big[B_j(0)\sigma_{\nu_\perp}^{~j}+{2\over s}B_{\nu_\perp}(0)\sigma_{\ast\bu}\big]\psi(x)\rangle_B
\label{onexi18}
\end{eqnarray}
It is easy to see that neither the projectile nor the target matrix element
in the r.h.s. of this equation gives $s$ so these terms can be neglected in comparison to Eqs. (\ref{onexi15})
and (\ref{onexi17}).

Thus, the two non-negligible terms in Eq. (\ref{onexi13}) give
\begin{eqnarray}
&&\hspace{-1mm}
\cheW^{(1)}_{1\mu\nu}(x)~=~{N_c\over s}\langle A,B|\big[\bar\psi_A(x)\gamma_\mu\psi_B(x)\big]\big[\bar\psi_B(0)\gamma_\nu\Xi_{1}(0)\big]
+\mu\leftrightarrow\nu|A,B\rangle+x\leftrightarrow 0
\nonumber\\
&&\hspace{5mm}
=~{p_{2\mu}\over s^3}\big[
\langle 
\bar\psi(x_\bu,x_\perp)\notp_2{1\over\alpha}\psi(0)\rangle_A
\langle\bar\psi\notB(0)\notp_1\gamma^\perp_\nu\psi(x_\ast,x_\perp)\rangle_B
\nonumber\\
&&\hspace{11mm}
+~i\langle\bar\psi(x_\bu,x_\perp)\sigma_{\ast\nu}{1\over\alpha}\psi(0)\rangle_A
\langle\bar\psi\notB(0)\hatp_1\psi(x_\ast,x_\perp)\rangle_B
\big]~+~\mu\leftrightarrow\nu
~+~x\leftrightarrow 0
\label{onexi19}
\end{eqnarray}
Using formulas (\ref{maelqg1}), (\ref{maelqg2}), (\ref{tw3mael1}), (\ref{11.36}), (\ref{11.41}),  and (\ref{11.43}) for quark-antiquark-gluon operators  and parametrizations 
from Sect. \ref{sec:paramlt} we get the contribution  to $W_{\mu\nu}$ in the form
\begin{eqnarray}
&&\hspace{-1mm}
W^{(1)}_{1\mu\nu}(q)~=~{1\over 16\pi^4}{1\over s}
\!\int\!dx_\bu dx_\ast d^2x_\perp~e^{-i\alpha x_\bu-i\beta x_\ast+i(q,x)_\perp}
\label{onexi20}\\
&&\hspace{33mm}
\times~
\langle A,B|\big[\bar\psi_A(x)\gamma_\mu\psi_B(x)\big]\big[\bar\psi_B(0)\gamma_\nu\Xi_{1}(0)\big]+x\leftrightarrow 0|A,B\rangle
~+~\mu\leftrightarrow\nu
\nonumber\\
&&\hspace{-1mm}
=~{1\over 64\pi^6N_c}{p_{2\mu}\over s^3}\!\int d^2k_\perp
\!\int\!dx_\bu d^2x_\perp~e^{-i\alpha x_\bu+i(k,x)_\perp}\!\int\!dx_\ast d^2x'_\perp e^{-i\beta x_\ast+i(q-k,x')_\perp}
\nonumber\\
&&\hspace{15mm}
\times~\big[
\langle 
\bar\psi(x_\bu,x_\perp)\notp_2{1\over\alpha}\psi(0)\rangle_A
\langle\bar\psi\notB(0)\notp_1\gamma^\perp_\nu\psi(x_\ast,x'_\perp)\rangle_B
\nonumber\\
&&\hspace{33mm}
+~i\langle\bar\psi(x_\bu,x_\perp)\sigma_{\ast\nu}{1\over\alpha}\psi(0)\rangle_A
\langle\bar\psi\notB(0)\hatp_1\psi(x_\ast,x'_\perp)\rangle_B
+x\leftrightarrow 0\big]~+~\mu\leftrightarrow\nu
\nonumber\\
&&\hspace{-1mm}
=~{p_{2\mu}\over \alpha_qsN_c}\!\int\! d^2k_\perp\Big\{(q-k)_\nu \big[f_1^f(\alpha_q,k_\perp)\barf _\perp^f(\beta_q,(q-k)_\perp)+
\barf_1^f(\alpha_q,k_\perp)f _\perp^f(\beta_q,(q-k)_\perp)\big]
\nonumber\\
&&\hspace{-1mm}
-~k_\nu {(q-k)_\perp^2\over m^2}\big[h_{1f}^\perp(\alpha_q,k_\perp)\barh _\perp^f(\beta_q,(q-k)_\perp)
+\barh_{1f}^\perp(\alpha_q,k_\perp)h _\perp^f(\beta_q,(q-k)_\perp)\big]\Big\}
~+~\mu\leftrightarrow\nu
\nonumber
\end{eqnarray}
where terms with replacement $f_1^f\leftrightarrow\barf_1^f$ and $h_{1f}^\perp\leftrightarrow\barh_{1f}^\perp$ come from $x\leftrightarrow 0$ contribution.

Next we consider the remaining $\sim p_{1\mu}$  term in Eq. (\ref{onexi9}) which can be rewritten as
\begin{eqnarray}
&&\hspace{-1mm}
{2p_{1\mu}\over s^2}N_c\langle A,B|[\bar\psi_A(x)\notp_2\psi_B(x)\big]\big[\bar\psi_B(0)\gamma_\nu\Xi_{1}(0)\big]|A,B\rangle
\nonumber\\
&&\hspace{-1mm}
=~{p_{1\mu}\over s^3}
\big\{\langle\bar\psi(x)\gamma_{\nu_\perp}\notp_2\gamma_i{1\over \alpha}\psi(0)\rangle_A
\langle\bar\psi B^i(0)\notp_2\psi(x)\rangle_B
\nonumber\\
&&\hspace{33mm}
~+~(\psi(0)\otimes\psi(x)\leftrightarrow\gamma_5\psi(0)\otimes\gamma_5\psi(x)]\big\}
\nonumber\\
&&\hspace{-1mm}
+~{p_{1\mu}\over s^3}
\langle\bar\psi(x)\sigma_{\ast i}{1\over \alpha}\psi(0)\rangle_A
\langle\bar\psi B_i(0)\sigma_{\ast \nu_\perp}\psi(x)\rangle_B
\nonumber\\
&&\hspace{33mm}
+~{ip_{1\mu}\over s^3}
\langle\bar\psi(x)\sigma_{\nu_\perp j}\sigma_{\ast i}{1\over \alpha}\psi(0)\rangle_A
\langle\bar\psi B_i(0)\sigma_\ast^{~j}\psi(x)\rangle_B
\nonumber\\
&&\hspace{-1mm}
=~{p_{1\mu}\over s^3}
\big\{\langle\bar\psi(x)\gamma_{\nu_\perp}\notp_2\gamma_i{1\over \alpha}\psi(0)\rangle_A
\langle\bar\psi B^i(0)\notp_2\psi(x)\rangle_B
\nonumber\\
&&\hspace{33mm}
~+~(\psi(0)\otimes\psi(x)\leftrightarrow\gamma_5\psi(0)\otimes\gamma_5\psi(x)]\big\}
\nonumber\\
&&\hspace{-1mm}
+~{p_{1\mu}\over s^3}
\langle\bar\psi(x)\sigma_{\ast i}{1\over \alpha}\psi(0)\rangle_A
\langle\bar\psi(0)[B_i(0)\sigma_{\ast \nu_\perp}+i\leftrightarrow \nu_\perp]\psi(x)\rangle_B
\nonumber\\
&&\hspace{33mm}
-~{p_{1\mu}\over s^3}
\langle\bar\psi(x)\sigma_{\ast \nu_\perp}{1\over \alpha}\psi(0)\rangle_A
\langle\bar\psi B_i(0)\sigma_\ast^{~i}\psi(x)\rangle_B
\label{onexi21}
\end{eqnarray}
where again we used formula (\ref{sigmasigmas}).
 Note that while the matrix elements between projectile states give contributions $\sim {s\over\alpha_q}k_\perp$, the target matrix elements cannot give $s$. Indeed, these target matrix elements know about $\hatp_1$ only through direction of Wilson lines so they should not change under rescaling $p_1\rightarrow \lambda p_1$, see the discussion in Sect. \ref{sec:paramlt}. Thus, the r.h.s. of Eq. (\ref{onexi21}) is 
 $\sim{p_{1\mu}\over\alpha_qs^2}k^\perp_\nu$ 
 which means that the $p_{1\mu}$  term in  Eq. (\ref{onexi12}) is
\begin{equation}
{2p_{1\mu}\over s^2}N_c\langle A,B|[\bar\psi_A(x)\notp_2\psi_B(x)\big]\big[\bar\psi_B(0)\gamma_\nu\Xi_{1}(0)\big]|A,B\rangle
~\sim~{p_{1\mu}q^\perp_\nu\over\alpha_qs^2}
\label{onexi22}
\end{equation}
so the corresponding contribution to $W_{\mu\nu}$ is $\sim {p_{1\mu}q^\perp_\nu\over\alpha_qs^2}$ which is $O\big({1\over s}\big)$ in comparison to that  of Eq. (\ref{XiA}).

Thus, the contribution of the first term in Eq. (\ref{kalw1}) to $W_{\mu\nu}$ is
\begin{eqnarray}
&&\hspace{-1mm}
W^{(1)}_{1\mu\nu}(q)~=~{1\over 16\pi^4}{1\over s}
\!\int\!dx_\bu dx_\ast d^2x_\perp~e^{-i\alpha x_\bu-i\beta x_\ast+i(q,x)_\perp}
\nonumber\\
&&\hspace{11mm}
\times~
\langle A,B|\big[\bar\psi_A(0)\gamma_\mu\psi_B(0)\big]\big[\bar\psi_B(x)\gamma_\nu\Xi_{1}(x)\big]+(x\leftrightarrow 0)|A,B\rangle
+\mu\leftrightarrow\nu
\nonumber\\
&&\hspace{-1mm}
=~{p_{2\mu}\over \alpha_qN_c}\!\int\! d^2k_\perp\Big[(q-k)^\perp_\nu F^f(q,k_\perp)
-~k_\nu {(q-k)_\perp^2\over m^2} H^f(q,k_\perp)\Big]
~+~\mu\leftrightarrow\nu
\label{XiA}
\end{eqnarray}
where $ F^f(q,k_\perp)$ and $H^f(q,k_\perp)$ are give by expressions (\ref{FH}) with $x_A\equiv\alpha_q$ and $x_B\equiv\beta_q$
\begin{eqnarray}
&&\hspace{-11mm}
F^f(q,k_\perp)~=~f_1^f(\alpha_q,k_\perp)\barf_1^f(\beta_q,(q-k)_\perp)~+~f_1^f\leftrightarrow\barf_1^f
\nonumber\\
&&\hspace{-11mm}
H^f(q,k_\perp)~=~h^\perp_{1f}(\alpha_q,k_\perp)\barh^\perp_{1f}(\beta_q,(q-k)_\perp) ~+~h_{1f}^\perp\leftrightarrow\barh_{1f}^\perp
\label{FH1}
\end{eqnarray}

Let us consider now the second term in Eq. (\ref{kalw1}). The calculation repeats that of the first term so we will indicate here main steps and pay attention to non-negligible terms only. If one of the indices (say, $\mu$) is longitudinal and the other transverse, we get
\begin{eqnarray}
&&\hspace{-1mm}
\Big({2p_{1\mu} p_2^{\mu'}\over s}+p_1\leftrightarrow p_2\Big)
{N_c\over s}\langle A,B|\big[\Bxi_{1}(x)\gamma_{\mu'}\psi_B(x)\big]\big[\bar\psi_B(0)\gamma_\nu\psi_A(0)\big]~+~\mu'\leftrightarrow\nu
|A,B\rangle
\nonumber\\
&&\hspace{-1mm}
=~\langle A,B|p_{2\mu}\big(\big[\Bxi_{1}(x)\notp_1\psi_B(x)\big]\big[\bar\psi_B(0)\gamma_\nu\psi_A(0)\big]~
+\big[\Bxi_{1}(x)\gamma_\nu\psi_B(x)\big]\big[\bar\psi_B(0)\notp_1\psi_A(0)\big]\big)
\nonumber\\
&&\hspace{-1mm}
+~p_{1\mu}\big[\Bxi_{1}(x)\gamma_\nu\psi_B(x)\big]\big[\bar\psi_B(0)\notp_2\psi_A(0)\big]
|A,B\rangle {2N_c\over s^2}
\label{onexi24}
\end{eqnarray}
where we used $\Bxi_{1}=-\big(\bar\psi_A{1\over\alpha}\big)\gamma^iB_i{\slashed{p}_2\over s}$.
The most important terms are those proportional to $p_{2\mu}$. Using Fierz transformation and separating color singlets, they can be rewritten as
(cf. Eq. (\ref{onexi13}))
\begin{eqnarray}
&&\hspace{-1mm}
{N_c\over s}\langle A,B|\big[\Bxi_{1}(x)\gamma_\mu\psi_B(x)\big]\big[\bar\psi_B(0)\gamma_\nu\psi_A(0)\big]
+\mu\leftrightarrow\nu|A,B\rangle
\nonumber\\
&&\hspace{-1mm}
=~{p_{2\mu}\over s^3}\Big\{\big[\langle\bsi{1\over \alpha}(x)\gamma^i\notp_2\notp_1 \psi(0)\rangle_A
\langle\bsi(0)\gamma_\nu^\perp B_i(x)\psi(x)\rangle_B
\nonumber\\
&&\hspace{-1mm}
+~\langle\bsi{1\over \alpha}(x)\gamma^i\notp_2\gamma^\perp_\nu\psi(0)\rangle_A
\langle\bsi(0)\notp_1B_i(x)\psi(x)\rangle_B
+~(\psi(0)\otimes\psi(x)\leftrightarrow\gamma_5\psi(0)\otimes\gamma_5\psi(x)\big]
\nonumber\\
&&\hspace{-1mm}
-~\langle\bsi{1\over \alpha}(x)\gamma^i\notp_2\sigma_{\bu\xi}\psi(0)\rangle_A
\langle\bsi(0)\sigma_{\nu_\perp}^{~\xi}B_i(x)\psi(x)\rangle_B
\nonumber\\
&&\hspace{22mm}
-~\langle\bsi{1\over \alpha}(x)\gamma^i\notp_2\sigma_{\nu_\perp\xi}\psi(0)\rangle_A
\langle\bsi(0)\sigma_\bu^{~\xi}B_i(x)\psi(x)\rangle_B\Big\}~+~\mu\leftrightarrow\nu
\label{onexi25}
\end{eqnarray}
After some algebra with $\gamma$-matrices this can be transformed to
\begin{eqnarray}
&&\hspace{-1mm}
W^{(1)}_{2\mu\nu}(x)~=~{N_c\over s}\langle A,B|\big[\Bxi_{1}(x)\gamma_\mu\psi_B(x)\big]\big[\bar\psi_B(0)\gamma_\nu\psi_A(0)\big]
+\mu\leftrightarrow\nu|A,B\rangle~+~x\leftrightarrow 0
\nonumber\\
&&\hspace{5mm}
=~{p_{2\mu}\over s^3}\Big\{\langle\bsi{1\over \alpha}(x)\notp_2\psi(0)\rangle_A
\langle\bsi(0)\gamma_\nu^\perp\notp_1\notB(x)\psi(x)\rangle_B
\nonumber\\
&&\hspace{11mm}
-~i\langle\bsi{1\over \alpha}(x)\sigma_{\ast\nu_\perp}\psi(0)\rangle_A
\langle\bsi(0)\notp_1\notB(x)\psi(x)\rangle_B\Big\}~+~\mu\leftrightarrow\nu~+~x\leftrightarrow 0
\label{onexi26}
\end{eqnarray}
plus terms small in comparison to ${p_{2\mu}q^\perp_\nu\over\alpha_qs}$.  
Using Eq. (\ref{maelqg2}) we can transform $\big(\bar\psi{1\over \alpha}\big)(x)$ to 
\begin{eqnarray}
&&\hspace{-1mm}
\!\int\!dx_\bu d^2x_\perp~e^{-i\alpha x_\bu+i(k,x)_\perp}\langle\big(\bar\psi{1\over \alpha}\big)(x)\Gamma\psi(0)\rangle_A
\nonumber\\
&&\hspace{-1mm}
=~i\!\int\!dx_\bu \!\int_{-\infty}^{x_\bu} \!dx'_\bu~d^2x_\perp~e^{-i\alpha x_\bu+i(k,x)_\perp}\langle\bar\psi(x'_\bu,x_\perp)\Gamma\psi(0)\rangle_A
\nonumber\\
&&\hspace{-1mm}
=~-{1\over\alpha_q}\!\int\!dx_\bu ~d^2x_\perp~e^{-i\alpha x_\bu+i(k,x)_\perp}\langle\bar\psi(x_\bu,x_\perp)\Gamma\psi(0)\rangle_A
\label{onexi27}
\end{eqnarray}
Using QCD equation of motion and other formulas from
sections \ref{sec:paramlt} and \ref{sec:qqgparam} one gets
\begin{eqnarray}
&&\hspace{-1mm}
{1\over 16\pi^4}{1\over s}
\!\int\!dx_\bu dx_\ast d^2x_\perp~e^{-i\alpha x_\bu-i\beta x_\ast+i(q,x)_\perp}
\nonumber\\
&&\hspace{33mm}
\times~
\langle A,B|\big[\Bxi_{1}(x)\gamma_\mu\psi_B(x)\big]\big[\bar\psi_B(0)\gamma_\nu\psi_A(0)\big]+\mu\leftrightarrow\nu|A,B\rangle
\nonumber\\
&&\hspace{-1mm}
=~{p_{2\mu}\over \alpha_qs N_c}\Big[(q-k)_\nu f_1^f(\alpha_q,k_\perp)\barf _\perp^f(\beta_q,(q-k)_\perp)
\nonumber\\
&&\hspace{33mm}
-~k_\nu {(q-k)_\perp^2\over m^2}h_{1f}^\perp(\alpha_q,k_\perp)\barh _\perp^f(\beta_q,(q-k)_\perp)\Big]
~+~\mu\leftrightarrow\nu
\label{BxiA}
\end{eqnarray}
so the contribution of Eq. (\ref{XiA}) is effectively doubled. Again, the term with $x\leftrightarrow 0$ exchange
leads to Eq. (\ref{BxiA}) with  $f_1\leftrightarrow \barf_1$ and $h_1^\perp\leftrightarrow\barh_1^\perp$ replacement.

Thus, the sum of first and second terms in Eq. (\ref{kalw1}) leads to twice Eq. (\ref{XiA})
\begin{equation}
\hspace{-0mm}
W^{(1)\mu\nu}_{1+2}~=~
{2p_2^\mu\over \alpha_qs} {1\over N_c}\!\int\! d^2k_\perp\Big[(q-k)_\perp^\nu 
F^f(q,k_\perp)
-~k_\perp^\nu {(q-k)_\perp^2\over m^2}H^f(q,k_\perp)\Big]
~+~\mu\leftrightarrow\nu
\label{BXiA}
\end{equation}
%

\subsubsection{Term with $\Xi_{2}$}
In this Section we calculate the third term in Eq. (\ref{kalw1}).
\begin{equation}
\hspace{-0mm}
\cheW^{(1)}_{3\mu\nu}~=~{N_c\over s}\langle A,B|[\bar\psi_A(x)\gamma_\mu\Xi_{2}(x)\big]\big[\bar\psi_B(0)\gamma_\nu\psi_A(0)\big]
+\mu\leftrightarrow\nu|A,B\rangle~+~x\leftrightarrow 0
\label{onexi29}
\end{equation}
 Again, main contribution correspond to one index (e.g. $\mu$) being longitudinal and the other transverse so we need
\begin{eqnarray}
&&\hspace{-1mm}
\Big({2p_1^\mu p_2^{\mu'}\over s}+\mu\leftrightarrow\mu'\Big)
{N_c\over s}\langle A,B|\big[\bar\psi_A(x)\gamma_\mu\Xi_{2}(x)\big]\big[\bar\psi_B(0)\gamma_{\nu_\perp}\psi_A(0)\big]~+~\mu'\leftrightarrow\nu
|A,B\rangle
\nonumber\\
&&\hspace{-1mm}
=~\langle A,B|p_{1\mu}\big(\big[\bar\psi_A(x)\notp_2\Xi_{2}(x)\big]\big[\bar\psi_B(0)\gamma_{\nu_\perp}\psi_A(0)\big]~
+\big[\bar\psi_A(x)\gamma_{\nu_\perp}\Xi_{2}(x)\big]\big[\bar\psi_B(0)\notp_2\psi_A(0)\big]\big)
\nonumber\\
&&\hspace{-1mm}
+~p_{2\mu}\big[\bar\psi_A(x)\gamma_{\nu_\perp}\Xi_{2}(x)\big]\big[\bar\psi_B(0)\notp_1\psi_A(0)\big]
|A,B\rangle {2N_c\over s^2}
\label{onexi30}
\end{eqnarray}
(recall that   $\Xi_{2}(x)~=~-{\notp_1\over s}\gamma^iA_i{1\over\beta}\psi_B(x)$ so $\notp_1\Xi_{2}=0$).

Let us consider first the term proportional to $p_{1\mu}$.
Performing Fierz transformation (\ref{fierz})  and sorting out the color-singlet contributions we get (cf. Eq. (\ref{maelqg2}))
\begin{eqnarray}
&&\hspace{-1mm}
{2p_{1\mu}\over s}p_2^{\mu'}{N_c\over s}
\langle A,B|[\bar\psi_B(0)\gamma_{\mu'}\psi_A(0)\big]\big[\bar\psi_A(x)\gamma_{\nu_\perp}\Xi_{2}(x)\big]
+\mu'\leftrightarrow\nu|A,B\rangle
\nonumber\\
&&\hspace{-1mm}
=~{p_{1\mu}\over s^2}\Big\{{1\over 2}\big\{
\langle\bar\psi\gamma_\nu\brA_i(x)\psi(0)\rangle_A\langle\bar\psi(0)\gamma^i{1\over \beta}\psi(x)\rangle_B
~+~(\psi(0)\otimes\psi(x)\leftrightarrow \gamma_5\psi(0)\otimes\gamma_5\psi(x)\big\}
\nonumber\\
&&\hspace{-1mm}
-~{1\over s}\langle\bar\psi\notp_2\brA_{\nu}(x)\psi(0)\rangle_A\langle\bar\psi(0)\notp_1{1\over \beta}\psi(x)\rangle_B
~+~(\psi(0)\otimes\psi(x)\leftrightarrow \gamma_5\psi(0)\otimes\gamma_5\psi(x)\big\}
\nonumber\\
&&\hspace{-1mm}
-~{1\over s}
\langle\bar\psi A^j(x)\sigma_{\ast\nu_\perp}\psi(0)\rangle_A\langle\bar\psi(0)\sigma_{\bu j}{1\over \beta}\psi(x)\rangle_B
-~{2i\over s^2}
\langle\bar\psi A^i(x)\sigma_{\ast\bu}\psi(0)\rangle_A
\langle\bar\psi(0)\sigma_{\ast\nu_\perp}\sigma_{\bu i}{1\over \beta}\psi(x)\rangle_B
\nonumber\\
&&\hspace{-1mm}
+~{i\over s}\langle\bar\psi(0)\sigma_{\ast j}\sigma_{\bu i}{1\over \beta}\psi(x)\rangle_B
\langle\bar\psi A^i(x)\sigma_\nu^{~j}\psi(0)\rangle_A
+{i\over s}\langle\bar\psi(0)\sigma_{\nu_\perp j}\sigma_{\bu i}{1\over \beta}\psi(x)\rangle_B
\langle\bar\psi A_i(x)\sigma_\ast^{~j}\psi(0)\rangle_A\Big\}
\nonumber\\
\label{onexi31}
\end{eqnarray}
It is clear that  matrix elements in the first line in the r.h.s. can produce only transverse factors $\sim q_\perp$ so the corresponding contribution 
$\sim{p_{1\mu}q^\perp_\nu m_\perp^2\over\beta_q s^2}$ can be neglected. Also, matrix element 
$\langle\bar\psi(0)\notp_1{1\over \beta}\psi(x)\rangle_B$ vanishes as seen from Eq. (\ref{mael5}). The remaining terms can be rewritten as
\begin{eqnarray}
&&\hspace{-1mm}
{2p_{1\mu}\over s}p_2^{\mu'}{N_c\over s}
\langle A,B|[\bar\psi_B(0)\gamma_{\mu'}\psi_A(0)\big]\big[\bar\psi_A(x)\gamma_{\nu_\perp}\Xi_{2}(x)\big]
+\mu'\leftrightarrow\nu|A,B\rangle
\nonumber\\
&&\hspace{-1mm}
=~{p_{1\mu}\over s^3}\Big\{-\langle\bar\psi\notp_2\brA_{\nu}(x)\psi(0)\rangle_A\langle\bar\psi(0)\notp_1{1\over \beta}\psi(x)\rangle_B
-\langle\bar\psi A^j(x)\sigma_{\ast j}\psi(0)\rangle_A
\langle\bar\psi(0)\sigma_{\bu \nu_\perp}{1\over \beta}\psi(x)\rangle_B
\nonumber\\
&&\hspace{11mm}
+~
\langle\bar\psi(x)\big[A_{\nu_\perp}(x)\sigma_{\ast j}-A_j(x)\sigma_{\ast\nu_\perp}\big]\psi(0)\rangle_A
\langle\bar\psi(0)\sigma_\bu^{~j}{1\over \beta}\psi(x)\rangle_B
\nonumber\\
&&\hspace{11mm}
+~{is\over 2}\langle\bar\psi A^i(x)\sigma_\nu^{~j}\psi(0)\rangle_A
\langle\bar\psi(0)\big[g_{ij}-i\epsilon_{ij}\gamma_5-i\sigma_{ji}+{2\over s}g_{ij}\sigma_{\ast\bu}\big]{1\over \beta}\psi(x)\rangle_B
\nonumber\\
&&\hspace{11mm}
-~i\langle\bar\psi A^i(x)\sigma_{\ast\bu}\psi(0)\rangle_A
\langle\bar\psi(0)\big[g_{i\nu}-i\epsilon_{i\nu_\perp}\gamma_5+i\sigma_{i\nu_\perp}+{2\over s}g_{i\nu}\sigma_{\ast\bu}\big]{1\over \beta}\psi(x)\rangle_B\Big\}
\label{onexi32}
\end{eqnarray}
where we have used Eq. (\ref{sigmasigmas}).
It is clear that only the first line in the r.h.s. can give the non-negligible contribution to $W_{\mu\nu}$. Indeed,
matrix element $\langle\bar\psi(x)\big[A_{\nu_\perp}(x)\sigma_{\ast j}-A_j(x)\sigma_{\ast\nu_\perp}\big]\psi(0)\rangle_A $ 
vanishes for unpolarized hadrons due to parity, see Eq. (\ref{formula2}).  In the third line in r.h.s., neither matrix element can produce $s$
so the corresponding contribution is again $\sim{p_{1\mu}q^\perp_\nu m_\perp^2\over\alpha_q s^2}$ while contribution from the last line 
is even smaller,   of order of ${p_{1\mu}q^\perp_\nu m_\perp^4\over\alpha_q s^3}$. Thus, we get
\begin{eqnarray}
&&\hspace{-1mm}
{2p_{1\mu}\over s}p_2^{\mu'}{N_c\over s}
\langle A,B|[\bar\psi_B(0)\gamma_{\mu'}\psi_A(0)\big]\big[\bar\psi_A(x)\gamma_{\nu_\perp}\Xi_{2}(x)\big]
+\mu'\leftrightarrow\nu|A,B\rangle
\label{onex33}\\
&&\hspace{-1mm}
=~{p_{1\mu}\over s^3}\Big\{\langle\bar\psi\notA(x)\notp_2\gamma_{\nu_\perp}\psi(0)\rangle_A\langle\bar\psi(0)\notp_1{1\over \beta}\psi(x)\rangle_B
+i\langle\bar\psi\notA(x)\notp_2\psi(0)\rangle_A
\langle\bar\psi(0)\sigma_{\bu \nu_\perp}{1\over \beta}\psi(x)\rangle_B
\nonumber
\end{eqnarray}
It remains to prove that the last term in Eq. (\ref{onexi30}) proportional to $p_{2\mu}$ is small. One can rewrite that term
similarly to Eq. (\ref{onexi21}) with replacement 
$p_1\leftrightarrow p_2$ and (projectile matrix elements) $\leftrightarrow$ 
(target ones). After that, the proof repeats arguments after Eq. (\ref{onexi21}) and one obtains the estimate
\begin{equation}
{2p_{2\mu}\over s^2}N_c\langle A,B\big[\bar\psi_A(x)\gamma_{\nu_\perp}\Xi_{2}(x)\big]\big[\bar\psi_B(0)\notp_1\psi_A(0)\big]|A,B\rangle
~\sim~p_{2\mu}q^\perp_\nu{m_\perp^2\over\beta_qs^2}
\label{onexi34}
\end{equation}
Similarly, by repeating arguments from  Section \ref{sec:onexifirst} with replacement $p_1\leftrightarrow p_2$ and projectile matrix 
elements $\leftrightarrow$ target ones, one can demonstrate that terms in Eq. (\ref{onexi29}) with $\mu,\nu$ both longitudinal or both transverse 
are small in comparison to terms listed in Eq. (\ref{pc}).

Thus, 
\begin{eqnarray}
&&\hspace{-1mm}
\cheW^{(1)}_{3\mu\nu}(x)~=~{N_c\over s}\langle A,B|[\bar\psi_A(x)\gamma_\mu\Xi_{2}(x)\big]\big[\bar\psi_B(0)\gamma_\nu\psi_A(0)\big]
+\mu\leftrightarrow\nu|A,B\rangle~+~x\leftrightarrow 0
\nonumber\\
&&\hspace{2mm}
=~{p_{1\mu}\over s^2}
\big[
\langle\bar\psi\notA(x_\bu,x_\perp)\notp_2\gamma_{\nu_\perp}\psi(0)\rangle_A\langle\bar\psi(0)\notp_1{1\over \beta}\psi(x_\ast,x'_\perp)\rangle_B
\nonumber\\
&&\hspace{5mm}
+~i\langle\bar\psi\notA(x_\bu,x_\perp)\notp_2\psi(0)\rangle_A
\langle\bar\psi(0)\sigma_{\bu \nu_\perp}{1\over \beta}\psi(x_\ast,x'_\perp)\rangle_B~+~\mu\leftrightarrow\nu\Big]~+~x\leftrightarrow 0
\label{chew13}
\end{eqnarray}
Using QCD equation of motion and formulas from Appendix, we obtain the corresponding contribution to $W_{\mu\nu}$ in the form
\begin{eqnarray}
&&\hspace{-1mm}
{1\over 16\pi^4}{1\over s}
\!\int\!dx_\bu dx_\ast d^2x_\perp~e^{-i\alpha x_\bu-i\beta x_\ast+i(q,x)_\perp}
\nonumber\\
&&\hspace{22mm}
\times~
\langle A,B|\big[\bar\psi_A (x)\gamma_\mu\Xi_{2}(x)\big]
[\bar\psi_B(0)\gamma_\nu\psi_A(0)\big]|A,B\rangle+\mu\leftrightarrow\nu
\nonumber\\
&&\hspace{-1mm}
=~{1\over 64\pi^6N_c}{p_{1\mu}\over s^3}\!\int\! dx_\ast dx_\bu\!\int\! d^2k_\perp
\!\int\!dx_\bu d^2x_\perp~e^{-i\alpha x_\bu+i(k,x)_\perp}\!\int\!dx_\ast d^2x'_\perp e^{-i\beta x_\bu+i(q-k,x')_\perp}
\nonumber\\
&&\hspace{15mm}
\times~\big[
\langle\bar\psi\notA(x_\bu,x_\perp)\notp_2\gamma_{\nu_\perp}\psi(0)\rangle_A\langle\bar\psi(0)\notp_1{1\over \beta}\psi(x_\ast,x'_\perp)\rangle_B
\nonumber\\
&&\hspace{22mm}
+~i\langle\bar\psi\notA(x_\bu,x_\perp)\notp_2\psi(0)\rangle_A
\langle\bar\psi(0)\sigma_{\bu \nu_\perp}{1\over \beta}\psi(x_\ast,x'_\perp)\rangle_B~+~\mu\leftrightarrow\nu\Big]
\nonumber\\
&&\hspace{-1mm}
=~{p_{1\mu}\over \beta_qs N_c}\!\int\! d^2k_\perp\Big[k_\nu f_1^f(\alpha_q,k_\perp)\barf _\perp^f(\beta_q,(q-k)_\perp)
\nonumber\\
&&\hspace{33mm}
-~(q-k)_\nu {k_\perp^2\over m^2}h_{1f}^\perp(\alpha_q,k_\perp)\barh _\perp^f(\beta_q,(q-k)_\perp)\Big]
~+~\mu\leftrightarrow\nu
\label{onexi35}
\end{eqnarray}
Same as in previous Section, the term with $x\leftrightarrow 0$ exchange
leads to Eq. (\ref{onexi35}) with  $f_1\leftrightarrow \barf_1$ and $h_1^\perp\leftrightarrow\barh_1^\perp$ replacement so we get
\begin{eqnarray}
&&\hspace{-1mm}
W^{(1)}_{3\mu\nu}~=~
{p_{1\mu}\over \beta_qs N_c}\!\int\! d^2k_\perp\Big[k_\nu F^f(q,k_\perp)-~(q-k)_\nu {k_\perp^2\over m^2}H^f(q,k_\perp)\Big]
~+~\mu\leftrightarrow\nu 
\label{XiB}
\end{eqnarray}
Repeating arguments from previous Section, it is possible to show that the
contribution of the fourth term
\begin{eqnarray}
&&\hspace{-1mm}
\cheW^{(1)}_{4\mu\nu}(x)~=~{N_c\over s}\langle A,B|[\bar\psi_A(x)\gamma_\mu\Xi_{2}(x)\big]\big[\bar\psi_B(0)\gamma_\nu\psi_A(0)\big]
+\mu\leftrightarrow\nu|A,B\rangle~+~x\leftrightarrow 0
\nonumber\\
&&\hspace{2mm}
=~{p_{1\mu}\over s^2}
\big[
\langle\bar\psi\gamma_{\nu_\perp}\notp_2\notA(0)\psi(0)\rangle_A\langle\big(\bar\psi{1\over \beta}\big)(0)\notp_1\psi(x_\ast,x'_\perp)\rangle_B
\nonumber\\
&&\hspace{5mm}
-~i\langle\bar\psi(x_\ast,x'_\perp\notp_2\notA(0)\psi(0)\rangle_A
\langle\big(\bar\psi{1\over \beta}\big)(0)\sigma_{\bu \nu_\perp}\psi(x_\ast,x'_\perp)\rangle_B~+~\mu\leftrightarrow\nu\Big]~+~x\leftrightarrow 0
\label{chew14}
\end{eqnarray}
doubles that of the third term so we get the full contribution of the terms with one quark-antiquark-gluon operators
in the form
\begin{eqnarray}
&&\hspace{-1mm}
W^{(1)}_{\mu\nu}~=~
{2\over N_c}\!\int\! d^2k_\perp\Big[\Big({p_{1\mu}k^\perp_\nu\over\beta_qs}+{p_{2\mu}(q-k)^\perp_\nu\over\alpha_qs}
\Big)F^f(q,k_\perp)
\nonumber\\
&&\hspace{-1mm}
-~\Big({p_{1\mu}(q-k)^\perp_\nu\over\beta_qs}{k_\perp^2\over m^2}+{p_{2\mu} k_\perp^\nu\over\alpha_qs} {(q-k)_\perp^2\over m^2}\Big) 
H^f(q,k_\perp)\Big]
~+~\mu\leftrightarrow\nu 
\label{XiBxi}
\end{eqnarray}
This result agrees with the corresponding $1/Q$ terms in Ref. \cite{Mulders:1995dh}.

\subsection{Term  with two quark-quark-gluon operators coming from $\Xi_{1}$ and $\Xi_{2}$\label{sec:2qqGa}}
Let us start with the first term in the r.h.s. of Eq. (\ref{kalw2a}).
Performing Fierz transformation (\ref{fierz}) we obtain
\begin{equation}
\hspace{1mm}
{N_c\over s}\langle A,B|(\bsi_A^m(x)\gamma_\mu\Xi_{2}^m(x))(\bsi_B^n(0)\gamma_\nu\Xi_{1}^n(0))
+\mu\leftrightarrow\nu|A,B\rangle+x\leftrightarrow 0
~=~g_{\mu\nu}\cheV_1+\cheV_{2\mu\nu}+\cheV_{3\mu\nu}
\label{kalvs}
\end{equation}
where 
\begin{eqnarray}
&&\hspace{-3mm}
\cheV_1~
=~{N_c\over 2s}\langle A,B|-[\bsi_A^n(x)\Xi_{1}^m(0)][\bsi_B^n(0)\Xi_{2}^m(x)]
+[\bsi_A^m(x)\gamma_5\Xi_{1}^n(0)][\bsi_B^n(0)\gamma_5\Xi_{2}^m(x)]
\label{kalv1}\\
&&\hspace{-3mm}
+~[\bsi_A^m(x)\gamma_\alpha\Xi_{1}^m(0)][\bsi_B^n(0)\gamma^\alpha\Xi_{2}^n(x)]
+[\bsi_A^m(x)\gamma_\alpha\gamma_5\Xi_{1}^m(0)]
[\bsi_B^n(0)\gamma^\alpha\gamma_5\Xi_{2}^n(x)]|A,B\rangle+x\leftrightarrow 0,
\nonumber\end{eqnarray}
\begin{eqnarray}
&&\hspace{-1mm}
\cheV_{2\mu\nu}~=~{N_c\over 2s}\langle A,B|
-~[\bsi_A^m(x)\gamma_\mu\Xi_{1}^n(0)][\bsi_B^n(0)\gamma_\nu\Xi_{2}^m(x)]
\nonumber\\
&&\hspace{11mm}
-~[\bsi_A^m(x)\gamma_\mu\gamma_5\Xi_{1}^n(0)][\bsi_B^n(0)\gamma_\nu\gamma_5\Xi_{2}^m(x)]
+\mu\leftrightarrow\nu|A,B\rangle+x\leftrightarrow 0,
\label{kalv2}
\end{eqnarray}
and
\begin{eqnarray}
&&\hspace{-1mm}
\cheV_{3\mu\nu}~=~{N_c\over 2s}\langle A,B|
[(\bsi_A^m(x)\sigma_{\mu\alpha}\Xi_{1}^n(0)][\bsi_B^n(0)\sigma_\nu^{~\alpha}\Xi_{2}^m(x)]+\mu\leftrightarrow\nu
\nonumber\\
&&\hspace{11mm}
-~g_{\mu\nu}[\bsi_A^m(x)\sigma^{\alpha\beta}\Xi_{1}^n(0)][\bsi_B^n(0)\sigma_{\alpha\beta}\Xi_{2}^m(x)
|A,B\rangle+x\leftrightarrow 0
\label{kalv3}
\end{eqnarray}
It is convenient to define $\cheV_{3\mu\nu}$ to be traceless. In next Sections, we will consider these terms in turn.

\subsubsection{Term propotional to $g_{\mu\nu}$}
Using $\Xi_{1}~=~-{g\slashed{p}_2\over s}\gamma^iB_i{1\over \alpha+i\epsilon}\psi_A$ and 
$\Xi_{2}~=~-{g\slashed{p}_1\over s}\gamma^iA_i{1\over\beta+i\epsilon}\psi_B$ from Eq. (\ref{fildz0}) and 
extracting color-singlet contributions one obtains
\begin{eqnarray}
&&\hspace{-1mm}
\cheV_1~=~{1\over 2s^3}
\nonumber\\
&&\hspace{-1mm}
\times~\Big\{-\Big[\langle \bsi A_i(x)\notp_2\gamma^j{1\over \alpha}\psi(0)\rangle_A
\langle\bsi B_j(0)\notp_1\gamma^i{1\over\beta}\psi(x)\rangle_B
-\psi(0)\otimes\psi(x)\leftrightarrow\gamma_5\psi(0)\otimes\gamma_5\psi(x)\Big]
\nonumber\\
&&\hspace{-1mm}
+~\Big[\langle\bsi A_i(x)\gamma_k\notp_2\gamma^j{1\over \alpha}\psi(0)\rangle_A
\langle\bsi B_j(0)\gamma^k\notp_1\gamma^i{1\over\beta}\psi(x)\rangle_B
+~\psi(0)\otimes\psi(x)\leftrightarrow\gamma_5\psi(0)\otimes\gamma_5\psi(x)\Big]
\nonumber\\
&&\hspace{14mm}
+~{2\over s}\Big[\langle\bsi A_i(x)\notp_1\notp_2\gamma^j{1\over \alpha}\psi(0)\rangle_A
\langle\bsi B_j(0)\notp_2\notp_1\gamma^i{1\over\beta}\psi(x)\rangle_B
\nonumber\\
&&\hspace{30mm}
+~\psi(0)\otimes\psi(x)\leftrightarrow\gamma_5\psi(0)\otimes\gamma_5\psi(x)\Big]\Big\}~+~x\leftrightarrow 0
\label{twoxia1}
\end{eqnarray}
Let us start with the first term. Using Eq. (\ref{gammas11}) and the fact that $\langle\bar\psi(x)\big[A_k\sigma_{\ast j}-A_j(x)\sigma_{\ast k}\big]\psi(0)\rangle_A=0$ (see the footnote \ref{zanulil}), 
we obtain
\begin{eqnarray}
&&\hspace{-1mm}
-{1\over 2s^3}\langle \bsi{\notA}(x)\notp_2{1\over \alpha}\psi(0)\rangle_A
\langle\bsi{\notB}(0)\notp_1{1\over\beta}\psi(x)\rangle_B
-{1\over 4s^2}\langle \bsi{\notA}(x)\gamma_i{1\over \alpha}\psi(0)\rangle_A
\langle\bsi{\notB}(0)
\gamma^i{1\over\beta}\psi(x)\rangle_B\Big]
\nonumber\\
&&\hspace{11mm}
-~{1\over 8s^2}[\langle \bsi A^i(x)\sigma_{jk}{1\over \alpha}\psi(0)\rangle_A
\langle\bsi B_i(0)\sigma^{jk}{1\over\beta}\psi(x)\rangle_B
\nonumber\\
&&\hspace{22mm}
=~-{1\over 2s^3}\langle \bsi{\notA}(x)\notp_2{1\over \alpha}\psi(0)\rangle_A
\langle\bsi{\notB}(0)\notp_1{1\over\beta}\psi(x)\rangle_B\Big[1+O\big({q_\perp^2\over s}\big)\Big]
\label{twoxia2}
\end{eqnarray}
where we used the fact that projectile and target matrix elements in the two last terms in the l.h.s. cannot produce factor of $s$.

Next, consider second term in Eq. (\ref{twoxia1}). Using Eqs. (\ref{gammas6a}) and (\ref{gammas11}), one can rewrite is as
\begin{eqnarray}
&&\hspace{-1mm}
{1\over 2s^3}\Big[\langle\bsi A_i(x)\gamma_k\notp_2\gamma^j{1\over \alpha}\psi(0)\rangle_A
\langle\bsi B_j(0)\gamma^k\notp_1\gamma^i{1\over\beta}\psi(x)\rangle_B
+~\psi(0)\otimes\psi(x)\leftrightarrow\gamma_5\psi(0)\otimes\gamma_5\psi(x)\Big]
\nonumber\\
&&\hspace{14mm}
=~{1\over s^3}\langle\bsi{\notA}(x)\notp_2\gamma^i{1\over \alpha}\psi(0)\rangle_A
\langle\bsi{\notB}(0)\notp_1\gamma_i{1\over\beta}\psi(x)\rangle_B
\label{twoxia3}
\end{eqnarray}
Similarly, from Eq. (\ref{gammas11}) we get the third term in the form
\begin{eqnarray}
&&\hspace{-1mm}
{1\over s^4}\Big[\langle\bsi A_i(x)\notp_1\notp_2\gamma^j{1\over \alpha}\psi(0)\rangle_A
\langle\bsi B_j(0)\notp_2\notp_1\gamma^i{1\over\beta}\psi(x)\rangle_B
+~\psi(0)\otimes\psi(x)\leftrightarrow\gamma_5\psi(0)\otimes\gamma_5\psi(x)\Big]
\nonumber\\
&&\hspace{-1mm}
=~{1\over 4s^2}\Big[\langle\bsi{\brA}_i(x)\gamma^j{1\over \alpha}\psi(0)\rangle_A
\langle\bsi{\breB}_j(0)\gamma^i{1\over\beta}\psi(x)\rangle_B
+~\psi(0)\otimes\psi(x)\leftrightarrow\gamma_5\psi(0)\otimes\gamma_5\psi(x)\Big]
\label{twoxia4}
\end{eqnarray}
Since both projectile and target matrix elements cannot give factor $s$ this contribution is $O\big({q_\perp^2\over s}\big)$ in comparison to that of 
the two  first  terms.
Thus, we get
\begin{eqnarray}
&&\hspace{-1mm}
\cheV_1~=~
{1\over s^3}\Big(\half\langle \bsi{\notA}(x)\notp_2{1\over \alpha}\psi(0)\rangle_A
\langle\bsi{\notB}(0)\notp_1{1\over\beta}\psi(x)\rangle_B
\nonumber\\
&&\hspace{5mm}
+~\langle\bsi{\notA}(x)\notp_2\gamma^i{1\over \alpha}\psi(0)\rangle_A
\langle\bsi{\notB}(0)\notp_1\gamma_i{1\over\beta}\psi(x)\rangle_B\Big)\Big[1+O\big({q_\perp^2\over s}\big)\Big]~+~x\leftrightarrow 0
\label{chve1}
\end{eqnarray}
Next, using QCD equations of motion (\ref{eqm2}), (\ref{11.43}) and formulas from Appendix \ref{sec:paramlt},  we obtain the contribution to $W_{\mu\nu}$ in the form
\begin{eqnarray}
&&\hspace{-1mm}
g_{\mu\nu} V_1(q)
~=~{g_{\mu\nu}\over 16\pi^4N_c}\!\int\! dx_\bu dx_\ast d^2x_\perp~e^{-i\alpha_qx_\bu-i\beta_qx_\ast+i(q,x)_\perp}\cheV_1(x) 
\nonumber\\
&&\hspace{5mm}
=~{g_{\mu\nu}\over \alpha_q\beta_qsN_c}\int\! d^2k_\perp\Big[(k,q-k)_\perp F^f(q,k_\perp)-~{1\over 2m^2} k_\perp^2(q-k)_\perp^2H^f(q,k_\perp)\Big]
\label{v1otvet}
\end{eqnarray}
where replacements $f_1^f\leftrightarrow\barf^f_1$ and $h_{1f}^\perp\leftrightarrow\barh_{1f}^\perp$ come from $x\leftrightarrow 0$ term.

\subsubsection{Term with TMD's $f_1$ \label{sec:kalv2}}
Separating color-singlet contributions one can rewrite Eq. (\ref{kalv2}) as
\begin{eqnarray}
&&\hspace{-1mm}
\cheV_{2\mu\nu}~=~-{1\over 2s^3}\big\{\langle\bsi A_i(x)\gamma_\mu\notp_2\gamma^j{1\over \alpha}\psi(0)\rangle_A
\langle\bsi B_j(0)\gamma_\nu\notp_1\gamma^i{1\over\beta}\psi(x)\rangle_B
\nonumber\\
&&\hspace{10mm}
+~\psi(0)\otimes\psi(x)\leftrightarrow\gamma_5\psi(0)\otimes\gamma_5\psi(x)+\mu\leftrightarrow\nu\}
~+~x\leftrightarrow 0
\label{kalve2}
\end{eqnarray}
We need to consider three cases: both $\mu$ and $\nu$ are transverse, both of them are longitudinal, and 
$\mu$ is longitudinal and $\nu$ transverse (plus {\it vice versa}). 

In the first case we can use formula (\ref{formula9}) and get
\begin{eqnarray}
&&\hspace{-1mm}
\cheV_{2\mu_\perp\nu_\perp}~
=~-{1\over 2s^3}\big\{\langle\bsi A_i(x)\gamma_{\mu_\perp}\notp_2\gamma^j{1\over \alpha}\psi(0)\rangle_A
\langle\bsi B_j(0)\gamma_{\nu_\perp}\notp_1\gamma^i{1\over\beta}\psi(x)\rangle_B
\nonumber\\
&&\hspace{10mm}
+~\psi(0)\otimes\psi(x)\leftrightarrow\gamma_5\psi(0)\otimes\gamma_5\psi(x)+\mu\leftrightarrow\nu\}
~+~x\leftrightarrow 0
\nonumber\\
&&\hspace{10mm}
=~-{g_{\mu\nu}^\perp\over s^3}\langle\bsi\notA(x)\notp_2\gamma_i{1\over \alpha}\psi(0)\rangle_A
\langle\bsi\notB(0)\notp_1\gamma^i{1\over\beta}\psi(x)\rangle_B~+~x\leftrightarrow 0
\label{kalvedvaperp}
\end{eqnarray}
which gives the contribution to $W_{\mu\nu}$ in the form
\begin{eqnarray}
&&\hspace{-1mm}
V_{2\mu_\perp\nu_\perp}~
=~{1\over 16\pi^4N_c}\!\int\! dx_\bu dx_\ast d^2x_\perp~e^{-i\alpha_qx_\bu-i\beta_qx_\ast+i(q,x)_\perp}\cheV_{2\mu_\perp\nu_\perp}(x) 
\nonumber\\
&&\hspace{-1mm}
=~-{g_{\mu\nu}^\perp\over \alpha_q\beta_qsN_c}\int\! d^2k_\perp (k,q-k)_\perp F^f(q,k_\perp)\label{kalv2glav}
\end{eqnarray}
where we again used formulas from  Appendices \ref{sec:paramlt} and \ref{sec:qqgparam}.

Next, if both $\mu$ and $\nu$ are longitudinal, we get
\begin{eqnarray}
&&\hspace{-1mm}
\cheV_{2\mu\nu}~=~-{2\over s^5}(p_{1\mu}p_{2\nu}+\mu\leftrightarrow\nu)
\big\{\langle\bsi A_i(x)\notp_1\notp_2\gamma^j{1\over \alpha}\psi(0)\rangle_A
\langle\bsi B_j(0)\notp_2\notp_1\gamma^i{1\over\beta}\psi(x)\rangle_B
\nonumber\\
&&\hspace{10mm}
+~\psi(0)\otimes\psi(x)\leftrightarrow\gamma_5\psi(0)\otimes\gamma_5\psi(x)\}
~+~x\leftrightarrow 0
\label{kalve2prodol}
\end{eqnarray}
Using formula (\ref{gammas10}) we rewrite r.h.s. of Eq. (\ref{kalve2}) as follows
\begin{eqnarray}
&&\hspace{-1mm}
\cheV_{2\mu\nu}~=~-{1\over 2s^3}(p_{1\mu}p_{2\nu}+\mu\leftrightarrow\nu)
\big\{\langle\bsi\brA_i(x)\gamma^j{1\over \alpha}\psi(0)\rangle_A
\langle\bsi\breB_j(0)\gamma^i{1\over\beta}\psi(x)\rangle_B
\nonumber\\
&&\hspace{10mm}
+~\psi(0)\otimes\psi(x)\leftrightarrow\gamma_5\psi(0)\otimes\gamma_5\psi(x)\}
~+~x\leftrightarrow 0
\label{kalve2prodo}
\end{eqnarray}
Since matrix elements in the r.h.s. cannot give factor $s$, the contribution of this term to $W_{\mu\nu}$ is $\sim{q_\perp^2\over s}$ times that
of Eq. (\ref{kalv2glav}).

Finally, let us consider the case when one index is longitudinal and the other transverse. Using Eq. (\ref{gammas16}) we get
\begin{eqnarray}
&&\hspace{-1mm}
\cheV_{2\mu\nu}^{\parallel\perp}~=~-{1\over s^4}\big\{p_{2\mu}\langle\bsi A_i(x)\notp_1\notp_2\gamma^j{1\over \alpha}\psi(0)\rangle_A
\langle\bsi B_j(0)\gamma_{\nu_\perp}\notp_1\gamma^i{1\over\beta}\psi(x)\rangle_B
\nonumber\\
&&\hspace{10mm}
+~p_{1\mu}\langle\bsi A_i(x)\gamma_{\nu_\perp}\notp_2\gamma^j{1\over \alpha}\psi(0)\rangle_A
\langle\bsi B_j(0)\notp_2\notp_1\gamma^i{1\over\beta}\psi(x)\rangle_B
\nonumber\\
&&\hspace{10mm}
+~\psi(0)\otimes\psi(x)\leftrightarrow\gamma_5\psi(0)\otimes\gamma_5\psi(x)+\mu\leftrightarrow\nu\}
~+~x\leftrightarrow 0
\nonumber\\
&&\hspace{-1mm} 
=~-{1\over 2s^3}\big\{p_{2\mu}\langle\bsi(x)\gamma^iA_\nu(x){1\over \alpha}\psi(0)\rangle_A
\langle\bsi{\notB}(0)\notp_1\gamma^i{1\over\beta}\psi(x)\rangle_B
\nonumber\\
&&\hspace{10mm}
+~p_{1\mu}\langle\bsi{\notA}(x)\notp_2\gamma^i{1\over \alpha}\psi(0)\rangle_A
\langle\bsi(0)\gamma^i B_\nu(0){1\over\beta}\psi(x)\rangle_B
\nonumber\\
&&\hspace{10mm}
+~\psi(0)\otimes\psi(x)\leftrightarrow\gamma_5\psi(0)\otimes\gamma_5\psi(x)+\mu\leftrightarrow\nu\}
~+~x\leftrightarrow 0
\label{kalve2perpro}
\end{eqnarray}
It is clear that $\langle\bsi{\notA}(x)\notp_2\gamma^i{1\over \alpha}\psi(0)\rangle_A$ and 
$\langle\bsi{\notB}(0)\notp_1\gamma^i{1\over\beta}\psi(x)\rangle_B$ bring one factor $s$ so  
\begin{eqnarray}
&&\hspace{-1mm}
\cheV_{2\mu\nu}^{\parallel\perp}~\sim~{p_{1\mu}q_{\perp_\nu}+\mu\leftrightarrow\nu\over\alpha_q\beta_q s^2}m_\perp^2
~~{\rm or}~~\sim~{p_{2\mu}q_{\perp_\nu}+\mu\leftrightarrow\nu\over\alpha_q\beta_q s^2}m_\perp^2
\label{v2small}
\end{eqnarray}
which is ${1\over \alpha_q s}$ or ${1\over \beta_q s}$ correction in comparison to Eq. (\ref{XiBxi}). Thus, the contribution to $W_{\mu\nu}$ is given by Eq. (\ref{kalv2glav})
\begin{equation}
\hspace{0mm}
V_{2\mu\nu}~
=~-{g_{\mu\nu}^\perp\over \alpha_q\beta_qsN_c}\int\! d^2k_\perp (k,q-k)_\perp F^f(q,k_\perp)
\label{v2otvet}
\end{equation}
%

\subsubsection{Term with TMD's $h_1^\perp$ \label{sec:v3}}
Let us consider now 
\begin{eqnarray}
&&\hspace{-1mm}
\cheV'_{3\mu\nu}~=~{N_c\over 2s}\langle A,B|
[(\bsi_A^m(x)\sigma_{\mu\alpha}\Xi_{1}^m(0)][\bsi_B^n(0)\sigma_\nu^{~\alpha}\Xi_{2}^n(x)]|A,B\rangle 
+\mu\leftrightarrow\nu+x\leftrightarrow 0
\label{kalv3a}
\end{eqnarray}
(the trace will be subtracted after the calculation). Separating color-singlet contributions, we get
\begin{equation}
\hspace{0mm}
\cheV'_{3\mu\nu}~=~{1\over 2s^3}\langle\bsi A_i(x)\sigma_{\mu\alpha}\notp_2\gamma^j{1\over \alpha}\psi(0)\rangle_A
\langle\bsi B_j(0)\sigma_\nu^{~\alpha}\notp_1\gamma^i{1\over\beta}\psi(x)\rangle_B+\mu\leftrightarrow\nu
+x\leftrightarrow 0
\label{kalv3b}
\end{equation}

First case is when $\mu$ and $\nu$ are transverse
\begin{eqnarray}
&&\hspace{-1mm}
\cheV'_{3\mu_\perp\nu_\perp}
~=~-{1\over 2s^3}\langle\bsi A^i(x)\sigma_{\mu_\perp k}\sigma_{\ast j}{1\over \alpha}\psi)0)\rangle_A
\langle\bsi B^j(0)\sigma_{\nu_\perp}^{~k}\sigma_{\bu i}{1\over\beta}\psi(x)\rangle_B
\nonumber\\
&&\hspace{-1mm}
-~{1\over s^4}\langle\bsi A^i(x)\sigma_{\bu\mu_\perp}\sigma_{\ast j}{1\over \alpha}\psi(0)\rangle_A
\langle\bsi B^j(0)\sigma_{\ast\nu_\perp}\sigma_{\bu i}{1\over\beta}\psi(x)\rangle_B
+\mu\leftrightarrow\nu
+x\leftrightarrow 0
\label{kalv3perp}
\end{eqnarray}
With the help of Eq.  (\ref{sigmasigmas}) the first term in the r.h.s. turns to
\begin{eqnarray}
&&\hspace{-3mm}
{1\over 2s^3}\langle\bsi A^i(x)[g_{\mu j}\sigma_{\ast k}-g_{jk}\sigma_{\ast\mu_\perp}]{1\over \alpha}\psi(0)\rangle_A
\langle\bsi B^j(0)[g_{\nu i}\sigma_\bu^{~k}-\delta_i^k\sigma_{\bu\nu_\perp}]{1\over\beta}\psi(x)\rangle_B
+\mu\leftrightarrow\nu+x\leftrightarrow 0
\nonumber\\
&&\hspace{-3mm}
=~{1\over 2s^3}\Big(\langle\bsi A_\nu(x)\sigma_{\ast k}{1\over \alpha}\psi(0)\rangle_A
\langle\bsi B_\mu(0)\sigma_\bu^{~k}{1\over\beta}\psi(x)\rangle_B
\nonumber\\
&&\hspace{-3mm}
-~\langle\bsi A_\nu(x)\sigma_{\ast\mu_\perp}{1\over \alpha}\psi(0)\rangle_A
\langle\bsi B^j(0)\sigma_{\bu j}{1\over\beta}\psi(x)\rangle_B
-~\langle\bsi A^i(x)\sigma_{\ast i}{1\over \alpha}\psi(0)\rangle_A
\langle\bsi B_\mu(0)\sigma_{\bu\nu_\perp}{1\over\beta}\psi(x)\rangle_B
\nonumber\\
&&\hspace{11mm}
+~\langle\bsi A^i(x)\sigma_{\ast\mu_\perp}{1\over \alpha}\psi(0)\rangle_A
\langle\bsi B_i(0)\sigma_{\bu\nu_\perp}{1\over\beta}\psi(x)\rangle_B\Big)
+\mu\leftrightarrow\nu+x\leftrightarrow 0
\label{kalv3perpe}
\end{eqnarray}
After some algebra, it can be rewritten as
\begin{eqnarray}
&&\hspace{-1mm}
-{g_{\mu\nu}^\perp\over 2s^3}\langle\bsi A^i(x)\sigma_{\ast i}{1\over \alpha}\psi(0)\rangle_A
\langle\bsi B^j(0)\sigma_{\bu j}{1\over\beta}\psi(x)\rangle_B
\nonumber\\
&&\hspace{11mm}
+~{1\over s^3}\big\{\langle\bsi\big(A_k\sigma_{\ast \mu_\perp}-\half g_{\mu k}\sigma_{\ast j}A^j\big)(x){1\over \alpha}\psi(0)\rangle_A
\nonumber\\
&&\hspace{22mm}
\times~\langle\bsi\big(B^k\sigma_{\bu \nu_\perp}-\half \delta_\nu^k\sigma_{\bu j}B^j\big)(0){1\over\beta}\psi(x)\rangle_B
+\mu\leftrightarrow\nu\big\}
+x\leftrightarrow 0
\label{kalv3perpee}
\end{eqnarray}
where again we used property (\ref{formula2}).  Using QCD equations of motion (\ref{11.41}), (\ref{11.43}) and parametrization (\ref{maelsa}) 
one can write the corresponding contribution to $W_{\mu\nu}$ as
\begin{eqnarray}
&&\hspace{-11mm}
V'_{3\mu_\perp\nu_\perp}~=~{g^\perp_{\mu\nu}\over 2\alpha_q\beta_qsN_c}\!\int d^2k_\perp {k_\perp^2(q-k)_\perp^2\over m^2}
H^f(q,k_\perp)
\label{kalvklad3perp}\\
&&\hspace{3mm}
+~{1\over \alpha_q\beta_qsN_c}\!\int {d^2k_\perp\over m^2} \big[(k^\perp_\mu(q-k)^\perp_\nu+\mu\leftrightarrow\nu)(k,q-k)_\perp
-k_\perp^2(q-k)^\perp_\mu(q-k)^\perp_\nu
\nonumber\\
&&\hspace{3mm}
-~(q-k_\perp)^2k^\perp_\mu k^\perp_\nu
-{g_{\mu\nu}^\perp\over 2}k_\perp^2(q-k_\perp)^2\big]
H_A^f(q,k_\perp)
\nonumber
\end{eqnarray}
where we introduced the notation
\begin{eqnarray}
&&\hspace{-11mm}
H_A^f(q,k_\perp)~\equiv~h_A^f (\alpha_q,k_\perp)\barh_A^f(\beta_q,(q-k)_\perp)+h_A^f\leftrightarrow\barh_A^f\
\label{HA}
\end{eqnarray}

The second term in Eq. (\ref{kalv3perp}) can be rewritten as
\begin{eqnarray}
&&\hspace{-1mm}
\cheV'_{3\mu_\perp\nu_\perp}
~=~-{1\over4 s^2}\langle\bsi A^i(x)\big(g^\perp_{\mu j}-i\epsilon_{\mu_\perp j}\gamma_5 
-i\sigma_{\mu_\perp j}-{2i\over s}g^\perp_{\mu j}\sigma_{\bu\ast}\big){1\over \alpha}\psi(0)\rangle_A
\nonumber\\
&&\hspace{-1mm}
\times~\langle\bsi B^j(0)\big(g^\perp_{i\nu}-i\epsilon_{i\nu_\perp}\gamma_5
+i\sigma_{i\nu_\perp}+{2\over s}g^\perp_{i\nu}\sigma_{\ast\bu}\big){1\over\beta}\psi(x)\rangle_B
+\mu\leftrightarrow\nu
+x\leftrightarrow 0
\label{kalv3perp2}
\end{eqnarray}
where we used Eq. (\ref{sigmasigmas}). 
It is clear that neither projectile no target matrix element in the r.h.s. can bring factor $s$ so
\begin{equation}
\hspace{1mm}
\cheV'_{3\mu_\perp\nu_\perp}
~\sim~{m_\perp^4\over\alpha_q\beta_q s^2}
\label{kalv3small}
 \end{equation}
which is $O\big({m_\perp^2\over s}\big)$ in comparison to Eq. (\ref{kalvklad3perp}). 

Next, consider the case when both $\mu$ and $\nu$ are longitudinal. The non-vanishing terms are
\begin{eqnarray}
&&\hspace{-1mm}\cheV'_{3\mu\nu}~=~
{4p_{1\mu}p_{2\nu}\over s^2}{1\over 2s^3}\langle\bsi A_i(x)\sigma_{\ast\alpha}\notp_2\gamma^j{1\over \alpha}\psi(0)\rangle_A
\langle\bsi B_j(0)\sigma_\bu^{~\alpha}\notp_1\gamma^i{1\over\beta}\psi(x)\rangle_B
\nonumber\\
&&\hspace{-1mm}
+~{4p_{1\nu}p_{2\mu}\over s^2}{1\over 2s^3}\langle\bsi A_i(x)\sigma_{\bu\alpha}\notp_2\gamma^j{1\over \alpha}\psi(0)\rangle_A
\langle\bsi B_j(0)\sigma_\ast^{~\alpha}\notp_1\gamma^i{1\over\beta}\psi(x)\rangle_B
\label{kalv3long}\\
&&\hspace{-1mm}
-~{4p_{2\mu}p_{1\nu}\over s^2}{1\over 2s^3}\langle\bsi A^i(x)\sigma_{\bu k}\sigma_{\ast j}{1\over \alpha}\psi(0)\rangle_A
\langle\bsi B^j(0)\sigma_\ast^{~k}\sigma_{\bu i}{1\over\beta}\psi(x)\rangle_B
+\mu\leftrightarrow\nu
+x\leftrightarrow 0
\nonumber
\end{eqnarray}
The first two terms in the r.h.s. can be rewritten as
\begin{eqnarray}
&&\hspace{-1mm}
\cheV'_{3\mu\nu}~=~
{g^\parallel_{\mu\nu}\over s^3}\langle\bsi A^i(x)\sigma_{\ast j}{1\over \alpha}\psi(0)\rangle_A
\langle\bsi B^j(0)\sigma_{\bu i}{1\over\beta}\psi(x)\rangle_B+x\leftrightarrow 0
\nonumber\\
&&\hspace{-1mm}
=~{g^\parallel_{\mu\nu}\over 2s^3}\langle\bsi A^i(x)\sigma_{\ast i}{1\over \alpha}\psi(0)\rangle_A
\langle\bsi B^j(0)\sigma_{\bu j}{1\over\beta}\psi(x)\rangle_B
\nonumber\\
&&\hspace{-1mm}
+~{g^\parallel_{\mu\nu}\over s^3}\langle\bsi\big(A^i(x)\sigma_{\ast j}-{g_{ij}\over 2}A^k(x)\sigma_{\ast k}\big) {1\over \alpha}\psi(0)\rangle_A
\langle\bsi B^j(0)\sigma_{\bu i}{1\over\beta}\psi(x)\rangle_B+x\leftrightarrow 0
\label{kalv3longi}
\end{eqnarray}
The corresponding contribution to $W_{\mu\nu}$ has the form
\begin{eqnarray}
&&\hspace{-1mm}
V'_{3\mu_\parallel\nu_\parallel}~=~-{g^\parallel_{\mu\nu}\over 2\alpha_q\beta_qsN_c}\!\int d^2k_\perp {k_\perp^2(q-k)_\perp^2\over m^2}
H^f(q,k_\perp)
\nonumber\\
&&\hspace{-1mm}
-~{g^\parallel_{\mu\nu}\over \alpha_q\beta_qsN_c}\!\int {d^2k_\perp\over m^2} \big[(k,q-k)_\perp^2-\half k_\perp^2(q-k)_\perp^2\big]
H_A^f(q,k_\perp)
\nonumber\\
&&\hspace{11mm}
\label{kalvklad3long}
\end{eqnarray}
where again we used QCD equations of motion (\ref{11.41}), (\ref{11.43}) and parametrization (\ref{maelsa}).

Next, it is easy to see that the third term in Eq. (\ref{kalv3long}) is small in comparison to Eq. (\ref{kalvklad3long}):
\begin{eqnarray}
&&\hspace{-1mm}
-~{4p_{2\mu}p_{1\nu}\over s^2}{1\over 2s^3}\langle\bsi A^i(x)\sigma_{\bu k}\sigma_{\ast j}{1\over \alpha}\psi(0)\rangle_A
\langle\bsi B^j(0)\sigma_\ast^{~k}\sigma_{\bu i}{1\over\beta}\psi(x)\rangle_B
+\mu\leftrightarrow\nu
+x\leftrightarrow 0
\nonumber\\
&&\hspace{-1mm}
=~-~{p_{2\mu}p_{1\nu}\over 2s^3}\langle\bsi A^i(x)\big[g_{jk}+i\epsilon_{jk}\gamma_5 
+i\sigma_{jk}-{2i\over s}g_{jk}\sigma_{\bu\ast}\big]{1\over \alpha}\psi(0)\rangle_A
\nonumber\\
&&\hspace{-1mm}
\times~\langle\bsi B^j(0)\big[g_{ik}-i\epsilon_{ik}\gamma_5
+i\sigma_{ik}+{2\over s}g_{ik}\sigma_{\ast\bu}\big]{1\over\beta}\psi(x)\rangle_B~\sim~{q_\perp^2\over s}\times{g^\parallel_{\mu\nu}\over\alpha_q\beta_qs}
\label{v3longsmall}
\end{eqnarray}
because neither projectile no target matrix element can bring factor $s$.

Finally, take one of the indices (say, $\mu$) longitudinal and the other transverse. From Eq. (\ref{kalv3b}) we get
\begin{eqnarray}
&&\hspace{-1mm}
\cheV_{3\mu_\parallel\nu_\perp}~=~-{p_{1\mu}\over s^4}\langle[\bsi A^i(x)\sigma_{\bu k}\sigma_{\ast j}{1\over \alpha}\psi(0)\rangle_A
\langle\bsi B^j(0)\sigma_{\nu_\perp}^{~k}\sigma_{\bu i}{1\over\beta}\psi(x)\rangle_B
\nonumber\\
&&\hspace{-1mm}
-~{ip_{1\mu}\over 2s^3}\langle[\bsi A^i(x)\sigma_{\bu\nu_\perp}\sigma_{\ast j}{1\over \alpha}\psi(0)\rangle_A
\langle\bsi B^j(0)\sigma_{\bu i}{1\over\beta}\psi(x)\rangle_B+\mu\leftrightarrow\nu
+x\leftrightarrow 0
\label{kalv3bprodperp}
\end{eqnarray}
Using formulas (\ref{sigmasigmas}) this can be rewritten as follows
\begin{eqnarray}
&&\hspace{-1mm}
\cheV_{3\mu_\parallel\nu_\perp}~=~{ip_{1\mu}\over 2s^3}\langle\bsi A^i(x)
\big[g_{jk}+i\epsilon_{jk}\gamma_5 
+i\sigma_{jk}-{2i\over s}g_{jk}\sigma_{\bu\ast}\big]{1\over \alpha}\psi(0)\rangle_A
\nonumber\\
&&\hspace{33mm}
\times~
\langle\bsi B^j(0)\big[g_{i\nu_\perp}\sigma_{\bu k}-g_{ik}\sigma_{\bu \nu_\perp}\big]{1\over\beta}\psi(x)\rangle_B
\nonumber\\
&&\hspace{-1mm}
-~{ip_{1\mu}\over 2s^3}\langle\bsi A^i(x)\big[g_{j\nu_\perp}+i\epsilon_{j\nu_\perp}\gamma_5 
+i\sigma_{j\nu_\perp}-{2i\over s}g_{j\nu_\perp}\sigma_{\bu\ast}\big]{1\over \alpha}\psi(0)\rangle_A
\nonumber\\
&&\hspace{33mm}
\times~
\langle\bsi B^j(0)\sigma_{\bu i}{1\over\beta}\psi(x)\rangle_B+\mu\leftrightarrow\nu
+x\leftrightarrow 0
\label{kalv3bprodperpe}
\end{eqnarray}
As we discussed above, projectile matrix elements in the r.h.s. like $\langle\bsi B^j(0)\sigma_{\bu i}{1\over\beta}\psi(x)\rangle_B$ 
can bring factor $s$ but the target matrix elements cannot so the corresponding contribution to $W_{\mu\nu}$ is of order
\begin{equation}
\hspace{1mm}
\cheV_{3\mu_\parallel\nu_\perp}
~\sim~\big(p_{1\mu}q^\perp_\nu+\mu\leftrightarrow\nu\big){m_\perp^2\over\alpha_q\beta_q s^2}
 \end{equation}
which is $O\big({q_\perp^2\over s}\big)$ in comparison to Eq. (\ref{XiBxi}).

Next, the sum of Eqs. (\ref{kalvklad3perp}) and (\ref{kalvklad3long}) is
\begin{eqnarray}
&&\hspace{-11mm}
V'_{3\mu\nu}~=~{g^\perp_{\mu\nu}-g^\parallel_{\mu\nu}\over 2\alpha_q\beta_qsN_c}\!\int d^2k_\perp {k_\perp^2(q-k)_\perp^2\over m^2}
H^f(q,k_\perp)
\label{kalvklad3perpmyy}\\
&&\hspace{-1mm}
+~{1\over \alpha_q\beta_qsN_c}\!\int d^2k_\perp {1\over m^2}\big\{[k^\perp_\mu(q-k)^\perp_\nu+\mu\leftrightarrow\nu](k,q-k)_\perp
-k_\perp^2(q-k)^\perp_\mu(q-k)^\perp_\nu
\nonumber\\
&&\hspace{-1mm}
-~(q-k_\perp)^2k^\perp_\mu k^\perp_\nu
-{g_{\mu\nu}^\perp\over 2}k_\perp^2(q-k_\perp)^2-g^\parallel_{\mu\nu}\big[(k,q-k)_\perp^2-\half k_\perp^2(q-k)_\perp^2\big]\big\}
H_A^f(q,k_\perp)
\nonumber
\end{eqnarray}
so subtracting trace we obtain
\begin{eqnarray}
&&\hspace{-1mm}
V_{3\mu\nu}~=~V'_{3\mu\nu}-g_{\mu\nu}{V'_3}_\xi^{~\xi} 
\label{v3otvet}\\
&&\hspace{-1mm}
=~{g^\perp_{\mu\nu}-g^\parallel_{\mu\nu}\over 2\alpha_q\beta_qsN_c}\!\int d^2k_\perp {1\over m^2}k_\perp^2(q-k)_\perp^2
H^f(q,k_\perp)
\nonumber\\
&&\hspace{-1mm}
+~{1\over \alpha_q\beta_qsN_c}\!\int d^2k_\perp {1\over m^2}\big\{[k^\perp_\mu(q-k)^\perp_\nu+\mu\leftrightarrow\nu](k,q-k)_\perp
-k_\perp^2(q-k)^\perp_\mu(q-k)^\perp_\nu
\nonumber\\
&&\hspace{-1mm}
-~(q-k_\perp)^2k^\perp_\mu k^\perp_\nu+g_{\mu\nu}^\perp(k,q-k)_\perp^2
-g_{\mu\nu}^\perp k_\perp^2(q-k_\perp)^2\big]\big\}
H_A^f(q,k_\perp)
\nonumber
\end{eqnarray}
As we will see in Sect. \ref{sec:results}, cancellation of terms $\sim g^\parallel_{\mu\nu}$ proportional to $h_A$ in the r.h.s of this equation is 
actually a consequence of (EM) gauge invariance.

Let us now assemble the contribution of terms (\ref{kalvs}) to $W_{\mu\nu}$. Summing Eqs. (\ref{v1otvet}),  (\ref{v2otvet}),  and (\ref{v3otvet}) we get
\begin{eqnarray}
&&\hspace{-1mm}
V_{\mu\nu}(q)~=~{1\over 32\pi^4}
\!\int\!dx_\bu dx_\ast d^2x_\perp~e^{-i\alpha x_\bu-i\beta x_\ast+i(q,x)_\perp}
\nonumber\\
&&\hspace{22mm}
\big[\langle A,B|(\bsi_A^m(x)\gamma_\mu\Xi_{2}^m(x))(\bsi_B^n(0)\gamma_\nu\Xi_{1}^n(0))
+\mu\leftrightarrow\nu|A,B\rangle+x\leftrightarrow 0\big]
\nonumber\\
&&\hspace{-1mm}
=~{g_{\mu\nu}^\parallel\over \alpha_q\beta_q sN_c}\int\! d^2k_\perp\Big\{
(k,q-k)_\perp F^f(q,k_\perp)
-~{1\over m^2} k_\perp^2(q-k)_\perp^2H^f(q,k_\perp)\Big\}
\nonumber\\
&&\hspace{-1mm}
+~{1\over \alpha_q\beta_qsN_c}\!\int d^2k_\perp {1\over m^2}\big\{[k^\perp_\mu(q-k)^\perp_\nu+\mu\leftrightarrow\nu](k,q-k)_\perp
-k_\perp^2(q-k)^\perp_\mu(q-k)^\perp_\nu
\nonumber\\
&&\hspace{11mm}
-~(q-k_\perp)^2k^\perp_\mu k^\perp_\nu+g_{\mu\nu}^\perp(k,q-k)_\perp^2
-g_{\mu\nu}^\perp k_\perp^2(q-k_\perp)^2\big]\big\}
H_A^f(q,k_\perp)
\label{votvet}
\end{eqnarray}

Finally, to get $W_{\mu\nu}^{(2a)}(q)$ of Eq. (\ref{kalw2a}) we need to add 
the contribution of the term $[\Bxi_{1}(x)\gamma_\mu\psi_B(x)\big]\big[\Bxi_{2}(0)\gamma_\nu\psi_A(0)\big]$.
Similarly to the case of one quark-quark-gluon operator considered in Sect. \ref{1qqG}, it can be demonstrated that this contribution doubles the result (\ref{votvet}) so we get
\begin{eqnarray}
&&\hspace{-1mm}
W_{\mu\nu}^{(2a)}(q)~=~{1\over 32\pi^4}
\!\int\!dx_\bu dx_\ast d^2x_\perp~e^{-i\alpha x_\bu-i\beta x_\ast+i(q,x)_\perp}
\big\{\langle A,B|[\bsi_A^m(x)\gamma_\mu\Xi_{2}^m(x)]
\nonumber\\
&&\hspace{5mm}
\times[\bsi_B^n(0)\gamma_\nu\Xi_{1}^n(0)]
+[\Bxi_{1}(x)\gamma_\mu\psi_B(x)\big]\big[\Bxi_{2}(0)\gamma_\nu\psi_A(0)\big]
+\mu\leftrightarrow\nu|A,B\rangle+x\leftrightarrow 0\big\}
\nonumber\\
&&\hspace{-1mm}
=~{2g_{\mu\nu}^\parallel\over Q_\parallel^2N_c}\int\! d^2k_\perp\Big\{
(k,q-k)_\perp F^f(q,k_\perp)
-~{1\over m^2} k_\perp^2(q-k)_\perp^2H^f(q,k_\perp)\Big\}
\nonumber\\
&&\hspace{5mm}
+~{2\over Q_\parallel^2N_c}\!\int d^2k_\perp {1\over m^2}\big\{[k^\perp_\mu(q-k)^\perp_\nu+\mu\leftrightarrow\nu](k,q-k)_\perp
-k_\perp^2(q-k)^\perp_\mu(q-k)^\perp_\nu
\nonumber\\
&&\hspace{11mm}
-~(q-k_\perp)^2k^\perp_\mu k^\perp_\nu+g_{\mu\nu}^\perp(k,q-k)_\perp^2
-g_{\mu\nu}^\perp k_\perp^2(q-k_\perp)^2\big]\big\}H_A^f(q,k_\perp)
\label{w2atvet}
\end{eqnarray}
where $Q_\parallel^2\equiv\alpha_q\beta_qs$
\subsection{Term  with two quark-quark-gluon operators coming from $\Bxi_{2}$ and  $\Xi_{2}$ \label{sec:2qqGb}}

Let us start with  the first term in Eq. (\ref{kalw2b}).
\begin{equation}
\hspace{-0mm}
\cheW^{(2b)}_{1\mu\nu}~=~{N_c\over s}\langle A,B|
\big[\bar\psi_A(x)\gamma_\mu\Xi_{2}(x)\big]\big[\Bxi_{2}(0)\gamma_\nu\psi_A(0)\big]
+\mu\leftrightarrow\nu|A,B\rangle~+~x\leftrightarrow 0
\label{w2b1munu}
\end{equation}
After Fierz transformation (\ref{fierz}) we obtain
\begin{eqnarray}
&&\hspace{-1mm}
\cheW^{(2b)}_{1\mu\nu}~=~-{N_c\over 2s}(\delta_\mu^\alpha\delta_\nu^\beta+\delta_\nu^\alpha\delta_\mu^\beta-g_{\mu\nu}g^{\alpha\beta})
\langle A,B|\big\{[\bar\psi_A^m(x)\gamma_\alpha\psi_A^n(0)][\Bxi_{2}^n(0)\gamma_\beta\Xi_{2}^m(x)]
\label{w2b1a}\\
&&\hspace{55mm}
+~\gamma_\alpha\otimes\gamma_\beta\leftrightarrow \gamma_\alpha\gamma_5\otimes\gamma_\beta\gamma_5\big\}|A,B\rangle
\nonumber\\
&&\hspace{-1mm}
+~{N_c\over 2s}(\delta_\mu^\alpha\delta_\nu^\beta+\delta_\nu^\alpha\delta_\mu^\beta-\half g_{\mu\nu}g^{\alpha\beta})
\langle A,B|[\bar\psi_A^m(x)\sigma_{\alpha\xi}\psi_A^n(0)][\Bxi_{2}^n(0)\sigma_\beta^{~\xi}\Xi_{2}^m(x)]|A,B\rangle
~+~x\leftrightarrow 0
\nonumber
\end{eqnarray}
(note that $\Bxi_{2}\Xi_{2}=\Bxi_{2}\gamma_5\Xi_{2}=0$). 
Using explicit expressions (\ref{fildz0}) for quark fields and separating color-singlet terms we get
\begin{equation}
\cheW^{(2b)}_{1\mu\nu}~=~\cheV^4_{\mu\nu}+\cheV^5_{\mu\nu}
\label{w2b1}
\end{equation}
where
\begin{eqnarray}
&&\hspace{-1mm}
\cheV^4_{\mu\nu}~=~-{1\over s^3}(\delta_\mu^\alpha p_{1\nu}+\delta_\nu^\alpha p_{1\mu}-g_{\mu\nu}p_1^\alpha)
\Big(\langle \bar\psi(x)A_j(x)\gamma_\alpha A_i(0)\psi(0)\rangle_A
\nonumber\\
&&\hspace{-1mm}
\times~\langle\big(\bar\psi{1\over \beta}\big)(0)\gamma^i\notp_1
\gamma^j{1\over\beta}\psi(x)\rangle_B
+\psi(0)\otimes\psi(x)\leftrightarrow \gamma_5\psi(0)\otimes\gamma_5\psi(x)\Big)~+~x\leftrightarrow 0
\label{v4}
\end{eqnarray}
and
\begin{eqnarray}
&&\hspace{-1mm}
\cheV^5_{\mu\nu}
=~
{1\over s^3}(\delta_\mu^\alpha\delta_\nu^\beta+\delta_\nu^\alpha\delta_\mu^\beta-\half g_{\mu\nu}g^{\alpha\beta})
\nonumber\\
&&\hspace{-1mm}
\times~\Big\{-p_{1\beta} \langle \bar\psi(x)A_j(x)\sigma_{\alpha k}A_i(0)\psi(0)\rangle_A
\langle\big(\bar\psi{1\over \beta}\big)(0)\gamma^i\sigma_\bu^{~k}
\gamma^j{1\over\beta}\psi(x)\rangle_B  
\nonumber\\
&&\hspace{-1mm} 
+~\langle \bar\psi(x)A_j(x)\sigma_{\alpha \bu}A_i(0)\psi(0)\rangle_A
\langle\big(\bar\psi{1\over \beta}\big)(0)\gamma^i\sigma_{\bu\beta_\perp}
\gamma^j{1\over\beta}\psi(x)\rangle_B \Big\}   ~+~x\leftrightarrow 0
\label{v5}
\end{eqnarray}
We will consider them in turn.

\subsubsection{Term proportional to $f_1\barf_1$ \label{sec:v4}}
Let us start with $g_{\mu\nu}$ term in Eq. (\ref{v4}).
\begin{equation}
\hspace{-0mm}
{g_{\mu\nu}\over s^3}
\langle \bar\psi(x)A_j(x)\notp_1A_i(0)\psi(0)\rangle_A
\langle\big(\bar\psi{1\over \beta}\big)(0)\gamma^i\notp_1
\gamma^j{1\over\beta}\psi(x)\rangle_B
+\notp_1\otimes\gamma_j\leftrightarrow \notp_1\gamma_5\otimes\gamma_j\gamma_5
\label{v4gmunu}
\end{equation}
It is obvious that the target matrix element can bring factor $s$. On the contrary, as we discussed above, the projectile
matrix element cannot produce $s$ since 
\begin{equation}
\langle \bar\psi(x)A_j(x)\gamma_\alpha A_i(0)\psi(0)\rangle_A~\sim~{p_{2\alpha}\over p_1\cdot p_2}
\times[g_{ij}\phi(x_\perp^2)+x_ix_j\xi(x_\perp^2)]~+~...
\label{AAsmall}
\end{equation}
Indeed, since projectile matrix elements know about $p_2$ only through the direction of Wilson lines, 
the l.h.s. can be proportional only to factor ${p_{2\alpha}\over p_1\cdot p_2}$ that does not change under rescaling of $p_2$. 
Also, due to Eq. (\ref{maelqg2}) $\langle\big(\bar\psi{1\over \beta}\big)(0)\otimes {1\over\beta}\psi(x)\rangle_B$ can be replaced by
 $-{1\over\beta_q^2}\langle\bar\psi(0)\otimes \psi(x)\rangle_B$.
Consequently, the r.h.s. of Eq. (\ref{v4gmunu}) is $\sim g_{\mu\nu}{m_\perp^4\over\beta_q^2s^2}$ 
\begin{equation}
\hspace{-0mm}
{g_{\mu\nu}\over s^3}
\langle \bar\psi(x)A_j(x)\notp_1A_i(0)\psi(0)\rangle_A
\langle\big(\bar\psi{1\over \beta}\big)(0)\gamma^i\notp_1
\gamma^j{1\over\beta}\psi(x)\rangle_B
+\notp_1\otimes\gamma_j\leftrightarrow \notp_1\gamma_5\otimes\gamma_j\gamma_5~\sim~ g_{\mu\nu}{m_\perp^4\over\beta_q^2s^2}
\label{v4gmunue}
\end{equation}
which is $O\big({m_\perp^2\alpha_q\over \beta_qs}\big)$ in comparison to Eq. (\ref{w2atvet}).

We get
\begin{eqnarray}
&&\hspace{-1mm}
\cheV^4_{\mu\nu}~=~-{p_{1\mu}\over s^3}\Big(
\langle \bar\psi(x)A_j(x)\gamma_\nu A_i(0)\psi(0)\rangle_A\langle\big(\bar\psi{1\over \beta}\big)(0)\gamma^i\notp_1
\gamma^j{1\over\beta}\psi(x)\rangle_B
\nonumber\\
&&\hspace{-1mm}
+~\psi(0)\otimes\psi(x)\leftrightarrow \gamma_5\psi(0)\otimes\gamma_5\psi(x)\Big)
~+~\mu\leftrightarrow \nu~+~x\leftrightarrow 0
\label{v4raz}
\end{eqnarray}
If the index $\nu$ is transverse, the contribution of this  equation to $W_{\mu\nu}$ is of order of
\begin{equation}
\cheV^4_{\mu\nu}~\sim~p_{1\mu}q^\perp_\nu{m_\perp^2\over \beta_q^2s^2}
\label{trlongmalo}
\end{equation}
which is $O\big({m_\perp^2\over\beta_qs}\big)$ in comparison to Eq. (\ref{XiBxi}). 

For the longitudinal indices $\mu$ and $\nu$ we get 
\begin{eqnarray}
&&\hspace{-1mm}
\cheV^4_{\mu\nu}~=~-{4p_{1\mu} p_{1\nu}\over s^4}\Big(
\langle \bar\psi(x)A_j(x)\notp_2A_i(0)\psi(0)\rangle_A\langle\big(\bar\psi{1\over \beta}\big)(0)\gamma^i\notp_1
\gamma^j{1\over\beta}\psi(x)\rangle_B
\nonumber\\
&&\hspace{-1mm}
+~\psi(0)\otimes\psi(x)\leftrightarrow \gamma_5\psi(0)\otimes\gamma_5\psi(x)\Big)
\nonumber\\
&&\hspace{-1mm}
-~{g^\parallel_{\mu\nu}\over s^3}\Big(
\langle \bar\psi(x)A_j(x)\notp_1A_i(0)\psi(0)\rangle_A\langle\big(\bar\psi{1\over \beta}\big)(0)\gamma^i\notp_1
\gamma^j{1\over\beta}\psi(x)\rangle_B
\nonumber\\
&&\hspace{-1mm}
+~\psi(0)\otimes\psi(x)\leftrightarrow \gamma_5\psi(0)\otimes\gamma_5\psi(x)\Big)
~+~x\leftrightarrow 0
\label{v4long}
\end{eqnarray}
Similarly to Eq. (\ref{v4gmunue}), the contribution of the second term to $W_{\mu\nu}$ is
\begin{equation}
\sim {g^\parallel_{\mu\nu}m_\perp^4\over\beta_q^2s^2}
~=~O\Big({\alpha_q m_\perp^2\over\beta_q s}\Big)\times \big[{\rm r.h.s.~of~Eq.~(\ref{w2atvet})}\big]
\label{gmunumalo}
\end{equation}
so we are left with the first term in the r.h.s. of Eq. (\ref{v4long}). Using Eq. (\ref{flagamma})
it can be rewritten as
\begin{eqnarray}
&&\hspace{-1mm}
\cheV^4_{\mu\nu}~=~-{4p_{1\mu} p_{1\nu}\over s^4}\Big(
\langle \bar\psi(x)\notA(x)\notp_2\notA(0)\psi(0)\rangle_A\langle\big(\bar\psi{1\over \beta}\big)(0)\notp_1{1\over\beta}\psi(x)\rangle_B
\nonumber\\
&&\hspace{-1mm}
+~\psi(0)\otimes\psi(x)\leftrightarrow \gamma_5\psi(0)\otimes\gamma_5\psi(x)\Big)
~+~x\leftrightarrow 0
\label{v4longa}
\end{eqnarray}
The corresponding contribution to $W_{\mu\nu}$ is obtained from QCD equation of motion (\ref{AA1}) and formula (\ref{maelqg2}) from 
Appendix \ref{sec:qqgparam}:
\begin{eqnarray}
&&\hspace{-1mm}
V^4_{\mu\nu}(q)~=~{4p_{1\mu} p_{1\nu}\over \beta_q^2s^2N_c}\int\! d^2k_\perp
k_\perp^2F^f(q,k_\perp)
\label{v4otvet}
\end{eqnarray}
%

\subsubsection{Term proportional to $h_1^\perp\barh_1^\perp$ \label{sec:v5}}
Let us start with $g_{\mu\nu}$ term in Eq. (\ref{v5}).
\begin{eqnarray}
&&\hspace{-1mm}
{g_{\mu\nu}\over s^3}
 \langle \bar\psi(x)A_j(x)\sigma_{\bu k}A_i(0)\psi(0)\rangle_A
\langle\big(\bar\psi{1\over \beta}\big)(0)\gamma^i\sigma_\bu^{~k}
\gamma^j{1\over\beta}\psi(x)\rangle_B  ~+~x\leftrightarrow 0
\label{v5gmunu}
\end{eqnarray}
The target matrix element is proportional to $s$ while the projectile one cannot bring $s$ due to Eq. (\ref{AAsmall}), so the contribution 
of the  r.h.s~of~Eq. (\ref{v5gmunu}) to $W_{\mu\nu}$ is of order
\begin{eqnarray}
\sim~{g_{\mu\nu}\over \beta_q^2s^2}m_\perp^4
~=~O\big({m_\perp^2\alpha_q\over s\beta_q}\Big)\times\big[{\rm r.h.s.~of~Eq.~(\ref{w2atvet})}\big]
\end{eqnarray}
similarly to Eq. (\ref{gmunumalo}). We get
\begin{eqnarray}
&&\hspace{-7mm}
\cheV^5_{\mu\nu}
=~
{1\over s^3}
\Big\{-p_{1\mu} \langle \bar\psi(x)A_j(x)\sigma_{\nu k}A_i(0)\psi(0)\rangle_A
\langle\big(\bar\psi{1\over \beta}\big)(0)\gamma^i\sigma_\bu^{~k}
\gamma^j{1\over\beta}\psi(x)\rangle_B  
\nonumber\\
&&\hspace{-7mm} 
+~\langle \bar\psi(x)A_j(x)\sigma_{\mu \bu}A_i(0)\psi(0)\rangle_A
\langle\big(\bar\psi{1\over \beta}\big)(0)\gamma^i\sigma_{\bu\nu_\perp}
\gamma^j{1\over\beta}\psi(x)\rangle_B \Big\} +\mu\leftrightarrow\nu  ~+~x\leftrightarrow 0
\label{v5dva}
\end{eqnarray}

Let us at first consider the second term in this formula:
\begin{eqnarray}
&&\hspace{-7mm}
{1\over s^3}
\langle \bar\psi(x)A_j(x)\sigma_{\mu_\perp \bu}A_i(0)\psi(0)\rangle_A
\langle\big(\bar\psi{1\over \beta}\big)(0)\gamma^i\sigma_{\bu\nu_\perp}
\gamma^j{1\over\beta}\psi(x)\rangle_B
\nonumber\\
&&\hspace{-7mm} 
+~{2p_{1\mu}\over s^4}\langle \bar\psi(x)A_j(x)\sigma_{\ast \bu}A_i(0)\psi(0)\rangle_A
\langle\big(\bar\psi{1\over \beta}\big)(0)\gamma^i\sigma_{\bu\nu_\perp}
\gamma^j{1\over\beta}\psi(x)\rangle_B \Big\} +\mu\leftrightarrow\nu  
\label{v5razdva}
\end{eqnarray}
Similarly to Eq. (\ref{AAsmall}), projectile matrix elements cannot give factor $s$ so the corresponding contribution to $W_{\mu\nu}$ is
of order of
\begin{equation}
(aq^\perp_\mu q^\perp_\nu+bq_\perp^2g^\perp_{\mu\nu}){m_\perp^2\over \beta_q^2s^2}~~~~{\rm or}~~~{2p_{1\mu}\over s^3}{m_\perp^4\over \beta_q^2s^2}
\label{v5smalls}
\end{equation}
that are $O\big({\alpha_qm_\perp^2\over\beta_qs}\big)$ in comparison to Eqs. (\ref{XiBxi}) and (\ref{w2atvet}), respectively. 

We are left with
\begin{eqnarray}
&&\hspace{-1mm}
\cheV^5_{\mu\nu}
=~
-{p_{1\mu}\over s^3}
\langle \bar\psi(x)A_j(x)\sigma_{\nu k}A_i(0)\psi(0)\rangle_A
\langle\big(\bar\psi{1\over \beta}\big)(0)\gamma^i\sigma_\bu^{~k}
\gamma^j{1\over\beta}\psi(x)\rangle_B  
 +\mu\leftrightarrow\nu+x\leftrightarrow 0
 \nonumber\\
&&\hspace{-1mm} 
=~
-{4p_{1\mu}p_{1\nu}\over s^4}
\langle \bar\psi(x)A_j(x)\sigma_{\ast k}A_i(0)\psi(0)\rangle_A
\langle\big(\bar\psi{1\over \beta}\big)(0)\gamma^i\sigma_\bu^{~k}
\gamma^j{1\over\beta}\psi(x)\rangle_B  
\nonumber\\
&&\hspace{-1mm} 
-~{g^\parallel_{\mu\nu}\over s^4}
\langle \bar\psi(x)A_j(x)\sigma_{\bu k}A_i(0)\psi(0)\rangle_A
\langle\big(\bar\psi{1\over \beta}\big)(0)\gamma^i\sigma_\bu^{~k}
\gamma^j{1\over\beta}\psi(x)\rangle_B
\nonumber\\
&&\hspace{-1mm}
-~\Big({p_{1\mu}\over s^3}
\langle \bar\psi(x)A_j(x)\sigma_{\nu k}A_i(0)\psi(0)\rangle_A
\langle\big(\bar\psi{1\over \beta}\big)(0)\gamma^i\sigma_\bu^{~k}
\gamma^j{1\over\beta}\psi(x)\rangle_B  
 +\mu\leftrightarrow\nu\Big)+x\leftrightarrow 0
\label{v5raz}
\end{eqnarray}
First, note that the two last terms are small, of order of Eq. (\ref{v5smalls}), for the same reason as Eq. (\ref{v5dva}) above.  As  to
the first term in r.h.s. of Eq. (\ref{v5raz}), using Eq. (\ref{sisigaga}) it can be rewritten as
\begin{equation}
\hspace{-0mm}
\cheV^5_{\mu\nu}
~=~
-{4p_{1\mu}p_{1\nu}\over s^4}
\langle \bar\psi(x)\notA(x)\sigma_{\ast k}\notA(0)\psi(0)\rangle_A
\langle\big(\bar\psi{1\over \beta}\big)(0)\sigma_\bu^{~k}
{1\over\beta}\psi(x)\rangle_B  
~+~x\leftrightarrow 0
\label{v5razotvet}
\end{equation}
so the corresponding contribution to $W_{\mu\nu}$ takes the form
\begin{equation}
\hspace{-0mm}
V^5_{\mu\nu}
~=~
-{4p_{1\mu}p_{1\nu}\over \beta_q^2s^2N_c}
\!\int\! d^2k_\perp
{1\over m^2} k_\perp^2(k,q-k)_\perp
H^f(q,k_\perp)
\label{v5otvet}
\end{equation}
where we used Eqs. (\ref{maelqg2}) and (\ref{AA2}).

The full result for $W_{\mu\nu}^{(2b)}$ is given by the sum of Eqs. (\ref{v4otvet}) and (\ref{v5otvet})
\begin{equation}
\hspace{-0mm}
W_{1\mu\nu}^{(2b)}~=~
{4p_{1\mu}p_{1\nu}\over \beta_q^2s^2N_c}
\!\int\! d^2k_\perp\Big[
k_\perp^2F^f(q,k_\perp)
-~{1\over m^2} k_\perp^2(k,q-k)_\perp H^f(q,k_\perp)\Big]
\label{w2b1otvet}
\end{equation}
%

\subsubsection{Second term in Eq. (\ref{kalw2b})}
Let us start now consider the second term in Eq. (\ref{kalw2b}).
\begin{equation}
\hspace{-0mm}
\cheW^{(2b)}_{2\mu\nu}~=~{N_c\over s}\langle A,B|
\big[\Bxi_{1}(x)\gamma_\mu\psi_B(x)\big]\big[\bar\psi_B(0)\gamma_\nu\Xi_{1}(0)\big]
+\mu\leftrightarrow\nu|A,B\rangle~+~x\leftrightarrow 0
\label{w2b2munu}
\end{equation}
After Fierz transformation (\ref{fierz}) we obtain
\begin{eqnarray}
&&\hspace{-1mm}
\cheW^{(2b)}_{2\mu\nu}~=~-{N_c\over 2s}(\delta_\mu^\alpha\delta_\nu^\beta+\delta_\nu^\alpha\delta_\mu^\beta-g_{\mu\nu}g^{\alpha\beta})
\langle A,B|\big\{[\Bxi_{1}^m(x)\gamma_\alpha\Xi_{1}^n(0)][\bsi_{B}^n(0)\gamma_\beta\psi_{B}^m(x)]
\\
&&\hspace{55mm}
+~\gamma_\alpha\otimes\gamma_\beta\leftrightarrow \gamma_\alpha\gamma_5\otimes\gamma_\beta\gamma_5\big\}|A,B\rangle
\nonumber\\
&&\hspace{-1mm}
+~{N_c\over 2s}(\delta_\mu^\alpha\delta_\nu^\beta+\delta_\nu^\alpha\delta_\mu^\beta-\half g_{\mu\nu}g^{\alpha\beta})
\langle A,B|[\Bxi_{1}^m(x)\sigma_{\alpha\xi}\Xi_{1}^n(0)][\psi_{1B}^n(0)\sigma_\beta^{~\xi}\psi_{1B}^m(x)]|A,B\rangle
~+~x\leftrightarrow 0
\nonumber
\end{eqnarray}
Sorting out color-singlet terms, we get similarly to sum of Eqs. (\ref{v4}) and  (\ref{v5})
\begin{eqnarray}
&&\hspace{-1mm}
\cheW^{(2b)}_{2\mu\nu}~=~-{1\over s^3}(\delta_\mu^\alpha p_{2\nu}+\delta_\nu^\alpha p_{2\mu}-g_{\mu\nu}p_2^\alpha)
\Big(\langle \bar\psi(x) B_j(x)\gamma_\alpha B_i(0)\psi(0)\rangle_B
\nonumber\\
&&\hspace{-1mm}
\times~\langle\big(\bar\psi{1\over \alpha}\big)(0)\gamma^i\notp_2
\gamma^j{1\over\alpha}\psi(x)\rangle_A
+\psi(0)\otimes\psi(x)\leftrightarrow \gamma_5\psi(0)\otimes\gamma_5\psi(x)\Big)~+~x\leftrightarrow 0
\nonumber\\
&&\hspace{-1mm}
+~
{1\over s^3}(\delta_\mu^\alpha\delta_\nu^\beta+\delta_\nu^\alpha\delta_\mu^\beta-\half g_{\mu\nu}g^{\alpha\beta})
\nonumber\\
&&\hspace{-1mm}
\times~\Big\{-p_{2\alpha}
\langle\big(\bar\psi{1\over \alpha}\big)(0)\gamma^i\sigma_\ast^{~k}
\gamma^j{1\over\alpha}\psi(x)\rangle_A  
 \langle \bar\psi(x) B_j(x)\sigma_{\beta k} B_i(0)\psi(0)\rangle_B
\nonumber\\
&&\hspace{-1mm} 
+~
\langle\big(\bar\psi{1\over \alpha}\big)(0)\gamma^i\sigma_{\ast\alpha_\perp}
\gamma^j{1\over\alpha}\psi(x)\rangle_A 
\langle \bar\psi(x) B_j(x)\sigma_{\beta \ast} B_i(0)\psi(0)\rangle_B \Big\}  ~+~x\leftrightarrow 0
\label{v45be}
\end{eqnarray}
Starting from this point, all calculations repeat those of Sections \ref{sec:v4} and \ref{sec:v5} with replacements of 
$p_1\leftrightarrow p_2$, $\alpha_q\leftrightarrow\beta_q$ and exchange of  projectile matrix elements and the target ones. The result is 
Eq. (\ref{w2botvet}) with these replacements so we finally get
\begin{eqnarray}
&&\hspace{-1mm}
W_{\mu\nu}^{(2b)}~=~
{4p_{1\mu}p_{1\nu}\over \beta_q^2s^2N_c}
\!\int\! d^2k_\perp\Big[
k_\perp^2 \big[f_1^f(\alpha_q,k_\perp)F^f(q,k_\perp)
-~{1\over m^2} k_\perp^2(k,q-k)_\perp H^f(q,k_\perp)\Big]
\nonumber\\
&&\hspace{-1mm}
+~{4p_{2\mu}p_{2\nu}\over \alpha_q^2s^2N_c}
\!\int\! d^2k_\perp\Big[
(q-k)_\perp^2F^f(q,k_\perp)
-~{1\over m^2} (q-k)_\perp^2(k,q-k)_\perp H^f(q,k_\perp)\Big]
\label{w2botvet}
\end{eqnarray}
%

\subsection{Third term  with two quark-quark-gluon operators \label{3qqGa}}
Let us consider the first term in the r.h.s. of Eq. (\ref{kalw2c}). After Fierz transformation it turns to
\begin{eqnarray}
&&\hspace{-1mm}
\half[(\Bxi_{1}^m(x)\gamma_\mu\Xi_{2}^m(x))(\bar\psi_B^n(0)\gamma_\nu\psi_A^n(0))+\mu\leftrightarrow\nu]
\nonumber\\
&&\hspace{-1mm}
=~-{g_{\mu\nu}\over 4}(\Bxi_{1}^m(x)\psi_A^n(0))(\bar\psi_B^n(0)\Xi_{2}^m(x))
+{g_{\mu\nu}\over 4}(\Bxi_{1}^m(x)\gamma_5\psi_A^n(0))(\bar\psi_B^n(0)\gamma_5\Xi_{2}(x))
\nonumber\\
&&\hspace{-1mm}
+~{g_{\mu\nu}\over 4}(\Bxi_{1}^m(x)\gamma_\alpha\psi_A^n(0))(\bar\psi_B(0)\gamma^\alpha\Xi_{2}^m(x))
+{g_{\mu\nu}\over 4}(\Bxi_{1}^m(x)\gamma_\alpha\gamma_5\psi_A(0))(\bar\psi_B^n(0)\gamma^\alpha\gamma_5\Xi_{2}(x))
\nonumber\\
&&\hspace{-1mm}
-~{1\over 4}[(\Bxi_{1}^m(x)\gamma_\mu\psi_A^n(0))(\bar\psi_B^n(0)\gamma_\nu\Xi_{2}^m(x))+\mu\leftrightarrow\nu]
\nonumber\\
&&\hspace{33mm}
-~{1\over 4}(\Bxi_{1}^m(x)\gamma_\mu\gamma_5\psi_A^n(0))(\bar\psi_B^n(0)\gamma_\nu\gamma_5\Xi_{2}^m(x))+\mu\leftrightarrow\nu]
\nonumber\\
&&\hspace{-1mm}
+~{1\over 4}[(\Bxi_{1}^m(x)\sigma_{\nu\alpha}\psi_A^n(0))(\bar\psi_B^n(0)\sigma_{\mu\alpha}\Xi_{2}^m(x))+\mu\leftrightarrow\nu]
\nonumber\\
&&\hspace{33mm}
-~{g_{\mu\nu}\over 8}(\Bxi_{1}^m(x)\sigma^{\alpha\beta}\psi_A^n(0))(\bar\psi_B^n(0)\sigma_{\alpha\beta}\Xi_{2}^m(x))
\label{fierzkalw2c}
\end{eqnarray}
Let us demonstrate that after sorting out color-singlet matrix elements the contribution $W_{\mu\nu}^{(2c)}$ is $O\big({1\over N_c^2}\big)$ in comparison to 
$W_{\mu\nu}^{(2a)}$ (and $W_{\mu\nu}^{(2b)}$). Consider a typical term in the r.h.s. of Eq. (\ref{fierzkalw2c})
\begin{eqnarray}
&&\hspace{-1mm}
{N_c\over s}\langle A,B|\Bxi_{1}^m(x)\Gamma_1\psi_A^n(0))(\bar\psi_B^n(0)\Gamma_2\Xi_{2}^m(x))|A,B\rangle
\label{kalw2craz}\\
&&\hspace{-1mm}
=~{N_c\over s}\langle A,B|\big(\bar\psi_A^k{1\over\alpha}\big)(x)A_j^{ml}(x)\gamma^i{\slashed{p}_2\over s}\Gamma_1\psi^n_A(0))(\bar\psi_B^n(0)\Gamma_2{\slashed{p}_1\over s}\gamma^jB^{km}_i(x){1\over\beta}\psi_B^l(x))|A,B\rangle
\nonumber
\end{eqnarray}
After separation of color singlet contributions
\begin{eqnarray}
&&\hspace{-1mm}
\langle A,B|(\bsi_A^k (A_i)^{ml}\psi_A^n)(\bsi_B^n (B_j)^{km} \psi_B^l)|A,B\rangle
\nonumber\\
&&\hspace{-1mm}
=~\langle \bsi_A^k (A_i)^{km}\psi_A^n\rangle_A\langle\bsi_B^n (B_j)^{ml} \psi_B^l)\rangle_B
-~if^{abc}\langle\bsi_A^k t^c_{kl}A_i^a\psi_A^n\rangle_A\langle\bsi_B^n B_j^b \psi_B^l\rangle_B
\nonumber\\
&&\hspace{-1mm}
=~{1\over N_c}\langle \bsi_A A_i\psi_A\rangle_A\langle\bsi_B B_j\psi_B\rangle_B
-2if^{abc}\langle A^a_i(\bsi_A t^ct^d\psi_A)\rangle_A\langle B^b_j(\bsi_B t^d\psi_B)\rangle_B
\nonumber\\
&&\hspace{44mm}
-~{2\over N_c}f^{abc}\langle A^a_i(\bsi_A t^c\psi_A)\rangle_A\langle B^b_j(\bsi_B \psi_B)\rangle_B
\nonumber\\
&&\hspace{-1mm}
=~{1\over N_c}\langle \bsi_A A_i\psi_A\rangle_A\langle\bsi_B B_j\psi_B\rangle_B
-2i{f^{abc}\over N_c^2-1}\langle A^a_i(\bsi_A t^ct^b\psi_A)\rangle_A\langle \bsi_B B_j\psi_B\rangle_B
\nonumber\\
&&\hspace{-1mm}
=~-{1\over N_c(N_c^2-1)}\langle\bsi_A A_i\psi_A\rangle_A\langle\bsi_B B_j\psi_B\rangle_B
\label{4.35}
\end{eqnarray}
we get
\begin{eqnarray}
&&\hspace{-1mm}
{N_c\over s}\langle A,B|\Bxi_{1}^m(x)\Gamma_1\psi_A^n(0))(\bar\psi_B^n(0)\Gamma_2\Xi_{2}^m(x))|A,B\rangle
\label{kalw2craze}\\
&&\hspace{-1mm} 
=~-{1\over (N_c^2-1)s^3}\langle\big(\bar\psi_A{1\over\alpha}\big)A_j(x)\gamma^i
\slashed{p}_2\Gamma_1\psi(0)\rangle_A
\langle\bar\psi(0)\Gamma_2\slashed{p}_1\gamma^jB_i(0){1\over\beta}\psi_B(x))\rangle_B
\nonumber
\end{eqnarray}
Since projectile and target matrix elements can bring $s$ each and ${1\over\alpha}$ and ${1\over\beta}$ convert to
${1\over\alpha_q}$ and ${1\over\beta_q}$, the typical contribution of (\ref{kalw2craze}) to $W^{\mu\nu}(q)$ is
\begin{equation}
 \sim ~{1\over N_c^2}\times\bigg(g_\perp^{\mu\nu}{q_\perp^2\over \alpha_q\beta_qs},~~ {q_\perp^\mu q_\perp^\nu\over \alpha_q\beta_qs},~~g_\parallel^{\mu\nu}{q_\perp^2\over \alpha_q\beta_qs}\bigg)
\label{kalw2csmall}
\end{equation}
 In Ref. \cite{Balitsky:2017gis} we calculated the sum of these structures
corresponding to convolution of $\mu$ and $\nu$. In principle, one can repeat that calculation and find contribution to
these structures separately. However, since the corresponding matrix elements of quark-quark-gluon operators are 
virtually unknown, in this paper we we will disregard such ${1\over N_c^2}$ terms. 

Thus, the contribution of Eq. (\ref{7lines}) to $W_{\mu\nu}(q)$ is given in the leading order in $N_c$ by the sum of equations 
(\ref{XiBxi}), (\ref{w2atvet}), and (\ref{w2botvet}). 

\section{Power corrections from $J_{A}^\mu(x)J_{B}^\nu(0)$ terms \label{2ndtype}}
Power corrections of the second type come from the terms
\begin{eqnarray}
&&\hspace{-1mm} 
\Bsi_1(x)\gamma_\mu\Psi_1(x)\Bsi_2(0)\gamma_\nu\Psi_2(0)~+~x\leftrightarrow 0
\end{eqnarray}
where $\Psi_1$ and $\Psi_2$ are given by Eq. (\ref{fildz0}). 
\footnote{In the appendix 8.3.2 to \cite{Balitsky:2017gis} it is demonstrated that  higher-order terms in the expansion Eq. (\ref{klfildz})
 (denoted by dots) are small in our kinematical region $s\gg Q^2\gg q_\perp^2$.} We get
\begin{eqnarray}
&&\hspace{-1mm}
\big[\big(\bar\psi_A +\Bxi_{1}\big)(x)\gamma_\mu\big(\psi_A+\Xi_{1}\big)(x)\big]
[\big(\bar\psi_B+\Bxi_{2}\big)(0)\gamma_\nu\big(\psi_B+\Xi_{2}\big)(0)\big]~+~x\leftrightarrow 0
\nonumber\\
&&\hspace{-1mm}
=~[\bar\psi_A(x)\gamma_\mu\psi_A(x)\big]\big[\bar\psi_B(0)\gamma_\nu\psi_B(0)\big]
\label{7newlines}\\
&&\hspace{-1mm}
+~[\Bxi_{1}(x)\gamma_\mu\psi_A(x)\big]\big[\bar\psi_B(0)\gamma_\nu\psi_B(0)\big]
+[\bar\psi_A(x)\gamma_\mu\Xi_{1}(x)\big]\big[\bar\psi_B(0)\gamma_\nu\psi_B(0)\big]
\nonumber\\
&&\hspace{-1mm}
+~
[\bar\psi_A(x)\gamma_\mu\psi_A(x)\big]\big[\Bxi_{2}(0)\gamma_\nu\psi_B(0)\big]
+[\bar\psi_A(x)\gamma_\mu\psi_A(x)\big]\big[\bar\psi_B(0)\gamma_\nu\Xi_{2}(0)\big]
\nonumber\\
&&\hspace{-1mm}
+~[\Bxi_{1}(x)\gamma_\mu\Xi_{1}(x)\big]\big[\bar\psi_B(0)\gamma_\nu\psi_B(0)\big]
+[\bar\psi_A(x)\gamma_\mu\psi_{A}(x)\big]\big[\Bxi_{2}(0)\gamma_\nu\Xi_{2}(0)\big]
\nonumber\\
&&\hspace{-1mm}
+~[\Bxi_{1}(x)\gamma_\mu\psi_A(x)\big]\big[\bar\psi_B(0)\gamma_\nu\Xi_{2}(0)\big]
+[\bar\psi_A(x)\gamma_\mu\Xi_{1}(x)\big]\big[\Bxi_{2}(0)\gamma_\nu\psi_B(0)\big]
\nonumber\\
&&\hspace{-1mm}
+~[\Bxi_{1}(x)\gamma_\mu\psi_A(x)\big]\big[\Bxi_{2}(0)\gamma_\nu\psi_B(0)\big]
+[\bar\psi_A(x)\gamma_\mu\Xi_{1}(x)\big]\big[\bar\psi_B(0)\gamma_\nu\Xi_{2}(0)\big]
~+~x\leftrightarrow 0.
\nonumber
\end{eqnarray}
First, let us demonstrate  that contributions to $W_{\mu\nu}$ from the second to fifth lines in eq. (\ref{7newlines}) vanish.
Obviously, matrix element of the operator in the second line vanishes. Formally,
\begin{eqnarray}
&&\hspace{-1mm}
\int\! dx_\bu~e^{-i\alpha_q x_\bu}\langle p_A|\hsi(x_\bu,x_\perp)\gamma_\mu\hsi(x_\bu,x_\perp)|p_A\rangle
~=~\delta(\alpha_q)\langle p_A|\hsi(0)\gamma_\mu\hsi(0)|p_A\rangle,
\nonumber\\
&&\hspace{-1mm}
\int\! dx_\ast~e^{-i\beta_q x_\ast}\langle p_B|\hsi(0)\gamma_\nu\hsi(0)|p_B\rangle
~=~\delta(\beta_q)\langle p_B|\hsi(0)\gamma_\nu\hsi(0)|p_B\rangle
\label{xxvanish}
\end{eqnarray}
and, non-formally, one hadron cannot produce the DY pair on its own. 

It is easy to see that  contributions to $\cheW_{\mu\nu}$ from the third and the fourth lines in Eq. (\ref{7newlines}) vanish
due to the absence of color-singlet structure. 
Indeed, let us consider for example the term
\begin{equation}
\hspace{-0mm}
[\Bxi_{1}(x)\gamma_\mu\psi_A(x)\big]\big[\bar\psi_B(0)\gamma_\nu\psi_B(0)\big]
~=~
-\big[\big(\bar\psi_A^m{1\over\alpha}\big)(x)\gamma^iB^{mn}_i{\slashed{p}_2\over s}\gamma_\mu\psi_A^n(x)\big]\big[\bar\psi_B^l(0)\gamma_\nu\psi_B^l(0)\big]
\end{equation}
The corresponding term in  $\cheW_{\mu\nu}$ is
\begin{eqnarray}
&&\hspace{-1mm}
-{N_c\over s}\langle\big(\bar\psi^m{1\over\alpha}\big)(x)\gamma^i{\slashed{p}_2\over s}\gamma_\mu\psi^n(x)\rangle_A
\langle\bar\psi^l(0)B^{mn}_i(0)\gamma_\nu\psi^l(0)\rangle_B~+~\mu\leftrightarrow \nu
\end{eqnarray}
which obviously does not have color-singlet contribution. Similarly, other three terms in the third and fourth lines in Eq. (\ref{7newlines}) vanish.

Next, let us demonstrate that the contribution of the fifth line in Eq. (\ref{7newlines}) vanishes for the same 
reason as in Eq. (\ref{xxvanish}). Let is consider for example the first term in the fifth line
\begin{eqnarray}
&&\hspace{-1mm}
[\Bxi_{1}(x)\gamma_\mu\Xi_{1}(x)\big]\big[\bar\psi_B(0)\gamma_\nu\psi_B(0)\big]
\nonumber\\
&&\hspace{-1mm}
=~{2p_{2\mu}\over s^2}\big[\big(\bar\psi_A{1\over\alpha}\big)(x)\gamma^iB_i(x)
\slashed{p}_2\gamma^jB_j(x){1\over \alpha}\psi_A(x)\big]\big[\bar\psi_B(0)\gamma_\nu\psi_B(0)\big]
\nonumber\\
&&\hspace{-1mm}
=~{p_{2\mu}\over N_cs^2}\big[\big(\bar\psi_A^m{1\over\alpha}\big)(x)\gamma^i
\slashed{p}_2\gamma^j{1\over \alpha}\psi_A^m(x)\big]\big[B^a_iB^a_j(x)\bar\psi^n_B(0)\gamma_\nu\psi^n_B(0)\big]
\end{eqnarray}
where we separated color-singlet contribution in the last line.  The corresponding term in $W_{\mu\nu}$ is
\begin{eqnarray}
&&\hspace{-1mm}
{1\over 64\pi^6N_c}{2p_{2\mu}\over s^3}\!\int d^2k_\perp
\!\int\!dx_\bu d^2x_\perp~e^{-i\alpha x_\bu+i(k,x)_\perp}\!\int\!dx_\ast d^2x'_\perp e^{-i\beta x_\bu+i(q-k,x')_\perp}
\\
&&\hspace{-1mm}
\times~\langle\big(\bar\psi{1\over\alpha}\big)(x_\bu,x_\perp)\gamma^i
\slashed{p}_2\gamma^j{1\over \alpha}\psi(x_\bu,x_\perp)\rangle_A
\langle B^a_i(x_\ast,x'_\perp)B^a_j(x_\ast,x'_\perp)\bar\psi(0)\gamma_\nu\psi(0)\rangle_B
\nonumber\\
&&\hspace{-1mm}
=~\delta(\alpha_q){1\over 32\pi^5N_c}{2p_{2\mu}\over s^3}
\langle\big(\bar\psi{1\over\alpha}\big)(0)\gamma^i
\slashed{p}_2\gamma^j{1\over \alpha}\psi(0)\rangle_A
\nonumber\\
&&\hspace{22mm}
\times~\!\int\!dx_\ast d^2x'_\perp e^{-i\beta x_\bu+i(q,x')_\perp}
\langle B^a_i(x_\ast,x'_\perp)B^a_j(x_\ast,x'_\perp)\bar\psi(0)\gamma_\nu\psi(0)\rangle_B~=~0
\nonumber
\end{eqnarray}
Similarly, the contribution of the second term in the fifth line of Eq. (\ref{7newlines}) will be proportional to $\delta(\beta_q)$ and hence 
vanish.

Let us now discuss the non-vanishing contributions coming from last two lines in Eq. (\ref{7newlines}). 
For example, the first term in the sixth line is
\begin{eqnarray}
&&\hspace{-1mm}
[\Bxi_{1}(x)\gamma_\mu\psi_A(x)\big]\big[\bar\psi_B(0)\gamma_\nu\Xi_{2}(0)\big]
~=~\big(\bar\psi_A^m{1\over\alpha}\big)\gamma^iB^{mn}_i{\slashed{p}_2\over s}
\gamma_\mu\psi^n_A(x)
\bar\psi^k_B\gamma_\nu{\slashed{p}_1\over s}\gamma^jA^{kl}_j{1\over\beta}\psi_B^l(0)
\nonumber\\
&&\hspace{-1mm}
=~\big[\big(\bar\psi_A^m{1\over\alpha}\big)\gamma^iA^{kl}_j(0){\slashed{p}_2\over s}
\gamma_\mu\psi^n_A(x)\big]
\big[\bar\psi^k_B\gamma_\nu{\slashed{p}_1\over s}\gamma^jB^{mn}_i(x){1\over\beta}\psi_B^l(0)\big]
\end{eqnarray}
Separating color-singlet contributions with the help of the formula 
\begin{equation}
\langle\bsi_m A^a_i\psi_n\rangle~=~{2t^a_{mn}\over N_c^2-1}\langle\bsi A_i\psi\rangle
\label{colors2}
\end{equation}
we get the corresponding term in $\cheW_{\mu\nu}$ in the form
\begin{eqnarray}
&&\hspace{-1mm}
{N_c\over N_c^2-1}{2\over s^3}\langle\big(\bar\psi{1\over\alpha}\big)(x)A_j(0)\gamma^i\slashed{p}_2
\gamma_\mu\psi(x)\rangle_A
\langle\bar\psi(0)\gamma_\nu\slashed{p}_1\gamma^jB_i(x){1\over\beta}\psi(0)\rangle_B
\end{eqnarray}
which is similar to Eq. (\ref{kalve2}) with exception of extra color factor ${N_c\over N_c^2-1}\simeq{1\over N_c}$. 
Consequently, as discussed in Sect. \ref{sec:kalv2}, non-negligible contributions come from transverse $\mu$ 
and $\nu$ only.  We calculate them in next Section.

\subsection{Last two lines  in eq. (\ref{7newlines}) \label{67lines}}
In this section we calculate the traceless part of  sixth and seventh lines  Eq. (\ref{7newlines}). Since we consider only transverse 
$\mu$ and $\nu$, to simplify notations we will call them $m$ and $n$ in this Section.
Using eq. (\ref{fildz0}) and separating color-singlet matrix elements with the help of Eq. (\ref{colors2}), we 
rewrite the traceless part of sixth and seventh lines in Eq. (\ref{7newlines}) as
\begin{eqnarray}
&&\hspace{-1mm}
\half\big([\Bxi_{1}(x)\gamma_m\psi_A(x)\big]\big[\bar\psi_B(0)\gamma_n\Xi_{2}(0)\big]
+[\bar\psi_A(x)\gamma_n\Xi_{1}(x)\big]\big[\Bxi_{2}(0)\gamma_n\psi_A(0)\big]
\nonumber\\
&&\hspace{-1mm}
+[\Bxi_{1}(x)\gamma_m\psi_A(x)\big]\big[\Bxi_{2}(0)\gamma_n\psi_B(0)\big]
+[\bar\psi_A(x)\gamma_m\Xi_{1}(x)\big]\big[\bar\psi_B(0)\gamma_n\Xi_{2}(0)\big]
+m\leftrightarrow n\big)
\nonumber\\
&&\hspace{-1mm}
=~{1\over 2(N_c^2-1)s^2}\Big(\big[\big(\bsi_A{1\over\alpha}\big)(x)\gamma^j\slashed{p}_2\gamma_m A_k(0)\psi_A(x)\big]
\big[\bsi_B(0)\gamma_n\slashed{p}_1\gamma^k B_j(x){1\over\beta}\psi_B(0)\big]
\nonumber\\
&&\hspace{-1mm}
+~
\big[\bsi_A(x)\gamma_n\slashed{p}_2\gamma^j A_k(0){1\over\alpha}\psi_A(x)\big]
\big[\big(\bsi_B{1\over\beta}\big)(0)\gamma^k\slashed{p}_1\gamma_m B_j(x)\psi_B(0)\big]
\nonumber\\
&&\hspace{-1mm}
+~\big[\big(\bsi_A{1\over\alpha}\big)(x)\gamma^j\slashed{p}_2\gamma_m A_k(0)\psi_A(x)\big]
\big[\big(\bsi_B{1\over\beta}\big)(0)\gamma^k\slashed{p}_1\gamma_n B_j(x)\psi_B(0)\big]
\nonumber\\
&&\hspace{-1mm}
+~
\big[\bsi_A(x)\gamma_m\slashed{p}_2\gamma^j A_k(0){1\over\alpha}\psi_A(x)\big]
\big[\bsi_B(0)\gamma_n\slashed{p}_1\gamma^k B_j(x){1\over\beta}\psi_B(0)\big]+m\leftrightarrow n\Big)
\label{flaxz1}
\end{eqnarray}

To save space, hereafter we do not display subtraction of trace with respect to $m,n$ indices but it is always assumed.
Using formulas (\ref{formulas67}) we can write down the contribution to $\cheW_{\mu\nu}$ from sixth and seventh lines in Eq. (\ref{7newlines}) in the form
\begin{eqnarray}
&&\hspace{-1mm}
\cheW^{\rm 6+7th}_{mn}=~{N_c\over (N_c^2-1)s^3}\Big(
\langle\big(\bsi{1\over\alpha}\big)(x)\slashed{p}_2\brA_m(0)\psi(x)\rangle_A
\langle\bar\psi(0)\breB_n(x)\slashed{p}_1{1\over\beta}\psi(0)\rangle_B
\nonumber\\
&&\hspace{-1mm}
+\langle\bar\psi(x)\brA_m(0)\slashed{p}_2{1\over \alpha}\psi(x)\rangle_A
\langle\big(\bar\psi{1\over\beta}\big)(0)\slashed{p}_1\breB_n(x)\psi(0)\rangle_B
\nonumber\\
&&\hspace{-1mm}
+~\langle\big(\bsi{1\over\alpha}\big)(x)\slashed{p}_2\brA_m(0)\psi(x)\rangle_A
\langle\big(\bar\psi{1\over\beta}\big)(0)\slashed{p}_1\breB_n(x)\psi(0)\rangle_B
\nonumber\\
&&\hspace{-1mm}
+~\langle\bar\psi(x)\brA_m(0)\slashed{p}_2\big({1\over \alpha}\psi\big)(x)\rangle_A
\langle\bar\psi(0)\breB_n\slashed{p}_1\big({1\over\beta}\psi\big)(0)\rangle_B+m\leftrightarrow n\Big)
~+~x\leftrightarrow 0,
\nonumber\\
&&\hspace{-1mm}
=~{N_c\over (N_c^2-1)s^3}\Big(
\langle\big(\bsi{1\over\alpha}\big)(x)\slashed{p}_2\brA_m(0)\psi(x)
+\bar\psi(x)\brA_m(0)\slashed{p}_2{1\over \alpha}\psi(x)\rangle_A
\nonumber\\
&&\hspace{-1mm}
\times~\langle\big(\bar\psi{1\over\beta}\big)(0)\slashed{p}_1\breB_n(x)\psi(0)\rangle_B
+\bar\psi(0)\breB_n(x)\slashed{p}_1{1\over\beta}\psi(0)\rangle_B
+m\leftrightarrow n\Big)
~+~x\leftrightarrow 0
\label{kalw67e}
\end{eqnarray}
Let us now consider  corresponding matrix elements. It is easy to see that
\begin{eqnarray}
&&\hspace{-1mm}
{1\over 8\pi^3s}\!\int\! d^2x_\perp dx_\bu~e^{-i\alpha x_\bu+i(k,x)_\perp}
\langle\bsi(x_\bu,x_\perp)\brA_i(0)\slashed{p}_2{1\over\alpha}\psi(x_\bu,x_\perp)\rangle_A
\nonumber\\
&&\hspace{-1mm}
=~{1\over\alpha}{1\over 4\pi^3s^2}\!\int\! d^2x_\perp dx_\bu~e^{-i\alpha x_\bu+i(k,x)_\perp}
\!\int_{-\infty}^{x_\bu}\! dx'_\bu\langle\bsi(x_\bu,x_\perp)\slashed{p}_2
[F_{\ast i}+i\gamma_5\tilF_{\ast i}](0)\psi(x'_\bu,x_\perp)\rangle_A
\nonumber\\
&&\hspace{33mm}
=~{k_i\over\alpha}j_1(\alpha,k_\perp),
\nonumber\\
&&\hspace{-1mm}
{1\over 8\pi^3s}\!\int\! d^2x_\perp dx_\bu~e^{-i\alpha x_\bu+i(k,x)_\perp}
\langle\big(\bsi {1\over\alpha}\big)(x_\bu,x_\perp)\slashed{p}_2\brA_i(0)\psi(x_\bu,x_\perp)\rangle_A
\nonumber\\
&&\hspace{-1mm}
=~-{1\over\alpha}{1\over 4\pi^3s^2}\!\int\! d^2x_\perp dx_\bu~e^{-i\alpha x_\bu+i(k,x)_\perp}
\!\int_{-\infty}^{x_\bu}\! dx'_\bu\langle\bsi (x'_\bu,x_\perp)\slashed{p}_2[F_{\ast i}-i\gamma_5\tilF_{\ast i}](0)\psi(x_\bu,x_\perp)\rangle_A,
\nonumber\\
&&\hspace{33mm}
=~-{k_i\over\alpha}\barj_1(\alpha,k_\perp),
\nonumber\\
&&\hspace{-1mm}
{1\over 8\pi^3s}\!\int\! dx_\bu~e^{-i\alpha x_\bu}
\langle\bsi(0)\brA_i(x_\bu,x_\perp)\slashed{p}_2{1\over\alpha}\psi(0)\rangle_A
\nonumber\\
&&\hspace{-1mm}
=~-{1\over\alpha}{1\over 4\pi^3s^2}\!\int\! d^2x_\perp dx_\bu~e^{-i\alpha x_\bu+i(k,x)_\perp}
\!\int_{-\infty}^0\! dx'_\bu\langle\bsi(0)\slashed{p}_2
[F_{\ast i}+i\gamma_5\tilF_{\ast i}](x_\bu,x_\perp)\psi(x'_\bu,0_\perp)\rangle_A
\nonumber\\
&&\hspace{33mm}
=~-{k_i\over\alpha}\barj^\star_1(\alpha,k_\perp),
\nonumber\\
&&\hspace{-1mm}
{1\over 8\pi^3s}\!\int\! d^2x_\perp dx_\bu~e^{-i\alpha x_\bu+i(k,x)_\perp}
\langle\big(\bsi {1\over\alpha}\big)(0)\slashed{p}_2\brA_i(x_\bu,x_\perp)\psi(0)\rangle_A
\nonumber\\
&&\hspace{-1mm}
=~{1\over\alpha}{1\over 4\pi^3s^2}\!\int\! d^2x_\perp dx_\bu~e^{-i\alpha x_\bu+i(k,x)_\perp}
\!\int_{-\infty}^{x_\bu}\! dx'_\bu\langle\bsi (x'_\bu,0_\perp)\slashed{p}_2[F_{\ast i}-i\gamma_5\tilF_{\ast i}](x_\bu,x_\perp)\psi(0)\rangle_A,
\nonumber\\
&&\hspace{33mm}
=~{k_i\over\alpha}j_1^\star(\alpha,k_\perp),
\label{paramjse}
\end{eqnarray}
where we used parametrization (\ref{paramjstar}). For the target matrix elements, we obtain
\begin{eqnarray}
&&\hspace{-1mm}
{1\over 8\pi^3s}\!\int\! d^2x_\perp dx_\ast~e^{-i\beta x_\ast+i(k,x)_\perp}\langle\bsi(x_\bu,x_\perp)
\breB_i(0)\slashed{p}_1{1\over\beta}\psi(x_\ast,x_\perp)\rangle_B
~=~{k_i\over\beta}j_1(\beta,k_\perp),
\nonumber\\
&&\hspace{-1mm}
{1\over 8\pi^3s}\!\int\! d^2x_\perp dx_\ast~e^{-i\alpha x_\ast+i(k,x)_\perp}
\langle\big(\bsi {1\over\beta}\big)(x_\ast,x_\perp)\slashed{p}_1\breB_i(0)\psi(x_\ast,x_\perp)\rangle_A
~=~-{k_i\over\beta}\barj_1(\beta,k_\perp),
\nonumber\\
&&\hspace{-1mm}
{1\over 8\pi^3s}\!\int\! d^2x_\perp dx_\ast~e^{-i\beta x_\ast+i(k,x)_\perp}
\langle\bsi(0)\breB_i(x_\ast,x_\perp)\slashed{p}_1{1\over\beta}\psi(0)\rangle_A~
=~-{k_i\over\beta}\barj^\star_1(\beta,k_\perp),
\nonumber\\
&&\hspace{-1mm}
{1\over 8\pi^3s}\!\int\! d^2x_\perp dx_\ast~e^{-i\beta x_\ast+i(k,x)_\perp}
\langle\big(\bsi {1\over\beta}\big)(0)\slashed{p}_1\breB_i(x_\ast,x_\perp)\psi(0)\rangle_A
~=~{k_i\over\beta}j_1^\star(\beta,k_\perp),
\label{paramjset}
\end{eqnarray}

The corresponding contribution to (traceless)
$W(\alpha_q,\beta_q,x_\perp)$ takes the form
\begin{eqnarray}
&&\hspace{-1mm}
W_{mn}^{\rm 6+7th}(q)-{\rm trace}~=~{s/2\over(2\pi)^4N_c}\int\!d^4x ~e^{-iqx} \big(\cheW_{mn}^{\rm 6+7th}(x_\perp)
-{\rm trace}\big),
\nonumber\\
&&\hspace{-1mm}
=~{1\over (N_c^2-1)\alpha_q\beta_qs}\!\int\! d^2k_\perp[(j_1-\barj_1)(\alpha_q,k_\perp)(j^\star_1-\barj^\star_1)
(\beta_q,(q-k)_\perp)+{\rm c.c.}]
\nonumber\\
&&\hspace{33mm}
\times~[k_m(q-k)_n+m\leftrightarrow n+g_{mn}(k,q-k)_\perp]
\label{w67traceless}
\end{eqnarray}
where we have recovered the subtraction of trace.

The trace part can be obtained in a similar way. Using Eq. (\ref{gammas11}) one gets
\begin{eqnarray}
&&\hspace{-1mm}
g^{mn}\cheW^{\rm 6+7th}_{mn}=~{2N_c\over (N_c^2-1)s^3}\Big(
\langle\big(\bsi{1\over\alpha}\big)(x)\brA_m(0)\slashed{p}_2\psi(x)\rangle_A
\langle\bar\psi(0)\slashed{p}_1\breB^m(x){1\over\beta}\psi(0)\rangle_B
\nonumber\\
&&\hspace{-1mm}
+\langle\bar\psi(x)\slashed{p}_2\brA_m(0){1\over \alpha}\psi(x)\rangle_A
\langle\big(\bar\psi{1\over\beta}\big)(0)\breB^m(x)\slashed{p}_1\psi(0)\rangle_B
\nonumber\\
&&\hspace{-1mm}
+~\langle\big(\bsi{1\over\alpha}\big)(x)\brA_m(0)\slashed{p}_2\psi(x)\rangle_A
\langle\big(\bar\psi{1\over\beta}\big)(0)\breB^m(x)\slashed{p}_1\psi(0)\rangle_B
\nonumber\\
&&\hspace{-1mm}
+~\langle\bar\psi(x)\slashed{p}_2\brA_m(0)\big({1\over \alpha}\psi\big)(x)\rangle_A
\langle\bar\psi(0)\slashed{p}_1\breB^m\big({1\over\beta}\psi\big)(0)\rangle_B+m\leftrightarrow n\Big)
~+~x\leftrightarrow 0,
\nonumber\\
&&\hspace{-1mm}
=~{2N_c\over (N_c^2-1)s^3}\Big(
\langle\big(\bsi{1\over\alpha}\big)(x)\brA_m(0)\slashed{p}_2\psi(x)
+\bar\psi(x)\slashed{p}_2\brA_m(0){1\over \alpha}\psi(x)\rangle_A
\nonumber\\
&&\hspace{-1mm}
\times~\langle\big(\bar\psi{1\over\beta}\big)(0)\breB^m(x)\slashed{p}_1\psi(0)\rangle_B
+\bar\psi(0)\slashed{p}_1\breB^m(x){1\over\beta}\psi(0)\rangle_B\Big)
~+~x\leftrightarrow 0
\label{kalw67}
\end{eqnarray}
The corresponding contribution to trace part of 
$W(\alpha_q,\beta_q,x_\perp)$ takes the form
\begin{eqnarray}
&&\hspace{-1mm}
g^{mn}W_{mn}^{\rm 6+7th}(q)~=~{s/2\over(2\pi)^4N_c}\int\!d^4x ~e^{-iqx} g^{mn}\cheW_{mn}^{\rm 6+7th}(x_\perp)
\label{w67trace}\\
&&\hspace{-1mm}
=~-{2\over (N_c^2-1)\alpha_q\beta_qs}\!\int\! d^2k_\perp (k,q-k)_\perp[(j_2-\barj_2)(\alpha_q,k_\perp)(j^\star_2-\barj^\star_2)
(\beta_q,(q-k)_\perp)+{\rm c.c.}]
\nonumber
\end{eqnarray}
which agrees with Eq. (6.2) from Ref. \cite{Balitsky:2017gis} after 
replacements $j_2=j_2^{\rm tw 3}-i\tilj_2^{\rm tw 3}$ and $\barj_2=j_1^{\rm tw 3}+i\tilj_1^{\rm tw 3}$.
It should be noted that the difference between $j_1$ and $j_2$ in traceless {\it vs} trace part is due to difference in formulas 
(\ref{formulas67}) and (\ref{gammas11}).

Thus, the result is the sum of Eqs. (\ref{w67traceless}) and 
(\ref{w67trace}) 
\begin{eqnarray}
&&\hspace{-1mm}
W_{\mu\nu}^{\rm 6+7th}(q)~=~{s/2\over(2\pi)^4N_c}\int\!d^4x ~e^{-iqx} \cheW_{mn}^{\rm 6+7th}(x_\perp)
\nonumber\\
&&\hspace{-1mm}
=~{1\over (N_c^2-1)\alpha_q\beta_qs}\!\int\! d^2k_\perp\Big([(j_1-\barj_1)(\alpha_q,k_\perp)(j^\star_1-\barj^\star_1)
(\beta_q,(q-k)_\perp)+{\rm c.c.}]
\nonumber\\
&&\hspace{33mm}
\times~[k_\mu(q-k)_\nu+\mu\leftrightarrow\nu+g^\perp_{\mu\nu}(k,q-k)_\perp]
\nonumber\\
&&\hspace{-1mm}
-~g^\perp_{\mu\nu} (k,q-k)_\perp[(j_2-\barj_2)(\alpha_q,k_\perp)(j^\star_2-\barj^\star_2)
(\beta_q,(q-k)_\perp)+{\rm c.c.}]
\label{w67otvet}
\end{eqnarray}
As we mentioned in the Introduction, in this paper we we will take into account only leading  and sub-leading terms in $N_c$ 
 and leave the ${1\over N_c^2}$ corrections discussed above for future publications.
 
 Finally, as proved in Appendix \ref{gluterms}, we can neglect contributions proportional to the product of quark and gluon TMDs. 

\section{Results and estimates \label{sec:results}}
\subsection{Results \label{sec:fotoresults}}
Assembling Eqs. (\ref{WLT}), (\ref{XiBxi}), (\ref{w2atvet}), (\ref{w2botvet}), and (\ref{w67otvet}) we get
the result for $W_{\mu\nu}(q)$ that consists of two parts:
\begin{equation}
W_{\mu\nu}(q)~=~W^1_{\mu\nu}(q)+W^2_{\mu\nu}(q)
\label{result}
\end{equation}
The first, gauge-invariant,  part is given by
\begin{eqnarray}
&&\hspace{-1mm}
W^1_{\mu\nu}(q)~=~W^{1F}_{\mu\nu}(q)+W^{1H}_{\mu\nu}(q),
\nonumber\\
&&\hspace{-1mm}
W^{1F}_{\mu\nu}(q)~=~\sum_f e_f^2W^{fF}_{\mu\nu}(q),~~~~
W^{fF}_{\mu\nu}(q)~=~{1\over N_c}\!\int\!d^2k_\perp F^f(q,k_\perp)\calW^F_{\mu\nu}(q,k_\perp),
\nonumber\\
&&\hspace{-1mm}
W^{1H}_{\mu\nu}(q)~=~\sum_f e_f^2W^{fH}_{\mu\nu}(q),~~~~W^{fH}_{\mu\nu}(q)~=~ {1\over N_c}\!\int\!d^2k_\perp H^f(q,k_\perp)\calW^H_{\mu\nu}(q,k_\perp)
\label{resultginv}
\end{eqnarray}
where $F^f$ and $H^f$ are given by Eq. (\ref{FH1}) and
\begin{eqnarray}
&&\hspace{-1mm}
\calW^{F}_{\mu\nu}(q,k_\perp)~=~
-g_{\mu\nu}^\perp +{1\over Q_\parallel^2}(q^\parallel_\mu q^\perp_\nu+q^\parallel_\nu q^\perp_\mu)
+{q_\perp^2\over Q_\parallel^4}q^\parallel_\mu q^\parallel_\nu+{\tilq_\mu\tilq_\nu\over Q_\parallel^2}[q_\perp^2-4(k,q-k)_\perp]
\nonumber\\
&&\hspace{20mm}
-~\Big[{\tilq_\mu\over  Q_\parallel^2}\Big(g^\perp_{\nu i}-{q^\parallel_\nu q_i\over Q_\parallel^2}\Big)(q-2k)_\perp^i
+\mu\leftrightarrow\nu\Big]
\label{WF}
\end{eqnarray}
%
\begin{eqnarray}
&&\hspace{-1mm}
m^2\calW_{\mu\nu}^H(q,k_\perp)~
\label{WH}\\
&&\hspace{-1mm}
=~-k^\perp_\mu(q-k)^\perp_\nu-k^\perp_\nu(q-k)^\perp_\mu-g_{\mu\nu}^\perp(k,q-k)_\perp
+2{\tilq_\mu\tilq_\nu-q^\parallel_\mu q^\parallel_\nu \over Q_\parallel^4}k_\perp^2(q-k)_\perp^2
\nonumber\\
&&\hspace{-1mm}
-~\Big({q^\parallel_\mu\over Q_\parallel^2}\big[k_\perp^2(q-k)^\perp_\nu+k^\perp_\nu(q-k)_\perp^2\big]
+~{\tilq_\mu\over Q_\parallel^2}\big[k_\perp^2(q-k)^\perp_\nu-k^\perp_\nu(q-k)_\perp^2\big]+\mu\leftrightarrow\nu\Big)
\nonumber\\
&&\hspace{-1mm}
-~{\tilq_\mu\tilq_\nu+q^\parallel_\mu q^\parallel_\nu \over Q_\parallel^4}\big[q_\perp^2-2(k,q-k)_\perp\big](k,q-k)_\perp
-~{q^\parallel_\mu\tilq_\nu+ \tilq_\mu q^\parallel_\nu \over Q_\parallel^4}(2k-q,q)_\perp(k,q-k)_\perp
\nonumber
\end{eqnarray}
where $q^\parallel_\mu\equiv \alpha_qp_1+\beta_qp_2$ and $\tilq_\mu\equiv \alpha_qp_1-\beta_qp_2$. These are the same expressions as in Eq. (\ref{wfwh}) if
one identifies $x_A$ with $\alpha_q$ and $x_B$ with $\beta_q$ and neglects $O\big({m^2\over s}\big)$ terms
in $p_A$ and $p_B$ and $O\big({q_\perp^4\over Q^4}\big)$ corrections due to difference between $Q^2$ and $Q_\parallel^2$. 
It is easy to see that $q^\mu W^F_{\mu\nu}~=~0$ and  $q^\mu W^H_{\mu\nu}~=~0$. Note that 
$q^\mu W^F_{\mu\nu}$ and  $q^\mu W^H_{\mu\nu}$ are exactly zero without any ${q_\perp^2\over Q^2}$ corrections. 
This is similar to usual ``forward'' DIS, but different from off-forward DVCS where the cancellations of right-hand sides of Ward identities 
involve infinite towers of twists \cite{Braun:2011dg,Braun:2011zr,Braun:2020zjm}
  
The second part is
\begin{eqnarray}
&&\hspace{-1mm}
W^2_{\mu\nu}(q)~
=~{1\over N_c}\sum_f e_f^2{1\over Q^2}\!\int d^2k_\perp
\bigg[ {1\over m^2}\big\{[k^\perp_\mu(q-k)^\perp_\nu+\mu\leftrightarrow\nu](k,q-k)_\perp
\nonumber\\
&&\hspace{-1mm}
-~k_\perp^2(q-k)^\perp_\mu(q-k)^\perp_\nu
-(q-k_\perp)^2k^\perp_\mu k^\perp_\nu+g_{\mu\nu}^\perp(k,q-k)_\perp^2
-g_{\mu\nu}^\perp k_\perp^2(q-k_\perp)^2\big]\big\}H_A^f(q,k_\perp)
\nonumber\\
&&\hspace{-1mm}
+~{N_c\over N_c^2-1}\Big([k^\perp_\mu(q-k)^\perp_\nu+\mu\leftrightarrow\nu+g^\perp_{\mu\nu}(k,q-k)_\perp]J_1^f(q,k_\perp)
\nonumber\\
&&\hspace{33mm}
-~g^\perp_{\mu\nu} (k,q-k)_\perp J_2^f(q,k_\perp)\Big)~+~O\big({1\over N_c^2}\big)\bigg]~+~O\big({Q_\perp^4\over Q^4}\big)
\label{resultnoninv}
\end{eqnarray}
where $H_A$ is given by Eq. (\ref{HA}) and 
\begin{eqnarray}
&&\hspace{-11mm}
J_1^f(q,k_\perp)~=~(j_1-\barj_1)(\alpha_q,k_\perp)(j^\star_1-\barj^\star_1)(\beta_q,(q-k)_\perp)+{\rm c.c.}
\nonumber\\
&&\hspace{-11mm}
J_2^f(q,k_\perp)~=~(j_2-\barj_2)(\alpha_q,k_\perp)(j^\star_2-\barj^\star_2)
(\beta_q,(q-k)_\perp)+{\rm c.c.}
\label{W2}
\end{eqnarray}
These terms are not gauge invariant: $q^\mu W^2_{\mu\nu}(q)\neq 0$. The reason is that gauge invariance is restored after adding terms like ${m_\perp^2\over Q^2}\times{\rm Eq. (\ref{pc})}$  which we do not calculate in this paper. Indeed, for example,
\begin{eqnarray}
&&\hspace{-1mm}
q^\mu W^2_{\mu\nu}(q)~\sim~{q^\perp_\nu q_\perp^2\over\alpha_q\beta_qs}~~~~
{\rm and}~~~q^\mu \times~{p_2^\mu q_\perp^\nu q_\perp^2\over \alpha_q^2\beta_qs^2}
~=~{q^\perp_\nu q_\perp^2\over\alpha_q\beta_qs}
\end{eqnarray}
They are of the same order so one should expect that gauge invariance is restored after calculation of the terms $\sim{p_2^\mu q_\perp^\nu q_\perp^2\over \alpha_q^2\beta_qs^2}$ which are beyond the scope of this paper.
For the same reason we see that all structures in Eq. (\ref{pc}) except 
${g_{\mu\nu}^\perp q_\perp^2\over\alpha_q\beta_qs}$
and ${q^\perp_\mu q^\perp_\nu \over\alpha_q\beta_qs}$ are determined by leading-twist TMDs $f_1$ and $h_1^\perp$.

Sometimes it is convenient to represent hadronic tensor in transverse coordinate space. Introducing 
\begin{eqnarray}
&&\hspace{-1mm}
\begin{array}{c}
f(\alpha,b_\perp)
\\
\barf(\alpha,b_\perp)
\end{array}\bigg\}~=~\int\!{d^2k_\perp\over 4\pi^2}~e^{i(k,b)_\perp}
\bigg\{\begin{array}{c}
f(\alpha,k_\perp)
\\
\barf(\alpha,k_\perp)
\end{array}
\end{eqnarray}
(and similarly for target TMDs) we get 
\begin{eqnarray}
&&\hspace{-1mm}
W^{1F}_{\mu\nu}(\alpha_q,\beta_q,b_\perp)~
\label{w1fcoord}\\
&&\hspace{-1mm}
=~4\pi^2\sum_f e_f^2 
\Big\{-g_{\mu\nu}^\perp +{i\over Q_\parallel^2}(q^\parallel_\mu\partial^\perp_\nu+q^\parallel_\nu\partial^\perp_\mu)
-{q^\parallel_\mu q^\parallel_\nu+\tilq_\mu\tilq_\nu\over Q_\parallel^2}\partial_\perp^2\Big]f_1\barf_1
-~{4\tilq_\mu\tilq_\nu\over Q_\parallel^2}(\partial_if_1^f)(\partial^i \barf_1^f) 
\nonumber\\
&&\hspace{11mm}
-~\Big[{\tilq_\mu\over  Q_\parallel^2}\Big(\delta_\nu^i-{q^\parallel_\nu \over Q_\parallel^2}i\partial^i\Big)
(f_1\partial_i\barf_1^f-\barf_1\partial_if_1^f)
+\mu\leftrightarrow\nu\Big]+f_1\leftrightarrow\barf_1\Big\}+O(\alpha_s)
\nonumber
\end{eqnarray}
where 
$f_1=f_1(\alpha_q,b_\perp)$, $\barf_1\equiv\barf_1(\beta_q,b_\perp)$ everywhere except 
$f\leftrightarrow\barf$ terms where it is opposite (the question about rapidity cutoffs for TMDs will be addressed in Sect. 
\ref{sec:match}). 

Similarly, we can write down $W^H$ contribution in coordinate space. For future use, however, it is convenient to define Fourier transform in a slightly different way. Introduce  $h_i(k_\perp)\equiv k_i h_1^\perp(k), ~\barh_i(k_\perp)\equiv k_i \barh_1^\perp(k)$ and 
\begin{eqnarray}
&&\hspace{-1mm}
\begin{array}{c}
h_i(\alpha,b_\perp)
\\
\barh_i(\alpha,b_\perp)
\end{array}\bigg\}~=~\int\!{d^2k_\perp\over 4\pi^2}~e^{i(k,b)_\perp}
\bigg\{\begin{array}{c}
h_i(\alpha,k_\perp)
\\
\barh_i(\alpha,k_\perp)
\end{array}
\label{defhi}
\end{eqnarray}
then $W^{1H}_{\mu\nu}$  can be represented as
\begin{eqnarray}
&&\hspace{-1mm}
m^2W_{\mu\nu}^{1H}(\alpha_q,\beta_q,b_\perp)~=~4\pi^2\sum_{\rm f} e_f^2\Big(g_{\mu\nu}^\perp h^f_j\barh^{fj}-h^f_\mu\barh^f_\nu-h_\nu\barh^f_\mu
+~2{q^\parallel_\mu q^\parallel_\nu -\tilq_\mu\tilq_\nu\over Q_\parallel^4}\partial^i h_i^f\partial^j\barh_j^f
\nonumber\\
&&\hspace{-1mm}
+~{q^\parallel_\mu\over Q_\parallel^2}\big[(i\partial^ih^f_i)\barh^f_\nu+h^f_\nu i\partial^i\barh^f_i\big]
+~{\tilq_\mu\over Q_\parallel^2}\big[(i\partial^ih^f_i)\barh^f_\nu-h^f_\nu i\partial^i\barh^f_i\big]~+~\mu\leftrightarrow\nu\Big)
-~{\tilq_\mu\tilq_\nu+q^\parallel_\mu q^\parallel_\nu \over Q_\parallel^4}
\nonumber\\
&&\hspace{-1mm}
\times~\big[\partial_\perp^2(h^f_i\barh_f^i)+\partial^i h^f_j\partial^i\barh^f_j\big]
-~{q^\parallel_\mu\tilq_\nu+ \tilq_\mu q^\parallel_\nu \over Q_\parallel^4}\partial_i[(h^f_j\partial\barh_f^j-\barh^f_j\partial h_f^j)]~+~h\leftrightarrow\barh
\label{w1hcoord}
\end{eqnarray}
where
$h_i=h_i(\alpha_q,k_\perp)$ and $\barh_i\equiv\barh_i(\beta_q,(q-k)_\perp)$ everywhere except 
$h\leftrightarrow\barh$ terms where it is opposite, cf. Eq. (\ref{w1fcoord}).

\subsection{Four Lorentz structures of hadronic tensor \label{sec:4DYs}}

The four Lorentz structures of hadronic tensor in Collins-Soper frame are given by Eq. (\ref{Ws}) where ($Q_\perp\equiv |q_\perp|$)
\begin{eqnarray}
&&\hspace{-11mm}
Z~=~{\tilq\over Q_\parallel}~=~{1\over Q_\parallel}(\alpha_qp_1-\beta_qp_2),~~~~~~~
X~=~\Big[{Q_\perp\over Q_\parallel Q}(\alpha_qp_1+\beta_qp_2)+{Q_\parallel\over Q_\perp Q}q_\perp\Big]
\end{eqnarray}
such that $q\cdot X=q\cdot Z=X\cdot Z=0$ and $X^2=Z^2=-1$.

First, let us check the structure corresponding to the total cross section of DY pair production. From Eq. (\ref{result}) we get
\begin{eqnarray}
&&\hspace{-1mm}
W^\mu_\mu(q)
~=~-{2\over N_c}\sum_fe_f^2\!\int\! d^2k_\perp\Big\{
\big[1-2{(k,q-k)_\perp\over Q^2}\big]
F^f(q,k_\perp)
\label{wmumu}\\
&&\hspace{-1mm}
+~2{k_\perp^2(q-k)_\perp^2\over m^2_NQ^2}H^f(q,k_\perp)
+{N_c\over N_c^2-1}(k,q-k)_\perp J_2^f(q,k_\perp)\Big\}
\Big[1+O\big({1\over N_c^2}\big)\Big]+O\big({q_\perp^4\over Q^4}\big)
\nonumber
\end{eqnarray}
which agrees with Eq. (6.2) from Ref. \cite{Balitsky:2017gis}. This equation gives the sum of structures  $W_\mu^\mu~=~-(2W_T+W_L)$.
 
\subsubsection{$W_L$}
The easiest structure to get is $W_L$.  Multiplying Eq. (\ref{result}) by $Z_\mu Z_\nu$ and comparing to Eq. (\ref{W}) we get
\begin{eqnarray}
&&\hspace{-1mm}
W_L(q)~=~Z^\mu Z^\nu W^1_{\mu\nu}(q)~
=~
\sum_fe_f^2{1\over Q^2N_c}\!\int\! d k_\perp\Big\{ (q-2k)_\perp^2F^f(q,k_\perp)
\label{wl}\\
&&\hspace{-1mm}
+~{1\over m^2}\big(2k_\perp^2(q-k)_\perp^2-[k_\perp^2+(q-k)_\perp^2](k,q-k)_\perp\big)
H^f(q,k_\perp)\Big\}
\Big[1+O\big({q_\perp^2\over Q^2}\big)+O\big({1\over N_c^2}\big)\Big]
\nonumber
\end{eqnarray}
Thus, one may say that  $W_L$ is known at LHC energies at $q_\perp^2\ll Q^2$ as far as $f_1$ and $h_1^\perp$ are known.

\subsubsection{$W_\Delta$}
Using formula
 $q_\perp^\mu Z^\nu W_{\mu\nu}~=~(X\cdot q)_\perp W_\Delta~=~-{Q_\parallel Q_\perp\over Q}W_\Delta$
 we get
\begin{eqnarray}
&&\hspace{-1mm}
W_\Delta~=~{Q\over Q_\parallel^2Q_\perp N_c}\sum_fe_f^2
\!\int\!d^2k_\perp\Big\{(q,q-2k)_\perp F^f(q,k_\perp)
\nonumber\\
&&\hspace{-1mm}
-~(q,q-2k){(k,q-k)_\perp\over m^2}H^f(q,k_\perp)\Big\}\Big[1+O\big({q_\perp^2\over Q^2}\big)+O\big({1\over N_c^2}\big)\Big]
\label{wd}
\end{eqnarray}
Again, we see that $W_\Delta$ is expressed via $f_1$ and $h_1^\perp$ with great accuracy.

\subsubsection{$W_T$}
Next, from Eqs. (\ref{wmumu}), (\ref{wl}) and $W_\mu^\mu~=~-(2W_T+W_L)$ one easily obtains
\begin{eqnarray}
&&\hspace{-1mm}
W_T(q)~=~{1\over N_c}\sum_fe_f^2\!\int\! d^2k_\perp\Big\{
\big[1-{q_\perp^2\over 2Q_\parallel^2}\big]F^f(q,k_\perp)
\nonumber\\
&&\hspace{-1mm}
+~{1\over 2m^2Q_\parallel^2}\big(2k_\perp^2(q-k)_\perp^2+[k_\perp^2+(q-k)_\perp^2](k,q-k)_\perp\big)H^f(q,k_\perp)
\nonumber\\
&&\hspace{-1mm}
+~{N_c\over N_c^2-1}(k,q-k)_\perp J_2^f(q,k_\perp)\Big\}
\Big[1+O\big({1\over N_c^2}\big)\Big]+O\big({q_\perp^4\over Q^4}\big)
\label{wt}
\end{eqnarray}
%

\subsubsection{$W_\dd$}
Finally, the easiest way to pick out $W_\dd$ is to multiply $W_{\mu\nu}$ by $q^\perp_\mu q^\perp_\nu/q_\perp^2$. One obtains from Eq. (\ref{W}) 
${q^\perp_\mu q^\perp_\nu\over q_\perp^2} W^{\mu\nu}(q)={Q_\parallel^2\over Q^2}(W_T-W_\dd)$. On the other hand, from Eqs. (\ref{WF}) and (\ref{WH}) 
one gets 
\begin{eqnarray}
&&\hspace{-1mm}
{q_\perp^\mu q_\perp^\nu\over q_\perp^2} W^1_{\mu\nu}(q)~
\nonumber\\
&&\hspace{-1mm}=~{1\over N_c}\sum_f e_f^2\!\int\!d^2k_\perp \Big\{F^f(q,k_\perp)
+~\Big[{(k,q-k)_\perp\over m^2}-{2(q,k)_\perp(q,q-k)_\perp\over m^2q_\perp^2}\Big] H^f(q,k_\perp)\Big\}
\nonumber\\
\label{vkladotw1}
\end{eqnarray}
and from Eq. (\ref{resultnoninv})
\begin{eqnarray}
&&\hspace{-1mm}
{q_\perp^\mu q_\perp^\nu\over q_\perp^2} W^2_{\mu\nu}(q)~
=~{1\over N_c}\sum_f e_f^2{1\over Q_\parallel^2q_\perp^2}\!\int d^2k_\perp\Big\{ {1\over m^2}
\big\{2(q,k)_\perp(q,q-k)_\perp(k,q-k)_\perp
\label{vkladotw2}\\
&&\hspace{-1mm}
-~k_\perp^2(q,q-k)_\perp^2
-(q-k_\perp)^2(q,k)_\perp^2-q_\perp^2(k,q-k)_\perp^2
+q_\perp^2 k_\perp^2(q-k_\perp)^2\big]\big\}H_A^f(q,k_\perp)
\nonumber\\
&&\hspace{-1mm}
+~{N_c\over N_c^2-1}\Big([2(q,k)_\perp(q,q-k)_\perp-q_\perp^2(k,q-k)_\perp]J_1^f(q,k_\perp)
+~q_\perp^2 (k,q-k)_\perp J_2^f(q,k_\perp)\Big)\Big\}.   
\nonumber
\end{eqnarray}
Thus, we get
\begin{eqnarray}
&&\hspace{-1mm}
W_\dd~=~W_T-{Q^2\over Q_\parallel^2}{q^\perp_\mu q_\perp^\nu\over q_\perp^2}W_{\mu\nu}~=~
{1\over N_c}\sum_f e_f^2\!\int\!d^2k_\perp \Big\{{q_\perp^2\over 2Q_\parallel^2}F^f(q,k_\perp)
\label{wdd}\\
&&\hspace{-1mm}
+~\Big({2(q,k)_\perp(q,q-k)_\perp\over q_\perp^2}-(k,q-k)_\perp
+{1\over Q^2}\big\{k_\perp^2(q-k)_\perp^2
\nonumber\\
&&\hspace{-1mm}
+~\half (k,q-k)_\perp[k_\perp^2+(q-k)_\perp^2]+q_\perp^2(k,q-k)_\perp-2(q,k)(q,q-k)_\perp\big\}
\Big){1\over m^2}H^f(q,k_\perp)
\nonumber\\
&&\hspace{-1mm}
+~{1\over m^2Q_\parallel^2}\Big({k_\perp^2\over q_\perp^2}(q,q-k)_\perp^2+{(q-k_\perp)^2\over q_\perp^2}(q,k)_\perp^2
- k_\perp^2(q-k_\perp)^2+(k,q-k)_\perp^2
\nonumber\\
&&\hspace{44mm}
-~2(q,k)_\perp(q,q-k)_\perp{(k,q-k)_\perp\over q_\perp^2}\Big)H_A^f(q,k_\perp)
\nonumber\\
&&\hspace{-1mm}
+~{N_c\over N_c^2-1}{1\over Q_\parallel^2}\Big[(k,q-k)_\perp-{2(q,k)_\perp(q,q-k)_\perp\over q_\perp^2}\Big]J_1^f(q,k_\perp)
~+~O\big({1\over N_c^2}\big)\Big\}+O\big({q_\perp^4\over Q^4}\big)
\nonumber
\end{eqnarray}
This is the only function which has a $O\big({1\over Q^2}\big)$, leading-$N_c$ contribution proportional to twist-three TMD $H_A$ not related to leading-twist TMDs
by equations of motion. The  functions $W_T,W_L$, and $W_\Delta$ do not have such contributions (although they have such contributions at the ${1\over N_c}$ level).

 \subsection{Estimates of $W_i(q)$ at $q_\perp^2\gg m^2$}
\subsubsection{Order-of-magnitude estimates \label{sec:omestimates}}
Following the analysis in Ref. \cite{Balitsky:2017gis}, let us estimate the relative strength of Lorentz structures $W_i$ at $q_\perp^2\gg m^2$.
First, we assume that ${1\over N_c}$ is a good parameter and leave only terms leading in $N_c$. 
Second,  at $q_\perp^2\gg m^2$ we probe the perturbative tails of TMD's $f_1\sim{1\over k_\perp^2}$ and 
$h_1^\perp\sim{1\over k_\perp^4}$ \cite{Zhou:2008fb}. 
So, as long as  $Q^2\gg q_\perp^2\gg m^2$ we can approximate
\begin{equation}
\hspace{-0mm}
f_1(\alpha_z,k_\perp^2)~\simeq~{f(\alpha_q)\over k_\perp^2},~~h_1^\perp(\alpha_q,k_\perp)~\simeq~{m^2_Nh(\alpha_q)\over k_\perp^4} ,
~~~~~~\barf_1\simeq{\barf(\alpha_q)\over k_\perp^2},~~\barh_1^\perp\simeq{m^2_N\barh(\alpha_q)\over k_\perp^4} 
\label{fhestimate1}
\end{equation}
(up to logarithmic corrections). Similarly, for the target we can use the estimate
\begin{equation}
f_1(\beta_z,k_\perp^2)~\simeq~{f(\beta_z)\over k_\perp^2},~~h_1^\perp(\beta_z,k_\perp^2)~\simeq~{m^2_N h(\beta_z)\over k_\perp^4} ,
~~~~~~\barf_1\simeq{\barf(\beta_z)\over k_\perp^2},~~\barh_1^\perp\simeq{m^2_N\barh(\beta_z)\over k_\perp^4} 
\label{fhestimate2}
\end{equation}
as long as $k_\perp^2\ll Q^2$.  Thus, we get an estimate
\begin{eqnarray}
&&\hspace{-1mm}
F^f(q,k_\perp)~\simeq~{F^f(\alpha_q,\beta_q)\over k_\perp^2(q-k)_\perp^2},
~~~~~F^f(\alpha_q,\beta_q)~\equiv~f^f(\alpha_q)\barf^f(\beta_q)+f^f\leftrightarrow\barf^f,
\nonumber\\
&&\hspace{-1mm}
H^f(q,k_\perp)~\simeq~m^4{H^f(\alpha_q,\beta_q)\over k_\perp^4(q-k)_\perp^4},
~~~~H^f(\alpha_q,\beta_q)~\equiv~h^f(\alpha_q)\barh^f(\beta_q)+h^f\leftrightarrow\barh^f
\label{fhestimate}
\end{eqnarray}

Note that due to the ``positivity constraint'' \cite{Bacchetta:1999kz}
\begin{equation}
h_1^\perp(x,k_\perp^2)\leq {m\over |k_\perp|}f_1^\perp(x,k_\perp^2)~~~
\label{poscons}
\end{equation}
we can safely assume that the functions $f(x)$ and $h(x)$ defined in 
Eqs. (\ref{fhestimate1}) and (\ref{fhestimate2}) are of the same order of magnitude.
Moreover,  both theoretical \cite{Scimemi_2018} and phenomenological \cite{Barone:2009hw,Barone:2010gk} analysis  indicate  that $h_1^\perp$ is several times smaller than $f_1$
 so in numerical estimates we will disregard the contribution of $h_1^\perp$.

\subsubsection{Power corrections for total DY cross section \label{sec:DYcross}}
Substituting the above approximations  to Eq. (\ref{wmumu}) we get
the following estimate  of the strength of power corrections for total DY cross section \cite{Balitsky:2017gis}
\begin{eqnarray}
&&\hspace{-1mm}
W^\mu_\mu(q)
~=~-{2\over N_c}\sum e_f^2\!\int\! d^2k_\perp\Big\{
\Big[1-2{(k,q-k)_\perp\over Q^2}\Big]F^f(q,k_\perp)
+~2{k_\perp^2(q-k)_\perp^2\over m^2_NQ^2}H^f(q,k_\perp)  \Big\}
\nonumber\\
&&\hspace{-1mm}
\simeq~-{2\over N_c}\sum e_f^2\!\int\! d^2k_\perp
\Big\{\Big[1-2{(k,q-k)_\perp\over Q^2}\Big]
{F^f(\alpha_q,\beta_q)\over k_\perp^2(q-k)_\perp^2}
+{2m^2\over Q^2}{H^f(\alpha_q,\beta_q)\over k_\perp^2(q-k)_\perp^2}\Big\}
\nonumber\\
&&\hspace{-1mm}
\simeq~-{2\over N_c}\sum e_f^2\!\int\! d^2k_\perp
\Big[1-2{(k,q-k)_\perp\over Q^2}\Big]
{F^f(\alpha_q,\beta_q)\over k_\perp^2(q-k)_\perp^2}
\label{dyx1}
\end{eqnarray}
where we used estimates (\ref{fhestimate}) and the fact that $(k,q-k)_\perp\sim q_\perp^2\gg m^2$. 
Thus, the relative weight of the leading term and power correction is determined by the factor
$1-2{(k,q-k)_\perp\over Q^2}$. Due to Eqs. (\ref{fhestimate1}) and (\ref{fhestimate2}), the integrals over $k_\perp$ are logarithmic and should be cut from below
by $m^2_N$ and from above by $Q^2$ so we get an estimate
\begin{eqnarray}
&&\hspace{-0mm}
\!\int\! d^2k_\perp{1\over k_\perp^2(q-k)_\perp^2}~\simeq~{2\pi\over q_\perp^2}\ln{q_\perp^2\over m^2},~~~~~~
\!\int\! d^2k_\perp{(k,q-k)_\perp\over k_\perp^2(q-k)_\perp^2}~\simeq~-\pi\ln{Q^2\over q_\perp^2}
\label{loginteg1}
\end{eqnarray}
where we assumed that the first integral is determined by the logarithmical region $q_\perp^2\gg k_\perp^2\gg m^2_N$ and the second by $Q^2\gg k_\perp^2\gg q_\perp^2$.
Taking these integrals to Eq. (\ref{dyx1})  one obtains
\begin{equation}
\hspace{-0mm}
W_\mu^\mu(q)
~=~-{4\pi \over N_c}\sum e_f^2\Big[{1\over q_\perp^2}\ln{q_\perp^2\over m^2_N}+{1\over Q^2}\ln{Q^2\over q_\perp^2}\Big]
F^f(\alpha_q,\beta_q)
\label{wmumuestimate}
\end{equation}
By this estimate, the power correction reaches the level of few percent at $q_\perp\sim {Q\over 4}$.

\subsubsection{Power corrections for $W_T$ \label{sec:wtpc}}
Let us now consider estimates described in  Sect. \ref{sec:omestimates}  for $W_T$ given by Eq. (\ref{wt}).  
At large $N_c$, we can omit the third line so
\begin{eqnarray}
&&\hspace{-0mm}
W_T(q)~=~{1\over N_c}\sum_fe_f^2\!\int\! \dhd^2k_\perp\Big[1-{q_\perp^2\over 2Q^2}\Big]F^f(q,k_\perp)
\\
&&\hspace{33mm}
+~{1\over 2m^2Q^2}\big(2k_\perp^2(q-k)_\perp^2+[k_\perp^2+(q-k)_\perp^2](k,q-k)_\perp\big)H^f(q,k_\perp)
\nonumber\\
&&\hspace{-1mm}
\simeq~{1\over N_c}\sum_fe_f^2\!\int\! \dhd^2k_\perp\Big\{\Big[1-{q_\perp^2\over 2Q^2}\Big]
{F^f(\alpha_q,\beta_q)\over k_\perp^2(q-k)_\perp^2}
\nonumber\\
&&\hspace{33mm}
+~{m^2\over Q^2}\Big(1+[k_\perp^2+(q-k)_\perp^2]{(k,q-k)_\perp\over 2k_\perp^2(q-k)_\perp^2}\Big)
{H^f(\alpha_q,\beta_q)\over k_\perp^2(q-k)_\perp^2}\Big\}
\label{wtestimat}
\end{eqnarray}
Again, due to $q_\perp^2\gg m^2$ the second term in braces can be neglected and we get 
\begin{eqnarray}
&&\hspace{-1mm}
W_T(q)~\simeq~{2\pi \over N_c}\Big[{1\over q_\perp^2}-{1\over 2Q^2}\Big]\ln{q_\perp^2\over m^2}\sum e_f^2F^f(\alpha_q,\beta_q)
\label{wtestimate}
\end{eqnarray}
Thus, for $W_T$ the power correction reaches 10\% level at $q_\perp\sim {Q\over 2}$.

\subsubsection{Estimate of $W_L$ \label{sec:wlest}}

Again, using estimates from Sect. \ref{sec:omestimates} one obtains
\begin{eqnarray}
&&\hspace{-1mm}
W_L(q)~=~
\sum_fe_f^2{1\over Q^2N_c}\!\int\! {d k_\perp\over k_\perp^2(q-k)_\perp^2}
\Big\{ (q-2k)_\perp^2F^f(\alpha_q,\beta_q)
\nonumber\\
&&\hspace{-1mm}
+~m^2\Big(2-[k_\perp^2+(q-k)_\perp^2]{(k,q-k)_\perp\over k_\perp^2(q-k)_\perp^2}\Big)
H^f(q,k_\perp)\Big\}
\label{wlestimat}
\end{eqnarray}
which gives approximately
\begin{eqnarray}
&&\hspace{-1mm}
W_L(q)~\simeq~
{2\pi\over Q^2N_c}\Big[\ln{q_\perp^2\over m^2_N}+2\ln{Q^2\over q_\perp^2}\Big]\sum_fe_f^2F^f(\alpha_q,\beta_q)
\label{wlestimate}
\end{eqnarray}
in agreement with Eqs. (\ref{wmumuestimate}) and (\ref{wtestimate}). 
The estimate of the ratio of $W_L/W_T$ is
\begin{eqnarray}
&&\hspace{-1mm}
{W_L(q)\over W_T(q)}~\simeq~
{q_\perp^2\over Q^2}\Big[1+2{\ln Q^2/q_\perp^2\over \ln q_\perp^2/m^2}\Big]
\label{wltestimate}
\end{eqnarray}
%

\subsubsection{Magnitude of $W_\Delta$ \label{sec:wdest}}
It is easy to see that $W_\Delta$ vanishes if one uses the estimates (\ref{fhestimate1}) and (\ref{fhestimate2}). Indeed, with these formulas
$F(q,k_\perp)$ and $H(q,k_\perp)$ are symmetric under replacement $k_\perp\leftrightarrow (q-k)_\perp$ whereas $(q,q-2k)_\perp$ in 
the integrand in Eq. (\ref{wd}) is antisymmetric. Moreover, this vanishing of $W_\Delta$ will occur for any factorizable model of TMDs $f_1$ and $h_1^\perp$: 
if $f_1(\alpha,k_\perp)~=~f(\alpha)\phi(k_\perp)$ and $h_1^\perp(\alpha,k_\perp)~=~h(\alpha)\psi(k_\perp)$ the integral (\ref{wd})
vanishes. 
On the other hand, $W_\Delta$ is only $\sim {Q_\perp\over Q}W_T$ so without better knowledge of TMDs it is impossible 
to tell whether $W_\Delta$ is smaller or bigger than, say,  $W_L$. Also, if the parameter $\alpha_q{Q\over Q_\perp}$ is not negligible, to compare
 $W_\Delta$ and  $W_L$ one needs to take into account $O(\alpha_q)$ 
corrections to $W_\Delta$ defined by TMDs other than $f_1$ and $h_1^\perp$. 

\subsubsection{Estimate of $W_\dd$ \label{sec:wddest}}
Let us consider the relative weight of the terms in the r.h.s. of Eq. (\ref{wdd}). As we mentioned, we assume that ${1\over N_c}$ is a valid
small parameter so we can omit the last $J_1$ term. Also,  it is natural to assume that $H_A^f(q,k_\perp)$ is of the same order of magnitude as 
$H^f(q,k_\perp)$ and,  since the term with $H_A$ is a power correction, it is not unreasonable to neglect this term in the first approximation. 
Using now estimates (\ref{fhestimate})  and the integrals 
\begin{eqnarray}
&&\hspace{-1mm}
\int\! d^2k {k_i(q-k)_j\over k_\perp^2(q-k)_\perp^4}\theta(k_\perp^2-m^2)\theta((q-k)_\perp^2-m^2)
~\simeq~{\pi\over 2q_\perp^2}\Big(g^\perp_{ik}+2{q_iq_k\over q_\perp^2}\Big)\ln{q_\perp^2\over m^2}
\nonumber\\
&&\hspace{-1mm}
\int\! d^2k {k_i(q-k)_j\over k_\perp^4(q-k)_\perp^4}\theta(k_\perp^2-m^2)\theta((q-k)_\perp^2-m^2)
~\simeq~{\pi\over q_\perp^4}\Big(g^\perp_{ik}+4{q_iq_k\over q_\perp^2}\Big)\ln{q_\perp^2\over m^2}
\label{loginteg2}
\end{eqnarray}
one gets an estimate
\begin{eqnarray}
&&\hspace{-1mm}
W_\dd~\simeq~
{1\over N_c}\sum_f e_f^2\!\int\!d^2k_\perp \Big\{{q_\perp^2\over 2Q_\parallel^2}F^f(q,k_\perp)
\nonumber\\
&&\hspace{-1mm}
+~\Big({2(q,k)_\perp(q,q-k)_\perp\over q_\perp^2}-(k,q-k)_\perp
+{1\over Q^2}\big[k_\perp^2(q-k)_\perp^2
\nonumber\\
&&\hspace{-1mm}
+~\half (k,q-k)_\perp[k_\perp^2+(q-k)_\perp^2]+q_\perp^2(k,q-k)_\perp-2(q,k)(q,q-k)_\perp\big]
\Big){1\over m^2}H^f(q,k_\perp)
\nonumber\\
&&\hspace{-1mm}
\simeq~{\pi\over Q^2N_c}\ln{q_\perp^2\over m^2}\sum_f e_f^2\Big[F^f(\alpha_q,\beta_q)
+{4m^2Q^2\over q_\perp^4}\Big(1-{q_\perp^2\over 2Q^2}\Big)H^f(\alpha_q,\beta_q)\Big]
\label{wddest}
\end{eqnarray}
%

\subsubsection{Lam-Tung relation}
It is easy to see that if one neglects $H$ in Eq. (\ref{wl}) the ratio of $W_L$ and $2W_\dd$ is approximately
\begin{eqnarray}
&&\hspace{-1mm}
{W_L\over 2W_\dd}~\simeq~1+2{\ln Q^2/q_\perp^2\over\ln{q_\perp^2/m^2}}
\label{ltungestim}
\end{eqnarray}
It seems like the Lam-Tung relation works better if we move closer to the domain of collinear factorization  $Q^2\sim Q_\perp^2\gg m^2$.
\subsubsection{Estimates of asymmetries \label{estasy}}
The differential cross section of DY process is parametrized as
\begin{equation}
\hspace{1mm}
\Big({d\sigma\over d^4q}\Big)^{-1}{d\sigma\over d\Omega d^4q}~=~{3\over 4\pi(\lambda+3)}\big(1+\lambda\cos^2\theta+\mu\sin 2\theta\cos\phi+
{\nu\over 2}\sin^2\theta\cos 2\phi\big)
\end{equation}
where  $\Omega$ 
is the solid angle of the lepton in terms of its polar and azimuthal angles in the center-of-mass system of the lepton pair. 
The angular coefficients $\lambda$, $\mu$, and $\nu$ can be expressed in terms of the hadronic tensor:
\begin{equation}
\hspace{1mm}
\lambda~=~{W_T-W_L\over W_T+W_L},~~~\mu~=~{W_\Delta\over W_T+W_L},~~~\nu~=~{2W_\dd\over W_T+W_L}
\label{lamunu}
\end{equation}
For an estimate, let us take $s$=8 TeV and $Q$=90 GeV so that $x_A\sim x_B\sim 0.1$ in central region of rapidity. 
Although we did not include the contribution of $Z$-boson, we can compare our order-of-magnitude estimates with experimental data
at this kinematics \cite{Sirunyan:2018owv,Aad:2019wmn}.
 Let us take $Q_\perp$= 20 GeV so
the power corrections $\sim{Q_\perp^2\over Q^2}$ are small but sizable, of order of few per cent.  At this kinematics, we obtain
\begin{equation}
\hspace{1mm}
1-\lambda~=~2{W_L\over W_T+W_L}~\simeq~ 2{1+2{\ln Q^2/q_\perp^2\over \ln q_\perp^2/m^2}\over 
{Q^2\over q_\perp^2}-\half+2{\ln Q^2/q_\perp^2\over \ln q_\perp^2/m^2}} ~\simeq~0.19
\end{equation}
from Eq. (\ref{wlestimate}) which agrees with estimates in Ref. \cite{Lambertsen:2016wgj}. Next, in our kinematics the expression
in square brackets in the r.h.s. in Eq. (\ref{wddest})  is approximately $F+0.17 H$. Since  the Boer-Mulders function 
seem to be of order of few percent of $f_1$ (see the discussion in Sect. \ref{sec:omestimates}), the term with $H$ can be safely neglected and we get
\begin{equation}
\hspace{1mm}
\nu~=~{2W_\dd\over W_T+W_L}~\simeq~{1\over {Q^2\over q_\perp^2}-\half+2{\ln Q^2/q_\perp^2\over\ln{q_\perp^2/m^2}}}~\simeq~0.05
\end{equation}
As to $\mu$ coefficient, as we mentioned, we cannot estimate it since with factorization hypothesis
for TMDs it vanishes. Reversing the argument, if $\mu$ will be measured to be much smaller than $\nu$, it will be an argument in favor
of factorization hypothesis for TMDs $f_1$ and $h_1^\perp$. Actually, there are  experiments at much lower $q_\perp$ and $Q$ $\sim$ few GeV
which indicate that $\mu$ is very small \cite{Zhu:2008sj}.

Last but not least, let us estimate Lam-Tung relation. With our approximation in the above kinematics we get 
\begin{eqnarray}
&&\hspace{-1mm}
{W_L\over 2W_\dd}~\simeq~1+2{\ln Q^2/q_\perp^2\over\ln{q_\perp^2/m^2}}~\simeq~2.0
\label{ltungest}
\end{eqnarray}
so it seems to be violated at this kinematics. Again, these order-of-magnitude estimates do not include the contribution to DY cross section 
mediated by the $Z$-boson.

\section{Coefficient functions and matching of rapidity cutoffs\label{sec:match}}

The result (\ref{resultginv}) is a tree-level formula and to fully understand Eq. (\ref{TMDf}) we
should specify the rapidity cutoffs for $f_1$'s and $h_1^\perp$'s. As we discussed in section \ref{sec:funt},
the rapidity cutoff for longitudinal momenta in $f_1(\alpha_q,k_\perp)$ is $\beta\leq \sigma_p$ and for $f_1(\beta_q,k_\perp)$ $\alpha\leq\sigma_t$, 
where $\sigma_p$ and $\sigma_t$ are rapidity bounds for central fields. To avoid double counting, the region where both 
$\alpha<\sigma_t$ and $\beta<\sigma_p$ should give only small power corrections. This is achieved if one takes 
$\sigma_p,\sigma_t\sim{Q_\perp\over\sqrt{s}}$ so power corrections from double counting are ${Q_\perp\over Q}$. 
In this case, the region $\alpha_q>\alpha>\sigma_t,~\beta_q>\beta>\sigma_p$ gives Sudakov double-log 
factor
\begin{eqnarray}
&&\hspace{-1mm}
C\Big(q,k,{\alpha_q\over\sigma_t},{\beta_q\over\sigma_p}\Big)
~\sim ~e^{-{\alpha_sc_F\over\pi}\ln{\alpha_q\over\sigma_t}\ln{\beta_q\over\sigma_p}}
\label{doublog}
\end{eqnarray}
where the coefficient ${\alpha_sc_F\over\pi}$ is two times $\gamma_{\rm cusp}$ for quarks.
A more precise formula can be obtained from the requirement
 that the product of two TMDs and the coefficient function (\ref{doublog}) does not depend on the ``rapidity divides'' 
$\sigma_p$ and $\sigma_t$. For simplicity, let us start with the leading-twist term $\sim g_{\mu\nu}^\perp F$. 
Rapidity evolution of the function $f_1(\alpha_q,b_\perp;\vigma_p)$ was found in Ref. \cite{Balitsky:2019ayf}
\footnote{As noted in Ref. \cite{Balitsky:2019ayf}, the factor $\sim \gamma_E$ depends on the exact way to cut integrals over $\alpha$
and $\beta$. Here the factor $-\gamma_E$ corresponds to ``smooth'' cutoffs $e^{-{\alpha\over\sigma_t}}$ and $e^{-{\beta\over\sigma_p}}$,
see the discussion in Ref. \cite{Balitsky:2019ayf}
}
\begin{eqnarray}
&&\hspace{-1mm}
{d\over d\ln\vigma_p}f_1(\alpha_q,b_\perp;\vigma_p)
~=~{\alpha_sC_F\over \pi}\big[-\ln\alpha_q\vigma_p-\half\ln {b_\perp^2s\over 4} -\gamma_E
+O(\alpha_s)\big]f_1(\alpha_q,b_\perp;\vigma_p)
\nonumber\\
&&\hspace{-1mm}
{d\over d\ln\vigma_t}f_1(\beta_q,b_\perp;\vigma_t)
~=~{\alpha_sC_F\over \pi}\big[-\ln\beta_q\vigma_t-\half\ln {b_\perp^2s\over 4} -\gamma_E+O(\alpha_s)\big]f_1(\beta_q,b_\perp;\vigma_t)
\label{evoleqs}
\end{eqnarray}
where $\vigma_p=\sigma_pb_\perp\sqrt{s}$, $\vigma_t=\sigma_tb_\perp\sqrt{s}$ are $b_\perp$-dependent cutoffs providing conformal invariance of the
leading-order TMD rapidity evolution (in the coordinate space) and $\gamma_E$ is Euler's constant. 
Similar equation holds true for $\barf_1$ since it is obtained from the evolution of the same operator.  

Looking at  Eqs. (\ref{doublog}) and (\ref{evoleqs}) one can guess that
the coefficient function $\sim g_\perp^{\mu\nu}$ times two TMDs $f_1$ in the coordinate space has the form
\begin{eqnarray}
&&\hspace{-1mm}
W_{g_\perp}(\alpha_q,\beta_q,b_\perp)~\sim~M(\alpha_q,\beta_q,b_\perp;\vigma_p,\vigma_t)
\big[f_1(\alpha_q,b_\perp;\vigma_p)\barf_1(\beta_q,b_\perp;\vigma_t) 
+f_1\leftrightarrow \barf_1\big]
\label{match}
\end{eqnarray}
where
\begin{equation}
\hspace{-0mm}
M(\alpha_q,\beta_q,b_\perp;\vigma_p,\vigma_t)~=~e^{-{\alpha_sc_F\over\pi}
\ln\big({\alpha_qb_\perp\sqrt{s}\over \vigma_t}e^{\gamma_E}\big)
\ln\big({\beta_qb_\perp\sqrt{s}\over \vigma_p}e^{\gamma_E}\big)
+{\alpha_sc_F\over 2\pi}\ln^2\vigma_p\vigma_t}\big[
1+O(\alpha_s)\big]
\label{M}
\end{equation}
It is easy to check that with $M$ given by Eq. (\ref{M})  we have ${d\over d\vigma_p}$(r.h.s. of Eq. (\ref{match}))~=~0 
and ${d\over d\vigma_t}$(r.h.s. of Eq. (\ref{match}))~=~0 so our guess (\ref{M}) for the coefficient function
is correct up to $O(\alpha_s)$ terms.  
\footnote{The Eq. (\ref{M}) is obtained in the leading order in $\alpha_s$ so
the argument of coupling constant is left undetermined.
 One should expect Sudakov formula with running coupling constant \cite{Parisi:1979se,Collins:1984kg} at the NLO level.
}

To write precise matching for other parts of $W_{\mu\nu}$ is a more complicated task. Let us start with  $W^{1F}_{\mu\nu}$ terms considered in 
the next Section.

\subsection{Matching for $W^{1F}$ terms}
We need to multiply Eq. (\ref{w1fcoord})  in coordinate space  by $M(\alpha_q,\beta_q,b_\perp;\vigma_p,\vigma_t)$. 
First, recall that 
\begin{equation}
\hspace{0mm} 
M(\alpha_q,\beta_q,b_\perp;\vigma_p,\vigma_t)
[f_1(\alpha_q,b_\perp;\vigma_p)\barf_1(\beta_q,b_\perp;\vigma_t) +f_1\leftrightarrow \barf_1\big]
\end{equation}
does not actually depend on the ``rapidity divides'' $\vigma_p$ and $\vigma_t$. However, the differentiation 
${\partial\over\partial b_i}$ affects evolution equations (\ref{evoleqs}). 
In this case we  modify the derivative with respect to $b_i$ as follows
\begin{eqnarray}
&&\hspace{-1mm}
\tartial_i f_1(\alpha_q,b_\perp;\vigma_p)~\equiv~\big(\partial_i-{\alpha_sc_F\over \pi}{b^i\over b_\perp^2}\ln\vigma_p\big)f_1(\alpha_q,b_\perp;\vigma_p),
\nonumber\\
&&\hspace{-1mm}
\tartial_i f_1(\beta_q,b_\perp;\vigma_t)~\equiv~\big(\partial_i-{\alpha_sc_F\over \pi}{b^i\over b_\perp^2}\ln\vigma_t\big)f_1(\beta_q,b_\perp;\vigma_t)
\label{tartial}
\end{eqnarray}
(and similarly for $\barf$'s) 
so that the l.h.s.' of these equations satisfy Eqs. (\ref{evoleqs}). 
\footnote{Strictly speaking, the difference between $\tartial_i f$ and $\partial_i f$ is 
$\sim O(\alpha_s)$  but since our matching is correct  at single-log limit (see Eq. (\ref{M})) we keep $\tartial_i f$
to avoid corrections $\sim O(\alpha_s\ln\vigma)$ in our matching formulas (\ref{w1fcoord}) (and (\ref{w1hcoord}) below).
}
Note also that $\partial_i(Mf\barf)=M(f\tartial_i\barf+\barf\tartial_i f)$.
With this definitions, one can write $W^{1F}_{\mu\nu}$ in the double-log approximation in the form
\begin{eqnarray}
&&\hspace{-1mm}
W^{1F}(\alpha_q,\beta_q,b_\perp)~=~\sum_{\rm flavors} e_f^2 W^{fF}(\alpha_q,\beta_q,b_\perp)~
\label{w1fcoordm}\\
&&\hspace{-1mm}
W^{fF}(\alpha_q,\beta_q,b_\perp)~=
\nonumber\\
&&\hspace{5mm}
=~ 4\pi^2
\Big\{-g_{\mu\nu}^\perp +{i\over Q_\parallel^2}(q^\parallel_\mu\partial^\perp_\nu+q^\parallel_\nu\partial^\perp_\mu)
-{q^\parallel_\mu q^\parallel_\nu+\tilq_\mu\tilq_\nu\over Q_\parallel^2}\partial_\perp^2\Big]Mf_1\barf_1
-~{4\tilq_\mu\tilq_\nu\over Q_\parallel^2}M(\tartial^if_1^f)(\tartial^i \barf_1^f) 
\nonumber\\
&&\hspace{5mm}
-~\Big[{\tilq_\mu\over  Q_\parallel^2}\Big(\delta_\nu^i-{q^\parallel_\nu \over Q_\parallel^2}i\partial^i\Big)
M(f_1\tartial_i\barf_1^f-\barf_1\tartial_if_1^f)
+\mu\leftrightarrow\nu\Big]+f_1\leftrightarrow\barf_1\Big\}+O(\alpha_s)
\nonumber
\end{eqnarray}
where $M=M(\alpha_q,\beta_q, b_\perp;\vigma_p,\vigma_t)$ and 
$f_1=f_1(\alpha_q,b_\perp,\vigma_p)$, $\barf_1\equiv\barf_1(\beta_q,b_\perp,\vigma_t)$ everywhere except 
$f\leftrightarrow\barf$ terms where it is opposite. 

It is easy to check gauge invariance:  $(\alpha_q p_1^\mu+\beta_q p_2^\mu+i\partial_\perp^\mu)W_{\mu\nu}^{1F}(\alpha_q,\beta_q,b_\perp)~=~0$.

\subsection{Matching for $W^{1H}$ terms}
\begin{figure}[htb]
\begin{center}
\includegraphics[width=111mm]{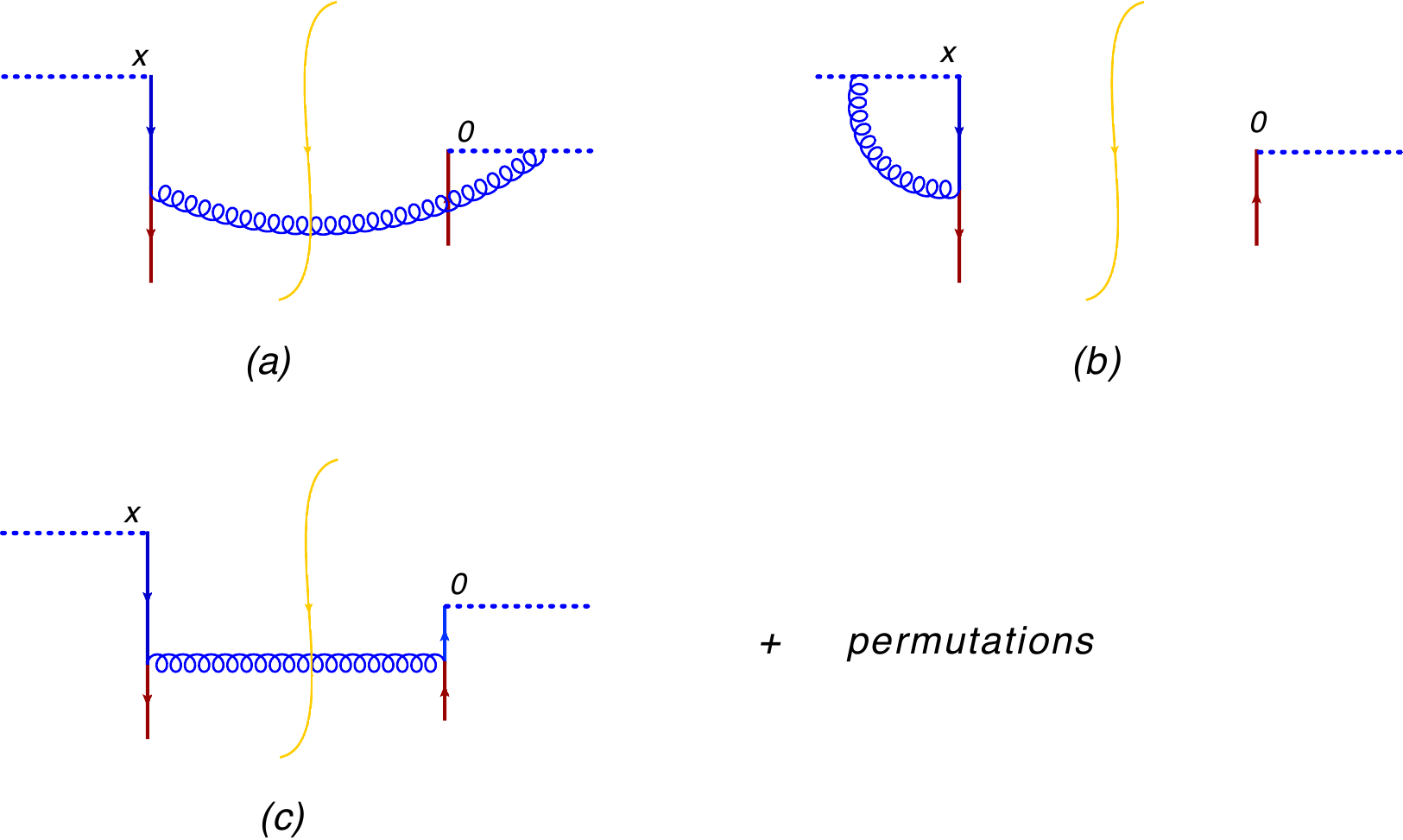}
\end{center}
\caption{Typical diagrams for the rapidity evolution of quark TMD in the Sudakov regime.   
\label{fig:tmdevolq}}
\end{figure}

First, with our definitions (\ref{defhi})  the  Eq. (\ref{hmael}) reads
\begin{eqnarray}
&&\hspace{-1mm}
{1\over 8\pi^3s}\!\int\!dx_\bu d^2x_\perp~e^{-i\alpha x_\bu+i(k,x)_\perp}
~\langle A|\bsi_f(x_\bu,x_\perp)\sigma_{i \bu}\psi_f(0)|A\rangle
=~{1\over m_N} h_i^f(\alpha,k_\perp)
\nonumber\\
&&\hspace{-1mm}
{1\over 8\pi^3s}\!\int\!dx_\bu d^2x_\perp~e^{-i\alpha x_\bu+i(k,x)_\perp}
~\langle A|\bsi_f(0)\sigma_{i \bu}\psi_f(x_\bu,x_\perp)|A\rangle
=~-{1\over m_N}\barh_i^f(\alpha,k_\perp)
\label{hmaelnew}
\end{eqnarray}
and similarly for the target matrix elements. With such definition, the evolution equation for 
$h_i(\alpha,b_\perp)\equiv {1\over 4\pi^2}\int\! d^2k_\perp e^{i(k,x)_\perp}h_i(\alpha,k_\perp)$ 
is the same as Eq. (\ref{evoleqs})
\begin{eqnarray}
&&\hspace{-1mm}
{d\over d\ln\vigma_p}h_i(\alpha_q,b_\perp;\vigma_p)
~=~{\alpha_sC_F\over \pi}\big[-\ln\alpha_q\vigma_p-\ln {|b_\perp|\sqrt{s}\over 2} -\gamma_E\big]h_i(\alpha_q,b_\perp;\vigma_p)
\nonumber\\
&&\hspace{-1mm}
{d\over d\ln\vigma_t}h_i(\beta_q,b_\perp;\vigma_t)
~=~{\alpha_sC_F\over \pi}\big[-\ln\beta_q\vigma_t-\half\ln {b_\perp^2s\over 4} -\gamma_E\big]h_i(\beta_q,b_\perp;\vigma_t)
\label{hevoleqs}
\end{eqnarray}
and similarly for $\barh_i$. The reason is that
 one-loop rapidity evolution for $\hbsi(x_\bu,x_\perp)\Gamma\hsi(0)$ in the 
 Sudakov region is the same for all matrices $\Gamma$  between $\hbsi(x_\bu,x_\perp)$ 
and $\hpsi_(0)$ due to the fact that the ``handbag''  diagram in Fig. \ref{fig:tmdevolq}c is small 
 and in two other diagrams (as well as self-energy corrections) the  matrix $\Gamma$  between $\hbsi(x_\bu,x_\perp)$ 
and $\hpsi(0)$ just multiplies the result of calculation. 

Next, to write down  the product of $m^2W_{\mu\nu}^H$ and the coefficient function
we need modified derivatives of $h_i$'s of Eq. (\ref{tartial}) type:
\begin{eqnarray}
&&\hspace{-1mm}
\tartial_i h_j(\alpha_q,b_\perp;\vigma_p)~\equiv~\big(\partial_i-{\alpha_sc_F\over \pi}{b^i\over b_\perp^2}\ln\vigma_p\big)h_j(\alpha_q,b_\perp;\vigma_p),
\nonumber\\
&&\hspace{-1mm}
\tartial_i h_j(\beta_q,b_\perp;\vigma_t)~\equiv~\big(\partial_i-{\alpha_sc_F\over \pi}{b^i\over b_\perp^2}\ln\vigma_t\big)h_j(\beta_q,b_\perp;\vigma_t)
\label{htartial}
\end{eqnarray}
(and similarly for $\barh$'s) so that  $\tartial_i h_j$ will satisfy same evolution equations (\ref{hevoleqs})  as $h_j$. We get then
\begin{eqnarray}
&&\hspace{-1mm}
W_{\mu\nu}^{1H}(\alpha_q,\beta_q,b_\perp)~=~\sum_{\rm flavors} e_f^2W_{\mu\nu}^{fH}(\alpha_q,\beta_q,b_\perp),
\nonumber\\
&&\hspace{-1mm}
{m^2\over 4\pi^2}W_{\mu\nu}^{fH}(\alpha_q,\beta_q,b_\perp)~=~M\Big(g_{\mu\nu}^\perp h^f_j\barh^{fj}-h^f_\mu\barh^f_\nu-h_\nu\barh^f_\mu
+~2{q^\parallel_\mu q^\parallel_\nu -\tilq_\mu\tilq_\nu\over Q_\parallel^4}\tartial^i h_i^f\tartial^j\barh_j^f
\nonumber\\
&&\hspace{-1mm}
+~{q^\parallel_\mu\over Q_\parallel^2}\big[(i\tartial^ih^f_i)\barh^f_\nu+h^f_\nu i\tartial^i\barh^f_i\big]
+~{\tilq_\mu\over Q_\parallel^2}\big[(i\tartial^ih^f_i)\barh^f_\nu-h^f_\nu i\tartial^i\barh^f_i\big]~+~\mu\leftrightarrow\nu\Big)
-~{\tilq_\mu\tilq_\nu+q^\parallel_\mu q^\parallel_\nu \over Q_\parallel^4}
\nonumber\\
&&\hspace{-1mm}
\times~\big[\partial_\perp^2(Mh^f_i\barh_f^i)+M\tartial^i h^f_j\tartial^i\barh^f_j\big]
-~{q^\parallel_\mu\tilq_\nu+ \tilq_\mu q^\parallel_\nu \over Q_\parallel^4}\partial_i[M(h^f_j\tartial\barh_f^j-\barh^f_j\tartial h_f^j)]~+~h\leftrightarrow\barh
\label{w1hcoordm}
\end{eqnarray}
where $M=M(\alpha_q,\beta_q;\vigma_p,\vigma_t)$ and 
$h_i=h_i(\alpha_q,k_\perp,\vigma_p)$, $\barh_i\equiv\barh_i(\beta_q,(q-k)_\perp,\vigma_t)$ everywhere except 
$h\leftrightarrow\barh$ terms where it is opposite, cf. Eq. (\ref{w1fcoord}).

Let us comment on  the choice of ``rapidity divides'' $\vigma_p$ and $\vigma_t$ in the product \\
$M(\alpha_q,\beta_q,b_\perp;\vigma_p,\vigma_t)f_1(\alpha_q,b_\perp;\vigma_p)\barf_1(\beta_q,b_\perp;\vigma_t)$ (and in similar $Mh\barh$ product). 
As we mentioned in the beginning of this Section, in order to avoid double counting one should write down  factorization of the amplitude in projectile, target and central fields at $\vigma_p,\vigma_t\sim 1$. After that, as discussed in Ref. \cite{Balitsky:2019ayf}, 
one can use the double-log Sudakov evolution (\ref{evoleqs}) until 
\begin{equation}
\vigma_p\geq \chvigma_p,~~~~ \chvigma_p\equiv{1\over\alpha_qb_\perp\sqrt{s}},~~~~~~~~~\vigma_t\geq \chvigma_t,~~~~\chvigma_t\equiv{1\over\beta_qb_\perp\sqrt{s}}
\label{vigmas}
\end{equation}
At this point, the result of Sudakov evolution is
\begin{eqnarray}
&&\hspace{-1mm}
M(\alpha_q,\beta_q,b_\perp;\chvigma_p,\chvigma_t)
\big[f_1(\alpha_q,b_\perp;\chvigma_p)\barf_1(\beta_q,b_\perp;\chvigma_t)+f_1\leftrightarrow \barf_1\big]
\nonumber\\
&&\hspace{-1mm}
=~e^{-{\alpha_s c_F\over 2\pi}\ln^2{Q^2b_\perp^2}}\big[f_1(\alpha_q,b_\perp;\chvigma_p)\barf_1(\beta_q,b_\perp;\chvigma_t)+f_1\leftrightarrow \barf_1\big]
\end{eqnarray}
so in the final result (\ref{w1fcoord}) one should take
\begin{equation}
Mf\barf ~\rightarrow~e^{-{\alpha_s N_c\over 2\pi}\ln^2{Q^2b_\perp^2}}\Big[1+O\big(\alpha_s\big)\Big]
f_1(\alpha_q,b_\perp;\chvigma_p)\barf_1(\beta_q,b_\perp;\chvigma_t)
\label{fsud}
\end{equation}
Similarly, for $W_{\mu\nu}^{1H}(\alpha_q,\beta_q,b_\perp)$ one should take
\begin{equation}
Mh_i\barh_j ~\rightarrow~e^{-{\alpha_s N_c\over 2\pi}\ln^2{Q^2b_\perp^2}}\Big[1+O\big(\alpha_s\big)\Big]
h_i(\alpha_q,b_\perp;\chvigma_p)\barh_j(\beta_q,b_\perp;\chvigma_t)
\label{hsud}
\end{equation}
at the end of Sudakov evolution (\ref{hevoleqs}). It should be emphasized that since factor $M$ is universal for (\ref{fsud}) and (\ref{hsud}),  
our estimates of asymmetries in Sect. \ref{estasy} are not affected by summation of Sudakov double logs. 

\subsubsection{Rapidity-only cutoff for TMD}
As discussed in Refs. \cite{Balitsky:2017flc,Balitsky:2017gis}, from the rapidity factorization (\ref{W5}) we get 
TMDs with rapidity-only cutoff $|\alpha|<\sigma_t$ or $|\beta|<\sigma_p$ (or with modifications (\ref{vigmas})). Such cutoff, relevant for small-$x$ physics, is 
different from the combination of UV and rapidity cutoffs for TMDs used by moderate-$x$ community, see the analysis in two  \cite{Echevarria:2015byo,Li:2016axz,Luebbert:2016itl}   
and three \cite{Echevarria:2016scs} loops. 
For the tree-level formulas of Sect. \ref{sec:results}, this difference in cutoffs does not matter, but if one uses the formulas from Sect. \ref{sec:match} 
and integrates models for TMDs with Sudakov factor $M$ of Eq. (\ref{M}), one has to relate TMDs with rapidity-only cutoffs
to the TMD models with conventional cutoffs. 
This requires calculations at the NLO level which are in progress.

\section{Conclusions and outlook \label{sec:coutlook}}
Main result of this paper is Eq. (\ref{result}) which gives the DY hadronic tensor for electromagnetic current at small $x$ 
with gauge invariance at the ${1\over Q^2}$ level. 
The part (\ref{resultginv}), determined by leading-twist TMDs $f_1$ and $h_1^\perp$, is manifestly gauge invariant. 
The only non-gauge invariant term at the ${1\over Q^2}$ level is Eq. (\ref{W2}) with  transverse $\mu$ and $\nu$ 
which is $\sim{q^\perp_\mu q^\perp_\nu\over Q^2}$ times twist-3 TMDs. Also, in the leading-$N_c$  approximation
the only structure affected by those terms is $W_\dd$, all other structures are calculated up to $O\big({q_\perp^4\over Q^4}\big)$ terms.
It is interesting to note that ${1\over Q^2}$ terms necessary for gauge invariance are calculated more than than two decades after
the calculation of ${1\over Q}$ corrections in Ref. \cite{Mulders:1995dh}.

It should be emphasized that,  as discussed above, our rapidity factorization is different 
from the standard factorization scheme for particle production in hadron-hadron 
scattering, namely splitting  the diagrams in collinear to projectile part, collinear to 
target part, hard factor, and soft factor \cite{Collins:2011zzd}. Here we factorize only in rapidity
and the $Q^2$ evolution arises from $k_\perp^2$ dependence of the rapidity evolution kernels, same
as in the BK (and NLO BK \cite{Balitsky:2008zza}) equations. Also, since matrix elements of TMD operators
with our rapidity cutoffs are UV-finite \cite{Balitsky:2015qba,Balitsky:2016dgz}, the only UV divergencies 
in our approach are usual
UV divergencies absorbed in the QCD running coupling. For the tree-level result (\ref{result}) this does not matter,
but if one intends to use the result like (\ref{w1fcoord}) with Sudakov logarithms for conventional TMDs with double
UV and rapidity cutoffs, one needs to relate our TMDs with rapidity-only cutoff to conventional TMDs. 
Needless to say,
the gauge-invariant tree-level result (\ref{resultginv}) should be correct for TMDs with any cutoffs.

An obvious outlook is to extend these results to the ``real'' DY process involving $Z$-boson contributions which are  relevant for
our kinematics.
Another outlook is the one-loop calculations in this rapidity-based factorization and comparison to resummations 
of large $\ln x$ and $\ln{Q^2/Q_\perp^2}$ based on usual 
collinear factorization, see e.g. Refs. \cite{Forte:2015gve,Marzani:2015oyb}.  The study is in progress.

The author is grateful to  V. Braun,  A. Prokudin,  A. Radyushkin, J. Qiu, and A. Vladimirov for valuable discussions. This  work is
 supported by Jefferson Science Associates, LLC under the U.S. DOE contract \#DE-AC05-06OR23177
 and by U.S. DOE grant \#DE-FG02-97ER41028.

\section{Appendix}
\subsection{Formulas with Dirac matrices \label{diracs}}
\subsubsection{Fierz transformation}
First, let us write down Fierz transformation for symmetric hadronic tensor
\begin{eqnarray}
&&\hspace{-1mm}
\half[(\bsi\gamma_\mu\chi)(\bhi\gamma_\nu\psi)+\mu\leftrightarrow\nu]
\label{fierz}\\
&&\hspace{-1mm}
=~
-{1\over 4}\big(\delta_\mu^\alpha\delta_\nu^\beta+\delta_\nu^\alpha\delta_\mu^\beta-g_{\mu\nu}g^{\alpha\beta}\big)
\big[(\bsi\gamma_\alpha\psi)(\bhi\gamma_\beta\chi)
+(\bsi\gamma_\alpha\gamma_5\psi)(\bhi\gamma_\beta\gamma_5\chi)\big]
\nonumber\\
&&\hspace{-1mm}
+~{1\over 4}\big(\delta_\mu^\alpha\delta_\nu^\beta+\delta_\nu^\alpha\delta_\mu^\beta-\half g_{\mu\nu}g^{\alpha\beta}\big)
(\bsi\sigma_{\alpha\xi}\psi)(\bhi\sigma_\beta^{~\xi}\chi)-{g_{\mu\nu}\over 4}(\bsi\psi)(\bhi\chi)+{g_{\mu\nu}\over 4}(\bsi\gamma_5\psi)(\bhi\gamma_5\chi)
\nonumber
\end{eqnarray}
%

\subsubsection{Formulas with $\sigma$-matrices \label{sigmaflas}}
It is convenient to define
\footnote{We use conventions from {\it Bjorken \& Drell} where $\epsilon^{0123}=-1$ and
$
\gamma^\mu\gamma^\nu\gamma^\lambda=g^{\mu\nu}\gamma^\lambda +g^{\nu\lambda}\gamma^\mu-g^{\mu\lambda}\gamma^\nu
-i\epsilon^{\mu\nu\lambda\rho}\gamma_\rho\gamma_5
$.
Also, with this convention $\tigma_{\mu\nu}\equiv\half \epsilon_{\mu\nu\lambda\rho}\sigma^{\lambda\rho}=i\sigma_{\mu\nu}\gamma_5$.
}
\begin{equation}
\epsilon_{ij}~\equiv~ {2\over s}\epsilon_{\ast\bu ij}~=~ {2\over s}p_2^\mu p_1^\nu\epsilon_{\mu\nu ij}
\label{eps2}
\end{equation}
such that $\epsilon_{12}~=~1$ and $\epsilon_{ij}\epsilon_{kl}~=~g_{ik}g_{jl}-g_{il}g_{jk}$. 
The frequently used formula is 
\begin{equation}
\hspace{-1mm}
\sigma_{\mu\nu}\sigma_{\alpha\beta}~=~(g_{\mu\alpha}g_{\nu\beta}-g_{\mu\beta}g_{\nu\alpha})-i\epsilon_{\mu\nu\alpha\beta}\gamma_5
-i(g_{\mu\alpha}\sigma_{\nu\beta}-g_{\mu\beta}\sigma_{\nu\alpha}-g_{\nu\alpha}\sigma_{\mu\beta}+g_{\nu\beta}\sigma_{\mu\alpha})
\label{sigmasigma}
\end{equation}
with variations
\begin{eqnarray}
&&\hspace{-1mm}
{2\over s}\sigma_{\bu i}\sigma_{\ast j}~=~g_{ij}-i\epsilon_{ij}\gamma_5 
-i\sigma_{ij}-{2i\over s}g_{ij}\sigma_{\bu\ast},~~~
{2\over s}\sigma_{\ast i}\sigma_{\bu j}~=~g_{ij}+i\epsilon_{ij}\gamma_5
-i\sigma_{ij}+{2\over s}g_{ij}\sigma_{\ast\bu},
\nonumber\\
&&\hspace{-1mm}
\sigma_{i j}\sigma_{\bu k}=-\sigma_{\bu k}\sigma_{i j}~=-ig_{ik}\sigma_{\bu j}+ig_{jk}\sigma_{\bu i},~~~
\sigma_{i j}\sigma_{\ast k}=-\sigma_{\ast k}\sigma_{i j}~=-ig_{ik}\sigma_{\ast j}+ig_{jk}\sigma_{\ast i}  
\nonumber\\
\label{sigmasigmas}
\end{eqnarray}
We need also the following formulas with $\sigma$-matrices in different matrix elements
\begin{eqnarray}
&&\hspace{-1mm}
\tigma_{\mu\nu}\otimes\tigma_{\alpha\beta}~
=~-\half(g_{\mu\alpha}g_{\nu\beta}-g_{\nu\alpha}g_{\mu\beta})\sigma_{\xi\eta}\otimes\sigma^{\xi\eta}
\nonumber\\
&&\hspace{-1mm}
+~g_{\mu\alpha}\sigma_{\beta\xi}\otimes\sigma_\nu^{~\xi}-g_{\nu\alpha}\sigma_{\beta\xi}\otimes\sigma_\mu^{~\xi}-g_{\mu\beta}\sigma_{\alpha\xi}\otimes\sigma_\nu^{~\xi}+g_{\nu\beta}\sigma_{\alpha\xi}\otimes\sigma_\mu^{~\xi}
-\sigma_{\alpha\beta}\otimes\sigma_{\mu\nu}
\label{tigmi}
\end{eqnarray}
and 
\begin{eqnarray}
&&\hspace{-1mm}
\tigma_{\mu\xi}\otimes\tigma_\nu^{~\xi}~=~-{g_{\mu\nu}\over 2}\sigma_{\xi\eta}\otimes\sigma^{\xi\eta}+\sigma_{\nu\xi}\otimes\sigma_\mu^{~\xi}
,~~~~\sigma_{\xi\eta}\otimes\tigma^{\xi\eta}~=~\tigma_{\xi\eta}\otimes\sigma^{\xi\eta}
\label{formulaxz}\\
&&\hspace{-1mm}
\sigma_{\mu\xi}\gamma_5\otimes\sigma_\nu^{~\xi}\gamma_5+\mu\leftrightarrow\nu-{g_{\mu\nu}\over 2}\sigma_{\xi\eta}\gamma_5\otimes\sigma^{\xi\eta}\gamma_5
~=~-[\sigma_{\mu\xi}\otimes\sigma_\nu^{~\xi}+\mu\leftrightarrow\nu-{g_{\mu\nu}\over 2}\sigma_{\xi\eta}\otimes\sigma^{\xi\eta}]
\nonumber
\end{eqnarray}
\begin{eqnarray}
&&\hspace{-1mm}
\sigma_\ast^{~k}\otimes\gamma_i\sigma_{\bu k}\gamma_j~=~\hatp_2\gamma^k\otimes\notp_1\gamma_i\gamma_k\gamma_j
~=~\hatp_2\gamma^k\otimes\notp_1(g_{ik}\gamma_j+g_{jk}\gamma_i-g_{ij}\gamma_k)
\nonumber\\
&&\hspace{-1mm}
=~\hatp_2(g_{ik}\gamma_j+g_{jk}\gamma_i-g_{ij}\gamma_k)\otimes \notp_1\gamma^k~=~(\gamma_j\sigma_\ast^{~k}\gamma_i)\otimes\sigma_{\bu k}
\label{sisigaga}
\end{eqnarray}
We will need also
\begin{eqnarray}
&&\hspace{-1mm}
\notp_2\otimes\gamma_i\notp_1\gamma_j+\notp_2\gamma_5\otimes\gamma_i\notp_1\gamma_j\gamma_5
~=~\gamma_j\notp_2\gamma_i\otimes\notp_1+\gamma_j\notp_2\gamma_i\gamma_5\otimes\notp_1\gamma_5
\label{flagamma}
\end{eqnarray}
\subsubsection{Formulas with $\gamma$-matrices and one gluon field}
In the gauge $A_\bu=0$ the field $A_i$ can be represented as 
\begin{equation}
A_i(x_\bu,x_\perp)~=~{2\over s}\!\int_{-\infty}^{x_\bu}\!dx'_\bu~A_{\ast i}(x'_\bu,x_\perp)
\label{ai}
\end{equation}
(see eq. (\ref{AfromF})). We define ``dual'' fields by
\begin{equation}
\tilde{A}_i(x_\bu,x_\perp)~=~{2\over s}\!\int_{-\infty}^{x_\bu}\!dx'_\bu~\tilde{A}_{\ast i}(x'_\bu,x_\perp),
~~~\tilde{B}_i(x_\ast,x_\perp)~=~{2\over s}\!\int_{-\infty}^{x_\ast}\!dx'_\ast~\tilde{B}_{\bu i}(x'_\ast,x_\perp),
\label{abefildz}
\end{equation}
 where $\tilF_{\mu\nu}=\half\epsilon_{\mu\nu\lambda\rho}F^{\lambda\rho}$ as usual. 
With this definition we have $\tilA_i=-\epsilon_{ij}A^j$  and $\tilB_i=\epsilon_{ij}B^j$ so 
\begin{equation}
\notp_2\brA_i~=~-\notA\notp_2\gamma_i,~~~~\brA_i\notp_2~=~-\gamma_i\notp_2\notA,~~~~
\notp_1\breB_i~=~-\notB\notp_1\gamma_i,~~~~\breB_i\notp_1~=~-\gamma_i\notp_1\notB
\label{glavla}
\end{equation}
where
\begin{equation}
\brA_i\equiv A_i-i\tilA_i\gamma_5,~~~\breB_i\equiv B_i-i\tilB_i\gamma_5
\end{equation}
We also use
\begin{eqnarray}
&&\hspace{-1mm}
A^i\notp_2\otimes\gamma_n\notp_1\gamma_i+~A^i\notp_2\gamma_5\otimes\gamma_n\notp_1\gamma_i\gamma_5
~=~-\notp_2\brA_n\otimes \notp_1-\notp_2\brA_n\gamma_5\otimes \notp_1\gamma_5
\nonumber\\
&&\hspace{-1mm}
A^i\notp_2\otimes\gamma_i\notp_1\gamma_n+~A^i\notp_2\gamma_5\otimes\gamma_i\notp_1\gamma_n\gamma_5
~=~-\brA_n\notp_2\otimes \notp_1-\brA_n\notp_2\gamma_5\otimes \notp_1\gamma_5
\nonumber\\
&&\hspace{-1mm}
\gamma_n\slashed{p}_2\gamma^i\otimes\slashed{p}_1 B_i
~+~ \gamma_n\slashed{p}_2\gamma^i\gamma_5 \otimes\slashed{p}_1\gamma_5 B_i
~=~-\slashed{p}_2\otimes \slashed{p}_1\breB_n -
\slashed{p}_2 \gamma_5\otimes \slashed{p}_1\breB_n\gamma_5
\nonumber\\
&&\hspace{-1mm}
\gamma^i\slashed{p}_2\gamma_n \otimes\slashed{p}_1B_i
~+~\gamma^i \slashed{p}_2\gamma_n\gamma_5 \otimes\slashed{p}_1\gamma_5 B_i
~=~-\slashed{p}_2\otimes \breB_n\slashed{p}_1 -
\slashed{p}_2 \gamma_5\otimes \breB_n\slashed{p}_1\gamma_5
\label{gammas1fild}
\end{eqnarray}
and
\begin{eqnarray}
&&\hspace{-1mm}
{2\over s}\big[\notp_1\notp_2\gamma_i\otimes B^i\gamma_n
+\notp_1\notp_2\gamma_i\gamma_5\otimes B^i\gamma_n\gamma_5\big]
~=~\gamma_i\otimes\gamma_n\breB_i+\gamma_i\gamma_5\otimes\gamma_n\breB_i\gamma_5
\nonumber\\
&&\hspace{-1mm}
{2\over s}\big[\gamma_i\notp_2\notp_1\otimes B^i\gamma_n
+\gamma_i\notp_2\notp_1\gamma_5\otimes B^i\gamma_n\gamma_5\big]
~=~\gamma_i\otimes\breB_i\gamma_n+\gamma_i\gamma_5\otimes\breB_i\gamma_n\gamma_5
\nonumber
\end{eqnarray}

\subsubsection{Formulas with $\gamma$-matrices and two gluon fields}

With definition (\ref{abefildz}), we have the following formulas
\begin{eqnarray}
&&\hspace{-1mm}
A_i\otimes \tilB_j=g_{ij}\tilA_k\otimes B^k-\tilA_j\otimes B_i,~~~\tilde{A}_i\otimes B_j=g_{ij}A_k\otimes \tilB^k-A_j\otimes \tilB_i
\label{abeznaki}\\
&&\hspace{-1mm}
\tilde{A}_i\otimes \tilB_j=-g_{ij}A_k\otimes B^k+A_j\otimes B_i, ~~~\Rightarrow~~~\tilde{A}_i\otimes\tilde{B}^i=-A_i\otimes B^i,~~\tilde{A}_i\otimes B^i= A_i\otimes\tilde{B}^i
\nonumber
\end{eqnarray}
Using these formulas, after some algebra one obtains
\begin{eqnarray}
&&\hspace{-2mm}
\gamma_m\notp_2\gamma_jA^i\otimes\gamma_n\notp_1\gamma_iB^j
+\gamma_m\notp_2\gamma_jA^i\gamma_5\otimes\gamma_n\notp_1\gamma_iB^j\gamma_5
=\notp_2\brA_n\otimes\notp_1\breB_m+\notp_2\brA_n\gamma_5\otimes\notp_1\breB_m\gamma_5
\nonumber\\
&&\hspace{-2mm}
\gamma_j\notp_2\gamma_mA^i\otimes\gamma_n\notp_1\gamma_iB^j
+\gamma_j\notp_2\gamma_mA^i\gamma_5\otimes\gamma_n\notp_1\gamma_iB^j\gamma_5
=\notp_2\brA_n\otimes\breB_m\notp_1+\notp_2\brA_n\gamma_5\otimes\breB_m\notp_1\gamma_5
\nonumber\\
&&\hspace{-2mm}
\gamma_m\notp_2\gamma_jA^i\otimes\gamma_i\notp_1\gamma_nB^j
+\gamma_m\notp_2\gamma_jA^i\gamma_5\otimes\gamma_i\notp_1\gamma_nB^j\gamma_5
=\brA_n\notp_2\otimes\notp_1\breB_m+\brA_n\notp_2\gamma_5\otimes\notp_1\breB_m\gamma_5
\nonumber\\
&&\hspace{-2mm}
\gamma_j\notp_2\gamma_mA^i\otimes\gamma_i\notp_1\gamma_nB^j
+\gamma_j\notp_2\gamma_mA^i\gamma_5\otimes\gamma_i\notp_1\gamma_nB^j\gamma_5
=\brA_n\notp_2\otimes\breB_m\notp_1+\brA_n\notp_2\gamma_5\otimes\breB_m\notp_1\gamma_5
\nonumber\\
\label{gammas7}
\end{eqnarray}
and
\begin{eqnarray}
&&\hspace{-1mm}
\notp_2\brA_m\otimes\notp_1\breB_n+\notp_2\brA_n\gamma_5\otimes\notp_1\breB_m\gamma_5~=~g_{mn}\notp_2\brA_k\otimes\notp_1\breB^k
\nonumber\\
&&\hspace{-1mm}
\notp_2\brA_m\otimes\breB_n\notp_1+\notp_2\brA_n\gamma_5\otimes\gamma_5\breB_m\notp_1~=~g_{mn}\notp_2\brA_k\otimes\breB^k\notp_1
\nonumber\\
&&\hspace{-1mm}
\brA_m\notp_2\otimes\notp_1\breB_n+\gamma_5\brA_n\notp_2\otimes\notp_1\breB_m\gamma_5~=~g_{mn}\brA_k\notp_2\otimes\notp_1\breB^k
\nonumber\\
&&\hspace{-1mm}
\brA_m\notp_2\otimes\breB_n\notp_1+\gamma_5\brA_n\notp_2\otimes\gamma_5\breB_m\notp_1~=~g_{mn}\brA_k\notp_2\otimes\breB^k\notp_1
\label{gammas6}
\end{eqnarray}
The corollary of Eq. (\ref{gammas6}) is
\begin{eqnarray}
&&\hspace{-11mm}
\notp_2\brA_k\gamma_5\otimes\notp_1\breB^k\gamma_5~=~\notp_2\brA_k\otimes\notp_1\breB^k,~~~~~~~~
\notp_2\brA_k\gamma_5\otimes\gamma_5\breB^k\notp_1~=~\notp_2\brA_k\otimes\breB^k\notp_1
\nonumber\\
&&\hspace{-11mm}
\gamma_5\brA_k\notp_2\otimes\notp_1\breB^k\gamma_5~=~\brA_k\notp_2\otimes\notp_1\breB^k,~~~~~~~~
\gamma_5\brA_k\notp_2\otimes\gamma_5\breB^k\notp_1~=~\brA_k\notp_2\otimes\breB^k\notp_1
\label{gammas6a}
\end{eqnarray}

From Eqs. (\ref{gammas7}) and (\ref{gammas6}) one easily obtains
\begin{equation}
\gamma_m\notp_2\gamma_jA^i\otimes\gamma_n\notp_1\gamma_iB^j
+\gamma_m\notp_2\gamma_jA^i\gamma_5\otimes\gamma_n\notp_1\gamma_iB^j\gamma_5
~+~m\leftrightarrow n~=~2g_{mn}\notp_2\brA_k\otimes\notp_1\breB^k
\label{formula9}
\end{equation}
and
\begin{eqnarray}
&&\hspace{-1mm}
\gamma_m\notp_2\gamma_jA^i\otimes\gamma_n\notp_1\gamma_iB^j
+\gamma_m\notp_2\gamma_jA^i\gamma_5\otimes\gamma_n\notp_1\gamma_iB^j\gamma_5
~-~m\leftrightarrow n~
\nonumber\\
&&\hspace{-1mm}
=~2\notp_2\brA_n\otimes\notp_1\breB_m~-~m\leftrightarrow n,
\nonumber\\
&&\hspace{-1mm}
\gamma_j\notp_2\gamma_mA^i\otimes\gamma_i\notp_1\gamma_nB^j
+\gamma_j\notp_2\gamma_mA^i\gamma_5\otimes\gamma_i\notp_1\gamma_nB^j\gamma_5
~-~m\leftrightarrow n~
\nonumber\\
&&\hspace{-1mm}
=~2\brA_n\notp_2\otimes\breB_m\notp_1~-~m\leftrightarrow n
\label{formula9a}
\end{eqnarray}
We need also formulas
\begin{eqnarray}
&&\hspace{-1mm}
{4\over s^2}A^i\notp_1\notp_2\gamma_j\otimes B^j\notp_1\notp_2\gamma_i
\nonumber\\
&&\hspace{11mm}
=~
A^i\gamma_j\otimes B^j\gamma_i-iA^i\gamma_j\gamma_5\otimes \tilB^j\gamma_i
+i\tilA^i \gamma_j\otimes  B^j\gamma_i\gamma_5+\tilA^i\gamma_j\gamma_5\otimes \tilB^j\gamma_i\gamma_5,
\nonumber\\
&&\hspace{-1mm}
{4\over s^2}\big(A^i\notp_1\notp_2\gamma_j\otimes B^j\notp_1\notp_2\gamma_i
+A^i\notp_1\notp_2\gamma_j\gamma_5\otimes B^j\notp_1\notp_2\gamma_i\gamma_5\big)
\nonumber\\
&&\hspace{11mm}
=~\gamma^j\brA_i\otimes\gamma^i\breB_j+\gamma^j\brA_i\gamma_5\otimes\gamma^i\breB_j\gamma_5,
\nonumber\\
&&\hspace{-1mm}
\gamma_i\brA_j\gamma_5\otimes\gamma_j\brA_i\gamma_5~=~\gamma_i\brA_j\otimes\gamma^i\breB^j-\gamma_i\brA^i\otimes\gamma_j\breB^j
\label{gammas10}
\end{eqnarray}
and
\begin{eqnarray}
&&\hspace{-1mm}
A_k\gamma_i\slashed{p}_2\gamma^j\otimes B_j\gamma^i\slashed{p}_1\gamma^k~=~
\slashed{p}_2\brA_i\otimes \slashed{p}_1\breB^i
~=~\notA\notp_2\gamma_i\otimes\notB\notp_1\gamma^i,
\nonumber\\
&&\hspace{-1mm}
A_k\gamma^j\slashed{p}_2\gamma_i\otimes B_j\gamma^k\slashed{p}_1\gamma^i~=~
\brA_i\slashed{p}_2\otimes \breB_i\slashed{p}_1
~=~\gamma_i\notp_2\notA\otimes\gamma^i\notp_1\notB,
\nonumber\\
&&\hspace{-1mm}
A_k\gamma_i\slashed{p}_2\gamma_j\otimes B^j\gamma^k\slashed{p}_1\gamma^i
~=~\slashed{p}_2\brA_i\otimes\breB^i\slashed{p}_1
~=~~\notA\notp_2\gamma_i\otimes\gamma^i\notp_1\notB,
\nonumber\\
&&\hspace{-1mm}
A_k\gamma_j\slashed{p}_2\gamma_i\otimes B^j\gamma^i\slashed{p}_1\gamma^k
~=~\brA_i\slashed{p}_2\otimes\slashed{p}_1\breB^i
~=~\gamma_i\notp_2\notA\otimes\notB\notp_1\gamma^i,
\label{gammas11}
\end{eqnarray}
\begin{eqnarray}
&&\hspace{-1mm}
A^k\gamma_m\slashed{p}_2\gamma_j\otimes B^j\gamma_n\slashed{p}_1\gamma_k +m\leftrightarrow n
-g_{mn}A^k\gamma_i\slashed{p}_2\gamma_j\otimes B^j\gamma^i\slashed{p}_1\gamma_k
\nonumber\\
&&\hspace{-1mm}
=~\brA_m\notp_2\otimes\breB_n\notp_1+m\leftrightarrow n-g_{mn}\brA_k\notp_2\otimes\breB^k\notp_1,
\nonumber\\
&&\hspace{-1mm}
A^k\gamma_j\slashed{p}_2\gamma_m\otimes B^j\gamma_k\slashed{p}_1\gamma_n+m\leftrightarrow n
-g_{mn}A^k\gamma_j\slashed{p}_2\gamma_i\otimes B^j\gamma_k\slashed{p}_1\gamma^i
\nonumber\\
&&\hspace{-1mm}
=~\notp_2\brA_m\otimes\notp_1\breB_n+m\leftrightarrow n-g_{mn}\notp_2\brA_k\otimes\notp_1\breB^k,
\nonumber\\
&&\hspace{-1mm}
A^k\gamma_m\slashed{p}_2\gamma_j\otimes B^j\gamma_k\slashed{p}_1\gamma_n+m\leftrightarrow n
-g_{mn}A^k\gamma_j\slashed{p}_2\gamma_i\otimes B^j\gamma_k\slashed{p}_1\gamma^i
\nonumber\\
&&\hspace{-1mm}
=~\brA_m\notp_2\otimes\notp_1\breB_n+m\leftrightarrow n-g_{mn}\brA_k\notp_2\otimes\notp_1\breB^k,
\nonumber\\
&&\hspace{-1mm}
A^k\gamma_j\slashed{p}_2\gamma_m\otimes B^j\gamma_n\slashed{p}_1\gamma_k+m\leftrightarrow n
-g_{mn}A^k\gamma_j\slashed{p}_2\gamma_i\otimes B^j\gamma_k\slashed{p}_1\gamma^i
~=~
\nonumber\\
&&\hspace{-1mm}
=~\notp_2\brA_m\otimes\breB_n\notp_1+m\leftrightarrow n-g_{mn}\notp_2\brA_k\otimes\breB^k\notp_1,
\label{formulas67}
\end{eqnarray}
\begin{eqnarray}
&&\hspace{-1mm}
{2\over s}\big[A_i\notp_1\notp_2\gamma^j\otimes B_j\gamma_n\notp_1\gamma^i
+A_i\notp_1\notp_2\gamma^j\gamma_5\otimes B_j\gamma_{\nu_\perp}\notp_1\gamma^i\gamma_5\big]
\label{gammas16}\\
&&\hspace{-1mm}
=~-\gamma_i\brA_n\otimes \notp_1\breB^i - \gamma_i\brA_n\gamma_5\otimes \notp_1\breB^i\gamma_5
~=~\gamma_i\brA_n\otimes\notB\notp_1\gamma^i+\gamma_i\brA_n\gamma_5\otimes\notB\notp_1\gamma^i\gamma_5,
\nonumber\\
&&\hspace{-1mm} 
{2\over s}\big[A_i\gamma_n\notp_2\gamma^j\otimes B_j\notp_2\notp_1\gamma^i
+A_i\gamma_n\notp_2\gamma^j\gamma_5\otimes B_j\notp_2\notp_1\gamma^i\gamma_5\big]
\nonumber\\
&&\hspace{-1mm} 
=-\notp_2\brA_i\otimes\gamma^i\breB_n-\notp_2\brA_i\gamma_5\otimes\gamma^i\breB_n\gamma_5 
~=~\notA\notp_2\gamma_i\otimes\gamma^i\breB_n+\notA\notp_2\gamma_i\gamma_5\otimes\gamma^i\breB_n\gamma_5. 
\nonumber
\end{eqnarray}
%

 \subsection{Parametrization of leading-twist matrix elements \label{sec:paramlt}}
 Let us  first consider matrix elements of operators without $\gamma_5$. The standard parametrization of quark TMDs reads
 (see e.g. Ref. \cite{Arnold:2008kf}))
\begin{eqnarray}
&&\hspace{-1mm}
{1\over 16\pi^3}\!\int\!dx_\bu d^2x_\perp~e^{-i\alpha x_\bu+i(k,x)_\perp}
~\langle A|\bsi_f(x_\bu,x_\perp)\gamma^\mu\psi_f(0)|A\rangle
\label{Amael}\\
&&\hspace{27mm}
=~p_1^\mu f_1^f(\alpha,k_\perp)
+k_\perp^\mu f_\perp^f(\alpha,k_\perp)+p_2^\mu{2m^2_N\over s}f_3^f(\alpha,k_\perp),
\nonumber\\
&&\hspace{-1mm}
{1\over 16\pi^3}\!\int\!dx_\bu d^2x_\perp~e^{-i\alpha x_\bu+i(k,x)_\perp}
~\langle A|\bsi_f(x_\bu,x_\perp)\psi_f(0)|A\rangle
~=~m_Ne^f(\alpha,k_\perp)
\nonumber
\end{eqnarray}
for quark distributions in the projectile and 
\begin{eqnarray}
&&\hspace{-1mm}
{1\over 16\pi^3}\!\int\!dx_\bu d^2x_\perp~e^{-i\alpha x_\bu+i(k,x)_\perp}
~\langle A|\bsi_f(0)\gamma^\mu\psi_f(x_\bu,x_\perp)|A\rangle
\label{baramael}\\
&&\hspace{27mm}
=~-p_1^\mu \barf_1^f(\alpha,k_\perp)
-k_\perp^\mu\barf _\perp^f(\alpha,k_\perp)-p_2^\mu{2m^2_N\over s}\barf_3^f(\alpha,k_\perp),
\nonumber\\
&&\hspace{-1mm}
{1\over 16\pi^3}\!\int\!dx_\bu d^2x_\perp~e^{-i\alpha x_\bu+i(k,x)_\perp}
~\langle A|\bsi_f(0)\psi_f(x_\bu,x_\perp)|A\rangle
~=~
m_N\bare^f(\alpha,k_\perp)
\nonumber
\end{eqnarray}
for the antiquark distributions. 
\footnote{In an arbitrary gauge, there are gauge links to $-\infty$ as displayed in  eq. (\ref{gaugelinks}).}

The corresponding matrix elements for the target are obtained by trivial replacements $p_1\leftrightarrow p_2$, $x_\bu\leftrightarrow x_\ast$
and $\alpha\leftrightarrow\beta$:
\begin{eqnarray}
&&\hspace{-1mm}
{1\over 16\pi^3}\!\int\!dx_\ast d^2x_\perp~e^{-i\beta x_\ast+i(k,x)_\perp}
~\langle B|\bsi_f(x_\ast,x_\perp)\gamma^\mu\psi_f(0)|B\rangle
\label{Bmael}\\
&&\hspace{27mm}
=~p_2^\mu f_1^f(\beta,k_\perp)+k_\perp^\mu f_\perp^f(\beta,k_\perp)
+p_1^\mu{2m^2_N\over s}f_3^f(\beta,k_\perp),
\nonumber\\
&&\hspace{-1mm}
{1\over 16\pi^3}\!\int\!dx_\ast d^2x_\perp~e^{-i\beta x_\ast+i(k,x)_\perp}
~\langle B|\bsi_f(x_\ast,x_\perp)\psi_f(0)|B\rangle
~=~m_Ne^f(\beta,k_\perp),
\nonumber
\end{eqnarray}
and
\begin{eqnarray}
&&\hspace{-1mm}
{1\over 16\pi^3}\!\int\!dx_\ast d^2x_\perp~e^{-i\beta x_\ast+i(k,x)_\perp}
~\langle B|\bsi_f(0)\gamma^\mu\psi_f(x_\ast,x_\perp)|B\rangle
\label{barbmael}\\
&&\hspace{27mm}
=~-p_2^\mu \barf_1^f(\beta,k_\perp)
-k_\perp^\mu\barf _\perp^f(\beta,k_\perp)-p_1^\mu{2m^2_N\over s}\barf_3^f(\beta,k_\perp),
\nonumber\\
&&\hspace{-1mm}
{1\over 16\pi^3}\!\int\!dx_\ast d^2x_\perp~e^{-i\beta x_\ast+i(k,x)_\perp}
~\langle B|\bsi_f(0)\psi_f(x_\ast,x_\perp)|B\rangle
~=~
m_N\bare^f(\beta,k_\perp).
\nonumber
\end{eqnarray}

Matrix elements of operators with $\gamma_5$ are parametrized as follows: 
\begin{eqnarray}
&&\hspace{-1mm}
{1\over 16\pi^3}\!\int\!dx_\bu d^2x_\perp~e^{-i\alpha x_\bu+i(k,x)_\perp}
~\langle A|\bsi_f(x_\bu,x_\perp)\gamma^\mu\gamma_5\psi_f(0)|A\rangle
~=~-i\epsilon_{\mu_\perp i}k^ig^\perp_f(\alpha,k_\perp),
\nonumber\\
&&\hspace{-1mm}
{1\over 16\pi^3}\!\int\!dx_\bu d^2x_\perp~e^{-i\alpha x_\bu+i(k,x)_\perp}
~\langle A|\bsi_f(0)\gamma^\mu\gamma_5\psi_f(x_\bu,x_\perp)|A\rangle
~=~-i\epsilon_{\mu_\perp i}k^i\barg^\perp_f(\alpha,k_\perp)
\nonumber\\
\label{mael5}
\end{eqnarray}
The corresponding matrix elements for the target are obtained by trivial replacements $p_1\leftrightarrow p_2$, $x_\bu\leftrightarrow x_\ast$
and $\alpha\leftrightarrow\beta$ similarly to eq. (\ref{barbmael}).

The parametrization of time-odd Boer-Mulders TMDs are
\begin{eqnarray}
&&\hspace{-1mm}
{1\over 16\pi^3}\!\int\!dx_\bu d^2x_\perp~e^{-i\alpha x_\bu+i(k,x)_\perp}
~\langle A|\bsi_f(x_\bu,x_\perp)\sigma^{\mu \nu}\psi_f(0)|A\rangle
\nonumber\\
&&\hspace{11mm}
=~{1\over m_N}(k_\perp^\mu p_1^\nu -\mu\leftrightarrow\nu)h_{1f}^\perp(\alpha,k_\perp)
+{2m_N\over s}(p_1^\mu p_2^\nu-\mu\leftrightarrow\nu)h_{f}(\alpha,k_\perp)
\nonumber\\
&&\hspace{33mm}
+~{2m_N\over s}(k_\perp^\mu p_2^\nu -\mu\leftrightarrow\nu)h_{3f}^\perp(\alpha,k_\perp),
\nonumber\\
&&\hspace{-1mm}
{1\over 16\pi^3}\!\int\!dx_\bu d^2x_\perp~e^{-i\alpha x_\bu+i(k,x)_\perp}
~\langle A|\bsi_f(0)\sigma^{\mu \nu}\psi_f(x_\bu,x_\perp)|A\rangle
\nonumber\\
&&\hspace{11mm}
=~-{1\over m_N}(k_\perp^\mu p_1^\nu -\mu\leftrightarrow\nu)\barh_{1f}^\perp(\alpha,k_\perp)
-{2m_N\over s}(p_1^\mu p_2^\nu-\mu\leftrightarrow\nu)\barh_{f}(\alpha,k_\perp)
\nonumber\\
&&\hspace{33mm}
-~{2m_N\over s}(k_\perp^\mu p_2^\nu -\mu\leftrightarrow\nu)\barh_{3f}^\perp(\alpha,k_\perp)
\label{hmael}
\end{eqnarray}
and similarly for the target with usual replacements   $p_1\leftrightarrow p_2$, $x_\bu\leftrightarrow x_\ast$
and $\alpha\leftrightarrow\beta$.

Note that
the coefficients in front of $f_3$,  $g^\perp_f$, $h$ and $h_3^\perp$ in eqs. (\ref{Amael}),  (\ref{Bmael}), (\ref{mael5}),  and  (\ref{hmael}) 
 contain an extra ${1\over s}$ since $p_2^\mu$ enters only through the direction
of gauge link so the result should not depend on rescaling $p_2\rightarrow\lambda p_2$. For this reason,  these functions do not contribute to $W(q)$ in our approximation.

Last but not least, an important point in our analysis is that any $f(x,k_\perp)$ may have only logarithmic dependence on Bjorken $x$ but
not the power dependence $\sim{1\over x}$. 
Indeed, the low-$x$ behavior of TMDs is determined by pomeron exchange with the nucleon.  The interaction of TMD with BFKL pomeron 
is specified by so-called impact factor and it is easy to check that the impact factors for all leading-twist TMDs are similar and do not give 
extra ${1\over x}$ factors. The only ${1\over x}$ may had come from some unfortunate definition of TMD which includes factor $s$ artificially, but 
from power counting (\ref{pc}) we see that all definitions of leading-twist TMDs do not have such factors.  

\subsection{Matrix elements of quark-quark-gluon operators \label{sec:qqgparam}}

 In this section we will demonstrate that matrix elements of quark-antiquark-gluon operators 
 from section \ref{sec:tw3first} can be expressed in terms of leading-power
 matrix elements from section \ref{sec:paramlt}. 
 
 First, let us note that operators ${1\over\alpha}$ and ${1\over\beta}$ in Eqs. (\ref{3.25}) are
 replaced by $\pm{1\over\alpha_q}$ and $\pm{1\over\beta_q}$ in forward matrix elements. Indeed,
\begin{eqnarray}
&&\hspace{-1mm}
\!\int\!dx_\bu ~e^{-i\alpha_q x_\bu}\langle\bar\Phi(x_\bu,x_\perp)\Gamma{1\over \alpha+\ie}\psi(0)\rangle_A
\label{maelqg1}\\
&&\hspace{-1mm}
=~{1\over i}\!\int\!dx_\bu \!\int_{-\infty}^0 \!\!\!dx'_\bu~e^{-i\alpha_q x_\bu}\langle\bar\Phi(x_\bu,x_\perp)\Gamma\psi(x'_\bu,0_\perp)\rangle_A
=~{1\over\alpha_q}\!\int\!dx_\bu ~e^{-i\alpha x_\bu}\langle\bar\Phi(x_\bu,x_\perp)\Gamma\psi(0)\rangle_A
\nonumber
\end{eqnarray}
where $\bar\Phi(x_\bu,x_\perp)$ can be $\bsi(x_\bu,x_\perp)$ or  $\bsi(x_\bu,x_\perp)A_i(x_\bu,x_\perp)$ and $\Gamma$ can be any $\gamma$-matrix.
Similarly,
\begin{eqnarray}
&&\hspace{-1mm}
\!\int\!dx_\bu ~e^{-i\alpha_q x_\bu}\langle\big(\bar\psi{1\over \alpha-\ie}\big)(x_\bu,x_\perp)\Gamma\Phi(0)\rangle_A
=~{1\over\alpha_q}\!\int\!dx_\bu ~e^{-i\alpha x_\bu}\langle\bar\psi(x_\bu,x_\perp)\Gamma\Phi(0)\rangle_A
\label{maelqg2}\\
&&\hspace{-1mm}
\!\int\!dx_\bu ~e^{-i\alpha_q x_\bu}\langle\big(\bar\psi{1\over \alpha-\ie}\big)(x_\bu,x_\perp)\Gamma{1\over \alpha+\ie}\psi(0)\rangle_A
=~{1\over\alpha_q^2}\!\int\!dx_\bu ~e^{-i\alpha_q x_\bu}\langle\bar\psi(x_\bu,x_\perp)\Gamma\psi(0)\rangle_A
\nonumber
\end{eqnarray}
where $\Phi(x_\bu,x_\perp)$ can be $\psi(x_\bu,x_\perp)$ or  $A_i(x_\bu,x_\perp)\psi(x_\bu,x_\perp)$. We need also
\begin{eqnarray}
&&\hspace{-7mm}
\!\int\!dx_\bu ~e^{-i\alpha_q x_\bu}\langle\big(\bar\psi{1\over \alpha-\ie}\big)(0)\Gamma\Phi(x_\bu,x_\perp)\rangle_A
=~-{1\over\alpha_q}\!\int\!dx_\bu ~e^{-i\alpha x_\bu}\langle\bar\psi(0)\Gamma\Phi(x_\bu,x_\perp)\rangle_A
\nonumber\\
&&\hspace{-7mm}
\!\int\!dx_\bu ~e^{-i\alpha_q x_\bu}\langle\bar\Phi(0)\Gamma{1\over \alpha+\ie}\psi(x_\bu,x_\perp)\rangle_A
=~-{1\over\alpha_q}\!\int\!dx_\bu ~e^{-i\alpha x_\bu}\langle\bar\Phi(0)\Gamma\psi(x_\bu,x_\perp)\rangle_A
\label{maelqg3}
\end{eqnarray}
The corresponding formulas for target matrix elements are obtained by substitution $\alpha\leftrightarrow\beta$ (and $x_\bu\leftrightarrow x_\ast$).

Next, we will use QCD equation of motion to reduce quark-quark-gluon TMDs to leading-twist TMDs (see Ref. \cite{Mulders:1995dh}).
Let us start with matrix element 
\begin{eqnarray}
&&\hspace{-1mm}
\!\int\! dx_\bu dx_\perp~e^{-i\alpha_qx_\bu+i(k,x)_\perp}\langle A|\bar\psi(x_\bu,x_\perp)\slashed{p}_2\brA_i(x_\bu,x_\perp)\psi(0)|A\rangle
\label{tw3mael1}\\
&&\hspace{-1mm}
=~-\!\int\! dx_\bu dx_\perp~e^{-i\alpha_qx_\bu+i(k,x)_\perp}\langle A|\bar\psi(x_\bu,x_\perp)\notA(x_\bu,x_\perp)\slashed{p}_2\gamma_i\psi(0)|A\rangle
\nonumber\\
&&\hspace{-1mm}
=~\int\! dx_\bu dx_\perp~e^{-i\alpha_qx_\bu+i(k,x)_\perp}
\nonumber\\
&&\hspace{31mm}
\times~\big[\langle A|\bsi(x_\bu,x_\perp)\slashed{k}_\perp\slashed{p}_2\gamma_i\psi(0)|A\rangle
+i
\langle A|\bsi(x_\bu,x_\perp)\!\stackrel{\leftarrow}{\notD}_\perp\!\gamma_j\slashed{p}_2\gamma_i\psi(0)|A\rangle\big].
\nonumber
\end{eqnarray}
Using QCD equations of motion (\ref{YMs}) we can rewrite the r.h.s. of eq. (\ref{tw3mael1}) as
\begin{eqnarray}
&&\hspace{-2mm}
\int\! dx_\bu dx_\perp~e^{-i\alpha_qx_\bu+i(k,x)_\perp}\big[\langle A|\bsi(x_\bu,x_\perp)\slashed{k}_\perp\slashed{p}_2\gamma_i\psi(0)|A\rangle
+\alpha_q\langle A|\bsi(x_\bu,x_\perp)\slashed{p}_1\slashed{p}_2\gamma_i\psi(0)|A\rangle\big]
\nonumber\\
&&\hspace{-1mm}
=~\int\! dx_\bu dx_\perp~e^{-i\alpha_qx_\bu+i(k,x)_\perp}
\Big[
-k_i\langle A|\bsi(x_\bu,x_\perp)\slashed{p}_2\psi(0)|A\rangle
+~\alpha_q{s\over 2} \langle A|\bsi(x_\bu,x_\perp)\gamma_i\psi(0)|A\rangle
\nonumber\\
&&\hspace{6mm}
-~i\epsilon_{ij}k^j\langle A|\bsi(x_\bu,x_\perp)\slashed{p}_2\gamma_5\psi(0)|A\rangle
+i{s\over 2}\alpha\epsilon_{ij}
\langle A|\bsi(x_\bu,x_\perp)\gamma^j\gamma_5\psi(0)|A\rangle\Big]
\nonumber\\
&&\hspace{14mm}
=~-k_i8\pi^3sf_1(\alpha_q,k_\perp)+8\pi^3s\alpha_q k_i\big[ f_\perp(\alpha_q,k_\perp)+g^\perp(\alpha_q,k_\perp)\big],
\label{tw3mael2}
\end{eqnarray}
where we used parametrizations (\ref{Amael}) and (\ref{mael5}) for the leading power matrix elements. 

Now, the second term in eq. (\ref{tw3mael2}) contains extra $\alpha_q$ with respect to the first term
\footnote{As discussed in the end of Sect. \ref{sec:paramlt}, all leading-twist TMDs can have only logarithmic dependence on 
Bjorken $x$ (which is here either $\alpha_q$ for the projectile or $\beta_q$ for the target matrix elements). 
}
, so
 it should be neglected in our kinematical region $s\gg Q^2\gg q_\perp^2$  and we get 
\begin{eqnarray}
&&\hspace{-1mm}
{1\over 8\pi^3s}\!\int\! dx_\bu dx_\perp~e^{-i\alpha_qx_\bu+i(k,x)_\perp}\langle A|\bar\psi^f(x_\bu,x_\perp)\slashed{p}_2\brA_i(x_\bu,x_\perp)\psi^f(0)|A\rangle
\label{11.42}\\
&&\hspace{-1mm}
=~-{1\over 8\pi^3s}\!\int\! dx_\bu dx_\perp~e^{-i\alpha_qx_\bu+i(k,x)_\perp}\langle A|\bar\psi^f\notA(x_\bu,x_\perp)\slashed{p}_2\gamma_i\psi^f(0)|A\rangle
~=~-k_if_1^f(\alpha_q,k_\perp)
\nonumber
\end{eqnarray}
By complex conjugation
\begin{eqnarray}
&&\hspace{-1mm}
{1\over 8\pi^3s}\!\int\! dx_\perp dx_\bu~e^{-i\alpha_q x_\bu+i(k,x)_\perp}
\langle A|\bsi _f(x_\bu,x_\perp)\brA_i(0)\slashed{p}_2\psi_f(0)|A\rangle
\label{11.36}\\
&&\hspace{-1mm}
=~-{1\over 8\pi^3s}\!\int\! dx_\bu dx_\perp~e^{-i\alpha_qx_\bu+i(k,x)_\perp}\langle A|\bar\psi^f(x_\bu,x_\perp)\gamma_i\slashed{p}_2\notA\psi^f(0)|A\rangle
=~-k_if_{1f}(\alpha_q,k_\perp).
\nonumber
\end{eqnarray}

For the  corresponding antiquark distributions we get 
\begin{eqnarray}
&&\hspace{0mm}
{1\over 8\pi^3s}\!\int\! dx_\perp dx_\bu~e^{-i\alpha x_\bu+i(k,x)_\perp}
\langle A|\bsi_f(0)\brA_i(x_\bu,x_\perp)\slashed{p}_2\psi_f(x_\bu,x_\perp)|A\rangle
\nonumber\\
&&\hspace{5mm}
=~{1\over 8\pi^3s}\!\int\! dx_\bu dx_\perp e^{-i\alpha_qx_\bu+i(k,x)_\perp}
\Big[-\langle A|\bsi(0)\gamma_i\slashed{p}_2\slashed{k}_\perp\psi(x_\bu,x_\perp)|A\rangle
\nonumber\\
&&\hspace{11mm}
-~i\langle A|\bsi(0)\gamma_i\slashed{p}_2\notD_\perp\psi(x_\bu,x_\perp)|A\rangle\Big]
~=~-k_i\barf_{1f}(\alpha_q,k_\perp)
\label{9.25}
\end{eqnarray}
and
\begin{equation}
\hspace{0mm}
{1\over 8\pi^3s}\!\int\! dx_\perp dx_\bu~e^{-i\alpha_q x_\bu+i(k,x)_\perp}
\langle A|\bsi _f(0)\slashed{p}_2\brA_i(0)\psi_f(x_\bu,x_\perp)|A\rangle
~=~
-k_i\barf_{1f}(\alpha_q,k_\perp) .
\label{11.45}
\end{equation}

The corresponding target matrix elements are obtained by trivial replacements 
$x_\ast\leftrightarrow x_\bu$, $\alpha_q\leftrightarrow\beta_q$ and
$\slashed{p}_2\leftrightarrow\slashed{p}_1$.

Next, let us consider
\begin{eqnarray}
&&\hspace{-1mm}
{1\over 8\pi^3s}\!\int\! dx_\bu dx_\perp~e^{-i\alpha_qx_\bu+i(k,x)_\perp}
\langle A|\bsi(x_\bu,x_\perp)\slashed{p}_2\notA(x_\bu,x_\perp)\psi(0)|A\rangle
\label{eqm1}\\
&&\hspace{11mm}
=~{1\over 8\pi^3s}\!\int\! dx_\bu dx_\perp~e^{-i\alpha_qx_\bu+i(k,x)_\perp}
\nonumber\\
&&\hspace{22mm}
\times~\Big[\langle A|\bsi(x_\bu,x_\perp)\slashed{k}_\perp\slashed{p}_2\psi(0)|A\rangle
+i\langle A|\bsi(x_\bu,x_\perp)\stackrel{\leftarrow}{\notD}_\perp\slashed{p}_2\psi(0)|A\rangle\Big].
\nonumber
\end{eqnarray}
Using QCD equation of motion and parametrization (\ref{hmael}), one can rewrite the r.h.s. of this equation as
\begin{eqnarray}
&&\hspace{-3mm}
{1\over 8\pi^3s}\!\int\! dx_\bu dx_\perp~e^{-i\alpha_qx_\bu+i(k,x)_\perp}
\Big[\langle A|\bsi(x_\bu,x_\perp)\slashed{k}_\perp\slashed{p}_2\psi(0)|A\rangle
+\alpha_q\langle A|\bsi(x_\bu,x_\perp)\slashed{p}_1\slashed{p}_2\psi(0)|A\rangle\Big]
\nonumber\\
&&\hspace{-1mm}
=~i{k_\perp^2\over m_N}h_{1}^\perp(\alpha_q,k_\perp)+\alpha_q m_N\big[e(\alpha,k_\perp)+i h(\alpha,k_\perp)\big].
\label{eqm2}
\end{eqnarray}
Again, only the first term contributes in our kinematical region so we finally get
\begin{eqnarray}
&&\hspace{-2mm}
{1\over 8\pi^3s}\!\int\! dx_\bu dx_\perp~e^{-i\alpha_qx_\bu+i(k,x)_\perp}
\langle A|\bsi^f(x_\bu,x_\perp)\slashed{p}_2\notA(x_\bu,x_\perp)\psi^f(0)|A\rangle
~=~i{k_\perp^2\over m}h_{1f}^\perp(\alpha_q,k_\perp).
\nonumber\\
\label{11.41}
\end{eqnarray}
By complex conjugation we obtain
\begin{eqnarray}
&&\hspace{-1mm}
{1\over 8\pi^3s}\!\int\! dx_\bu dx_\perp~e^{-i\alpha_qx_\bu+i(k,x)_\perp}
\langle A|\bsi^f(x_\bu,x_\perp)\slashed{p}_2\notA(0)\psi^f(0)|A\rangle
~=~i{k_\perp^2\over m}h_{1f}^\perp(\alpha_q,k_\perp).
\nonumber\\
\label{11.42}
\end{eqnarray}
For corresponding antiquark distributions one gets in a similar way
\begin{eqnarray}
&&\hspace{-2mm}
{1\over 8\pi^3s}\!\int\! dx_\bu dx_\perp~e^{-i\alpha_qx_\bu+i(k,x)_\perp}
\langle A|\bsi^f(0)\slashed{p}_2\notA(x_\bu,x_\perp)\psi^f(x_\bu,x_\perp)|A\rangle
=~
i{k_\perp^2\over m}\barh_{1f}^\perp(\alpha_q,k_\perp),
\nonumber\\
&&\hspace{-2mm}
{1\over 8\pi^3s}\!\int\! dx_\bu dx_\perp~e^{-i\alpha_qx_\bu+i(k,x)_\perp}
\langle A|\bsi^f(0)\slashed{p}_2\notA(0)\psi^f(x_\bu,x_\perp)|A\rangle
=~
i{k_\perp^2\over m}\barh_{1f}^\perp(\alpha_q,k_\perp).
\nonumber\\
\label{11.43}
\end{eqnarray}
The target matrix elements are obtained by usual replacements 
$x_\ast\leftrightarrow x_\bu$, $\alpha_q\leftrightarrow\beta_q$ and
$\slashed{p}_2\leftrightarrow\slashed{p}_1$.

Finally, we need
\begin{eqnarray}
&&\hspace{-1mm}
{1\over 8\pi^3s}\!\int\! dx_\bu dx_\perp~e^{-i\alpha_qx_\bu+i(k,x)_\perp}
\langle A|\bsi(x_\bu,x_\perp)\notA(x_\bu,x_\perp)\slashed{p}_2\notA(0)\psi(0)|A\rangle
\nonumber\\
&&\hspace{1mm}
=~{1\over 8\pi^3s}\!\int\! dx_\bu dx_\perp~e^{-i\alpha_qx_\bu+i(k,x)_\perp}
\langle A|\bsi(x_\bu,x_\perp)\Big(\slashed{k}_\perp
+i\stackrel{\leftarrow}{\notD}\big)\slashed{p}_2
\big(\slashed{k}_\perp
-i{\notD}\big)\psi(0)|A\rangle
\nonumber\\
&&\hspace{22mm}
=~{k_\perp^2\over 16\pi^3}f_1(\alpha_q,k_\perp)~+~O(\alpha_q,\beta_q)
\label{AA1}
\end{eqnarray}
and similarly
\begin{eqnarray}
&&\hspace{-1mm}
{1\over 8\pi^3s}\!\int\! dx_\bu dx_\perp~e^{-i\alpha_qx_\bu+i(k,x)_\perp}
\langle A|\bsi(x_\bu,x_\perp)\notA(x_\bu,x_\perp)\sigma_{\ast i}\notA(0)\psi(0)|A\rangle
\nonumber\\
&&\hspace{1mm}
=~{1\over 8\pi^3s}\!\int\! dx_\bu dx_\perp~e^{-i\alpha_qx_\bu+i(k,x)_\perp}
\langle A|\bsi(x_\bu,x_\perp)\Big(\slashed{k}_\perp
+i\stackrel{\leftarrow}{\notD}\big)\sigma_{\ast i}
\big(\slashed{k}_\perp
-i{\notD}\big)\psi(0)|A\rangle
\nonumber\\
&&\hspace{1mm}
=~{1\over 16\pi^3}{k_ik_\perp^2\over m}h_1^\perp(\alpha_q,k_\perp)~+~O(\alpha_q,\beta_q)
\label{AA2}
\end{eqnarray}

For corresponding antiquark distributions we get
\begin{eqnarray}
&&\hspace{-1mm}
{1\over 8\pi^3s}\!\int\! dx_\bu dx_\perp~e^{-i\alpha_qx_\bu+i(k,x)_\perp}
\langle A|\bsi(0)\notA(0)\slashed{p}_2\notA(x_\bu,x_\perp)\psi(x_\bu,x_\perp)|A\rangle
\nonumber\\
&&\hspace{22mm}
=~-{k_\perp^2\over 16\pi^3}\barf_1(\alpha_q,k_\perp)~+~O(\alpha_q,\beta_q)
\nonumber\\
&&\hspace{-1mm}
{1\over 8\pi^3s}\!\int\! dx_\bu dx_\perp~e^{-i\alpha_qx_\bu+i(k,x)_\perp}
\langle A|\bsi(x_\bu,x_\perp)\notA(x_\bu,x_\perp)\sigma_{\ast i}\notA(0)\psi(0)|A\rangle
\nonumber\\
&&\hspace{22mm}
=~-{1\over 16\pi^3}{k_ik_\perp^2\over m}h_1^\perp(\alpha_q,k_\perp)~+~O(\alpha_q,\beta_q)
\end{eqnarray}

Also, as we saw in Sect. \ref{sec:v3}, at the leading order in $N_c$ there is one quark-antiquark-gluon 
operator that does not reduce to twist-2 distributions. It can be parametrized as follows (cf. Eq. (\ref{11.43}))
\begin{eqnarray}
&&\hspace{-1mm}
{1\over 16\pi^3}{2\over s}\!\int\!dx_\bu d^2x_\perp~e^{-i\alpha x_\bu+i(k,x)_\perp}
~\langle A|\bsi_f(x_\bu,x_\perp)\big[A_i(x)\sigma_{\ast j}-\half g_{ij}A^k\sigma_{\ast k}(x)\big]\psi_f(0)|A\rangle~
\nonumber\\
&&\hspace{33mm}
=~-(k_ik_j+\half g_{ij}k_\perp^2){1\over m}h_A^f(\alpha,k_\perp),
\nonumber\\
&&\hspace{-1mm}
{1\over 16\pi^3}{2\over s}\!\int\!dx_\bu d^2x_\perp~e^{-i\alpha x_\bu+i(k,x)_\perp}
~\langle A|\bsi_f(0)[A_i(0)\sigma_{\bu j}-\half g_{ij}A^k\sigma_{\bu k}(0)]\psi_f(x_\bu,x_\perp)|A\rangle~
\nonumber\\
&&\hspace{33mm}
=~-(k_ik_j+\half g_{ij}k_\perp^2){1\over m}\barh_A^f(\alpha,k_\perp)
\label{maelsa}
\end{eqnarray}
 and similarly for the target matrix elements.

\subsection{Parametrization of TMDs from section \ref{67lines} \label{67param}}
We  parametrize TMDs from section \ref{67lines} as follows
\begin{eqnarray}
&&\hspace{-1mm}
{1\over 8\pi^3s}\!\int\! d^2x_\perp dx_\bu~e^{-i\alpha x_\bu+i(k,x)_\perp}
\!\int_{-\infty}^{x_\bu}\! dx'_\bu\langle A|\bsi (x_\bu,x_\perp){2\slashed{p}_2\over s}
\big[F_{\ast i}(0)+ i\gamma_5\tilF_{\ast i}(0)\big]\psi(x'_\bu,x_\perp)|A\rangle
\nonumber\\
&&\hspace{-1mm}
=~k_ij_1(\alpha,k_\perp),
\nonumber\\
&&\hspace{-1mm}
{1\over 8\pi^3s}\!\int\! d^2x_\perp dx_\bu~e^{-i\alpha x_\bu+i(k,x)_\perp}
\!\int_{-\infty}^{x_\bu}\! dx'_\bu\langle A|\bsi (x_\bu,x_\perp){2\slashed{p}_2\over s}
\big[F_{\ast i}(0)- i\gamma_5\tilF_{\ast i}(0)\big]\psi(x'_\bu,x_\perp)|A\rangle
\nonumber\\
&&\hspace{-1mm}
=~k_ij_2(\alpha,k_\perp),
\nonumber\\
&&\hspace{-1mm}
{1\over 8\pi^3s}\!\int\! d^2x_\perp dx_\bu~e^{-i\alpha x_\bu+i(k,x)_\perp}
\!\int_{-\infty}^{x_\bu}\! dx'_\bu\langle A|\bsi (x'_\bu,x_\perp){2\slashed{p}_2\over s}
\big[F_{\ast i}(0)- i\gamma_5\tilF_{\ast i}(0)\big]\psi(x_\bu,x_\perp)|A\rangle
\nonumber\\
&&\hspace{-1mm}
=~k_i\barj_1(\alpha,k_\perp),
\nonumber\\
&&\hspace{-1mm}
{1\over 8\pi^3s}\!\int\! d^2x_\perp dx_\bu~e^{-i\alpha x_\bu+i(k,x)_\perp}
\!\int_{-\infty}^{x_\bu}\! dx'_\bu\langle A|\bsi (x'_\bu,x_\perp){2\slashed{p}_2\over s}
\big[F_{\ast i}(0)+i\gamma_5\tilF_{\ast i}(0)\big]\psi(x_\bu,x_\perp)|A\rangle
\nonumber\\
&&\hspace{-1mm}
=~k_i\barj_2(\alpha,k_\perp)
\label{paramj}
\end{eqnarray}
By complex conjugation we get
\begin{eqnarray}
&&\hspace{-1mm}
{1\over 8\pi^3s}\!\int\! d^2x_\perp dx_\bu~e^{-i\alpha x_\bu+i(k,x)_\perp}
\!\int_{-\infty}^0\! dx'_\bu\langle A|\bar\psi(x'_\bu,0_\perp){2\slashed{p}_2\over s}
[F_{\ast i}(x)-i\gamma_5\tilF_{\ast i}(x)]\psi(0)|A\rangle,
\nonumber\\
&&\hspace{-1mm}
=~k_i j_1^\star(\alpha,k_\perp),
\nonumber\\
&&\hspace{-1mm}
{1\over 8\pi^3s}\!\int\! d^2x_\perp dx_\bu~e^{-i\alpha x_\bu+i(k,x)_\perp}
\!\int_{-\infty}^0\! dx'_\bu\langle A|\bar\psi(x'_\bu,0_\perp){2\slashed{p}_2\over s}
\big[F_{\ast i}(x)+i\gamma_5\tilF_{\ast i}(x)\big]\psi(0)|A\rangle
\nonumber\\
&&\hspace{-1mm}
=~k_ij_2^\star(\alpha,k_\perp),
\nonumber\\
&&\hspace{-1mm}
{1\over 8\pi^3s}\!\int\! d^2x_\perp dx_\bu~e^{-i\alpha x_\bu+i(k,x)_\perp}
\!\int_{-\infty}^0\! dx'_\bu\langle A|\bar\psi(0){2\slashed{p}_2\over s}
[F_{\ast i}(x)+i\gamma_5\tilF_{\ast i}(x)]\psi(x'_\bu,0_\perp)|A\rangle
\nonumber\\
&&\hspace{-1mm}
=~k_i \barj_1^\star(\alpha,k_\perp),
\nonumber\\
&&\hspace{-1mm}
{1\over 8\pi^3s}\!\int\! d^2x_\perp dx_\bu~e^{-i\alpha x_\bu+i(k,x)_\perp}
\!\int_{-\infty}^0\! dx'_\bu\langle A|\bar\psi(0){2\slashed{p}_2\over s}
\big[F_{\ast i}(x)-i\gamma_5\tilF_{\ast i}(x)\big]\psi(x'_\bu,0_\perp)|A\rangle
\nonumber\\
&&\hspace{-1mm}
=~k_i\barj_2^\star(\alpha,k_\perp).
\label{paramjstar}
\end{eqnarray}
Note that unlike two-quark matrix elements,
quark-quark-gluon ones may have  imaginary parts.

Target matrix elements are obtained by usual substitutions 
$\alpha\leftrightarrow\beta$, $\slashed{p}_2\leftrightarrow\slashed{p}_1$, $x_\bu\leftrightarrow x_\ast$, and $\hatF_{\ast i}\leftrightarrow \hatF_{\bu i}$.

For completeness let us present the explicit form of the gauge links in an arbitrary gauge:
\begin{eqnarray}
&&\hspace{-3mm}
\bsi (x'_\bu,x_\perp)
F_{\ast i}(0)\psi(x_\bu,x_\perp)
~\rightarrow~\bsi (x'_\bu,x_\perp)[x'_\bu,-\infty_\bu]_x[x_\perp,0_\perp]_{-\infty_\bu}
\\
&&\hspace{40mm}
\times~[-\infty_\bu,0]_{0_\perp}F_{\ast i}(0)
[0,-\infty_\bu]_{0_\perp}[0_\perp,x_\perp]_{-\infty_\bu}[-\infty_\bu,x_\bu]_x\psi(x_\bu,x_\perp).
\nonumber
\end{eqnarray}

\subsection{Gluon power corrections from $J_{A}^\mu(x)J_{A\mu}(0)$ terms \label{gluterms}}
There is one more type of contributions proportional to the product of quark and gluon TMDs 
\begin{eqnarray}
&&\hspace{-1mm}
J_{A}^\mu(x)J_A^\nu(0)~=~
\nonumber\\
&&\hspace{-1mm}
=~\sum_{\rm flavors}\Big(\big[\Bxi_{1}(x)\gamma^\mu\psi_A(x)\big]\big[\bar\psi_A(0)\gamma^\nu\Xi_{1}(0)\big]
+\big[\bar\psi_A(x)\gamma^\mu\Xi_{1}(x)\big]\big[\Bxi_{1}(0)\gamma^\nu\psi_A(0)\big]
\nonumber\\
&&\hspace{-1mm}
+~\big[\Bxi_{1}(x)\gamma^\mu\psi_A(x)\big]\big[\Bxi_{1}(0)\gamma^\nu\psi_A(0)\big]
+\big[\bar\psi_A(x)\gamma_\nu\Xi_{1}(x)\big]\big[\bar\psi_A(0)\gamma^\nu\Xi_{1}(0)\big]\Big),
\label{5.15}
\end{eqnarray}
where we neglected terms which cannot contribute to $W$ due to the reason discussed after eq. (\ref{xxvanish}), i.e.
that one hadron (``A'' or ``B'') cannot produce the DY pair on its own. 

Let us  consider the first term in the r.h.s of this equation
\begin{eqnarray}
&&\hspace{-1mm}
\cheW_{\mu\nu}(x)~=~\langle A,B|\big[\Bxi_{1}(x)\gamma_\mu\psi_A(x)\big]\big[\bar\psi_A(0)\gamma_\nu\Xi_{1}(0)\big]|A,B\rangle
\label{5.16}\\
&&\hspace{-1mm}
=~-{g^2\over s(N_c^2-1)}
\langle \big(\bsi{1\over\alpha}\big)(x)\gamma_i\slashed{p}_2\gamma_\mu\psi(x)
\bsi(0)\gamma_\nu\slashed{p}_2\gamma_j{1\over\alpha}\psi(0)\rangle_A
\langle A^{ai}(x)A^{aj}(0)\rangle_B
\nonumber
\end{eqnarray}
To estimate the magnitude of this contribution, first note that
\begin{eqnarray}
&&\hspace{-1mm}
\!\int\! dx_\ast ~e^{-i\beta_qx_\ast}~\langle B|A^a_i(x)A^{a}_j(0)|B\rangle
\label{gluonvklad}\\
&&\hspace{11mm}
=~
{4\over s^2}\int\! dx_\ast ~e^{-i\beta_qx_\ast}\!\int_{-\infty}^{x_\ast}\! dx'_\ast \!\int_{-\infty}^{0}\! dx''_\ast~
\langle B|F^a_{\bu i}(x'_\ast,x_\perp)F^a_{\bu j}(x''_\ast,0_\perp)|B\rangle
\nonumber\\
&&\hspace{11mm}
=~{4\over\beta_q^2s^2}\!\int\! dx_\ast ~e^{-i\beta_qx_\ast}\langle B|F^a_{\bu i}(x_\ast,x_\perp)F^a_{\bu j}(0)|B\rangle
\nonumber\\
&&\hspace{11mm}
=~-{1\over\beta_q}8\pi^2\alpha_s\Big[\cald_g(\beta_q,x_\perp)+{1\over m^2}(2\partial_i\partial_j+g_{ij}\partial_\perp^2\calh_g(\beta_q,x_\perp)\big]
\nonumber
\end{eqnarray}
where we used parametrization (3.26) from Ref. \cite{Balitsky:2017flc}. Since the gluon TMDs $\cald_g(x_B,x_\perp)$ and $\calh_g(\beta_q,x_\perp)$ behave
only logarithmically as $x_B\rightarrow 0$ \cite{Balitsky:2016dgz}, the contribution of eq. (\ref{5.16}) to $W(q)$ is of order of 
${m_\perp^2\over\beta_q s}\ll{m_\perp^2\over Q^2}$. (As discussed in Ref. \cite{Balitsky:2017flc},  the projectile TMD in the r.h.s. of eq. (\ref{5.16}) does {\it not} give ${1\over\alpha_q}$ after Fourier transformation).  Also, this contribution is $\sim{1\over N_c}$ with respect to our leading terms.

Similarly, all other terms in eq. (\ref{5.15}) are either 
${m_\perp^2\over\beta_q s}$ or ${m_\perp^2\over\alpha_q s}$ times ${1\over N_c}$ so they can be neglected.
\footnote{
It is worth mentioning that if the DY pair is produced in the region of rapidity close to the projectile,
the contribution (\ref{gluonvklad}) may be the most important since gluon parton densities  at small $x_B$ 
are larger than the quark ones.}

\bibliography{fact1}
\bibliographystyle{JHEP}

\end{document}